\PassOptionsToPackage{unicode}{hyperref}
\PassOptionsToPackage{hyphens}{url}
\PassOptionsToPackage{dvipsnames,svgnames,x11names}{xcolor}
\documentclass[
  12pt]{article}

\usepackage{amsmath,amssymb,amsthm}
\usepackage{iftex}
\ifPDFTeX
  \usepackage[T1]{fontenc}
  \usepackage[utf8]{inputenc}
  \usepackage{textcomp} 
\else 
  \usepackage{unicode-math}
  \defaultfontfeatures{Scale=MatchLowercase}
  \defaultfontfeatures[\rmfamily]{Ligatures=TeX,Scale=1}
\fi
\usepackage{lmodern}
\ifPDFTeX\else  
\fi
\IfFileExists{upquote.sty}{\usepackage{upquote}}{}
\IfFileExists{microtype.sty}{
  \usepackage[]{microtype}
  \UseMicrotypeSet[protrusion]{basicmath} 
}{}
\makeatletter
\@ifundefined{KOMAClassName}{
  \IfFileExists{parskip.sty}{%
    \usepackage{parskip}
  }{
    \setlength{\parindent}{0pt}
    \setlength{\parskip}{6pt plus 2pt minus 1pt}}
}{
  \KOMAoptions{parskip=half}}
\makeatother
\usepackage{xcolor}
\usepackage[table]{xcolor}
\usepackage{booktabs}
\usepackage{array}

\definecolor{caudate}{HTML}{0072B2}
\definecolor{putamen}{HTML}{E69F00}
\definecolor{tms}{HTML}{009E73}
\definecolor{stroopword}{HTML}{CC79A7}
\definecolor{cuhdr}{HTML}{D55E00}
\definecolor{stroopcolor}{HTML}{999999}

\newcommand{\caudatecell}[1]{\cellcolor{caudate!18}{\bfseries #1}}
\newcommand{\putamencell}[1]{\cellcolor{putamen!18}{\bfseries #1}}
\newcommand{\tmscell}[1]{\cellcolor{tms!18}{\bfseries #1}}
\newcommand{\stroopwordcell}[1]{\cellcolor{stroopword!18}{\bfseries #1}}
\newcommand{\cuhdrcell}[1]{\cellcolor{cuhdr!18}{\bfseries #1}}
\newcommand{\stroopcolorcell}[1]{\cellcolor{stroopcolor!18}{\bfseries #1}}
\makeatletter
\ifx\paragraph\undefined\else
  \let\oldparagraph\paragraph
  \renewcommand{\paragraph}{
    \@ifstar
      \xxxParagraphStar
      \xxxParagraphNoStar
  }
  \newcommand{\xxxParagraphStar}[1]{\oldparagraph*{#1}\mbox{}}
  \newcommand{\xxxParagraphNoStar}[1]{\oldparagraph{#1}\mbox{}}
\fi
\ifx\subparagraph\undefined\else
  \let\oldsubparagraph\subparagraph
  \renewcommand{\subparagraph}{
    \@ifstar
      \xxxSubParagraphStar
      \xxxSubParagraphNoStar
  }
  \newcommand{\xxxSubParagraphStar}[1]{\oldsubparagraph*{#1}\mbox{}}
  \newcommand{\xxxSubParagraphNoStar}[1]{\oldsubparagraph{#1}\mbox{}}
\fi
\makeatother

\usepackage{longtable,booktabs,array}
\usepackage{calc} 
\usepackage{etoolbox}
\makeatletter
\patchcmd\longtable{\par}{\if@noskipsec\mbox{}\fi\par}{}{}
\makeatother
\IfFileExists{footnotehyper.sty}{\usepackage{footnotehyper}}{\usepackage{footnote}}
\makesavenoteenv{longtable}
\usepackage{graphicx}
\makeatletter
\def\maxwidth{\ifdim\Gin@nat@width>\linewidth\linewidth\else\Gin@nat@width\fi}
\def\maxheight{\ifdim\Gin@nat@height>\textheight\textheight\else\Gin@nat@height\fi}
\makeatother
\setkeys{Gin}{width=\maxwidth,height=\maxheight,keepaspectratio}
\makeatletter
\def\fps@figure{htbp}
\makeatother

\makeatletter
\@ifpackageloaded{caption}{}{\usepackage{caption}}
\AtBeginDocument{%
\ifdefined\contentsname
  \renewcommand*\contentsname{Table of contents}
\else
  \newcommand\contentsname{Table of contents}
\fi
\ifdefined\listfigurename
  \renewcommand*\listfigurename{List of Figures}
\else
  \newcommand\listfigurename{List of Figures}
\fi
\ifdefined\listtablename
  \renewcommand*\listtablename{List of Tables}
\else
  \newcommand\listtablename{List of Tables}
\fi
\ifdefined\figurename
  \renewcommand*\figurename{Figure}
\else
  \newcommand\figurename{Figure}
\fi
\ifdefined\tablename
  \renewcommand*\tablename{Table}
\else
  \newcommand\tablename{Table}
\fi
}
\@ifpackageloaded{float}{}{\usepackage{float}}
\floatstyle{ruled}
\@ifundefined{c@chapter}{\newfloat{codelisting}{h}{lop}}{\newfloat{codelisting}{h}{lop}[chapter]}
\floatname{codelisting}{Listing}

\makeatother
\makeatletter
\@ifpackageloaded{caption}{}{\usepackage{caption}}
\@ifpackageloaded{subcaption}{}{\usepackage{subcaption}}
\makeatother

\ifLuaTeX
  \usepackage{selnolig}  
\fi
\usepackage[]{natbib}
\usepackage{bookmark}

\IfFileExists{xurl.sty}{\usepackage{xurl}}{} 
\hypersetup{
  pdftitle={Title},
  pdfauthor={Author 1; Author 2},
  pdfkeywords={3 to 6 keywords, that do not appear in the title},
  colorlinks=true,
  linkcolor={blue},
  filecolor={Maroon},
  citecolor={Blue},
  urlcolor={Blue},
  pdfcreator={LaTeX via pandoc}}

\newtheorem{Th}{\underline{\bf Theorem}}
\newtheorem{Pro}{\underline{\bf Proposition}}

\newtheorem{Lem}{\underline{\bf Lemma}}
\newtheorem{Cor}{\underline{\bf Corollary}}
\newtheorem{Rem}{\underline{\bf Remark}}

\usepackage{comment}
\usepackage[usenames,dvipsnames]{xcolor}
\usepackage{array,mathabx}
\usepackage{enumitem}
\usepackage{multirow}
\usepackage{hyperref}
\usepackage{xr}

\def\G{{\cal G}}
\def\H{{\cal H}}

\def\L{{\cal L}}

\def\M{{\bf{\cal  M}}}

\def\o{{\bf o}}
\def\S{{\bf S}}

\def\Z{{\bf Z}}
\def\a{{\bf a}}

\def\h{{\bf h}}

\def\z{{\bf z}}
\def\0{{\bf 0}}
\def\boxit#1{\vbox{\hrule\hbox{\vrule\kern6pt
          \vbox{\kern6pt#1\kern6pt}\kern6pt\vrule}\hrule}}
    \def\wt{\widetilde}
    \def\wc{\widecheck}
\def\sumi{\sum_{i=1}^n}

\def\wh{\widehat}

\def\log{\hbox{log}}

\def\var{\hbox{var}}
\def\cov{\hbox{cov}}

\def\Normal{\hbox{Normal}}
\def\TruncNormal{\hbox{TruncNormal}}

\def\bse{\begin{eqnarray*}}
\def\ese{\end{eqnarray*}}
\def\be{\begin{eqnarray}}
\def\ee{\end{eqnarray}}
\def\bq{\begin{equation}}
\def\eq{\end{equation}}
\def\bse{\begin{eqnarray*}}
\def\ese{\end{eqnarray*}}

\def\wh{\widehat}
\def\trans{^{\rm T}}

\def\bd{{\mathbf d}}
\def\bh{{\mathbf h}}

\def\b1e{{\mathbf e}}

\def\bu{{\mathbf u}}
\def\bS{{\mathbf S}}

\def\eff{_{\rm eff}}
\def\n{\nonumber}
\def\bb{{\boldsymbol\beta}}

\def\btheta{{\boldsymbol{\theta}}}

\newcommand{\bzeta}{{\boldsymbol\zeta}}
\newcommand{\bSigma}{{\boldsymbol\Sigma}}
\newcommand{\balpha}{{\boldsymbol\alpha}}

\def\bA{{\mathbf A}}
\def\ba{{\mathbf a}}
\def\bc{{\boldsymbol c}}
\def\bB{{\mathbf B}}
\def\bC{{\mathbf C}}

\def\bg{{\mathbf g}}

\def\b1e{{\mathbf e}}

\def\bI{{\mathbf I}}

\def\bo{{\mathbf o}}
\def\bO{{\mathbf O}}
\def\br{{\mathbf r}}

\def\bS{{\mathbf S}}

\def\bt{{\mathbf t}}

\def\bz{{\mathbf z}}

\def\bv{{\mathbf v}}

\def\bZ{{\mathbf Z}}

\newcommand{\bxi}{\mbox{\boldmath $\xi$}}

\def\bSigma{{\boldsymbol {\Sigma}}}

\def\bA{{\mathbf A}}
\def\ba{{\mathbf a}}
\def\bB{{\mathbf B}}
\def\bC{{\mathbf C}}

\def\bg{{\mathbf g}}
\def\b1e{{\mathbf e}}
\def\bI{{\mathbf I}}

\def\bB{{\mathbf B}}
\def\br{{\mathbf r}}
\def\bS{{\mathbf S}}

\def\bt{{\mathbf t}}

\def\bz{{\mathbf z}}

\def\bv{{\mathbf v}}

\def\bZ{{\mathbf Z}}

\def\bSigma{{\boldsymbol {\Sigma}}}

\def\G{{\cal G}}
\def\H{{\cal H}}

\def\L{{\cal L}}

\def\wh{\widehat}
\def\trans{^{\rm T}}

\def\bd{{\mathbf d}}
\def\bh{{\mathbf h}}
\def\b1e{{\mathbf e}}

\def\bq{{\mathbf q}}

\def\bu{{\mathbf u}}
\def\bS{{\mathbf S}}

\def\btheta{{\boldsymbol{\theta}}}

\newcommand{\bphi}{\mbox{\boldmath $\phi$}}

\def\bd{{\boldsymbol{d}}}
\def\bq{{\boldsymbol{q}}}

\newcommand{\indep}{\rotatebox[origin=c]{90}{$\models$}}

\begin{document}

\def\spacingset#1{\renewcommand{\baselinestretch}%
{#1}\small\normalsize} \spacingset{1}

\setcounter{page}{1} 

  \title{\bf SPYCE: A Doubly Robust Estimator for Trials Targeting 
Early Huntington Disease under Outcome-Dependent 
Censoring
  }
  
  \author{Kihyun Han$^{1}$\footnotemark[1], Yanyuan Ma$^{1}$\footnotemark[1], Karen Marder$^{2}$\footnotemark[1], and Tanya P. Garcia$^{3}$\thanks{The authors are supported by the National Institute of Neurological Disorders and Stroke under grant R01NS131225.
    }\hspace{.2cm}\\
    $^1$Department of Statistics, Pennsylvania State University\\
    $^2$Department of Neurology, Columbia University Medical Center\\
    $^3$Department of Biostatistics, Gillings School of Global Public Health, \\
    University of North Carolina at Chapel Hill
    }
     \renewcommand{\thefootnote}{\arabic{footnote}}
  \maketitle

\bigskip
\begin{abstract}
Clinical trials for neurodegenerative diseases must 
identify sensitive endpoints---outcomes that change 
rapidly enough to detect treatment effects. In 
Huntington disease, this requires measuring how 
outcomes change as participants approach Stage 1. Yet 
many participants exit studies before reaching this 
stage, making their time to Stage 1 right-censored. 
Estimating how outcomes change requires models for 
both time to Stage 1 and time to study exit. When 
participants with worse outcomes exit earlier, this
outcome-dependent censoring causes existing estimators 
to produce contradictory results: for the same 
cognitive outcome, one estimator suggests improvement 
while another shows decline. Existing estimators 
either ignore outcome-dependent censoring or require 
one model to be correctly specified, with no 
protection when it is not. We introduce SPYCE, a 
doubly robust estimator (consistent when either model 
is correctly specified) that achieves the smallest 
possible variance and allows both models to be 
estimated nonparametrically without sacrificing 
efficiency. Applied to data from PREDICT-HD, an 
observational Huntington disease study, SPYCE resolves 
current contradictions, identifies caudate and putamen 
volume ratios as the most promising sensitive endpoints, 
and shows that as few as 241 participants per arm are 
needed to detect treatment effects, versus hundreds of 
thousands under estimators that cannot handle 
outcome-dependent censoring.
\end{abstract}

\noindent%

\spacingset{1}

\section{Introduction}\label{sec:intro}

\begin{figure}[!ht] 
    \centering
   \includegraphics[width=0.6\linewidth]{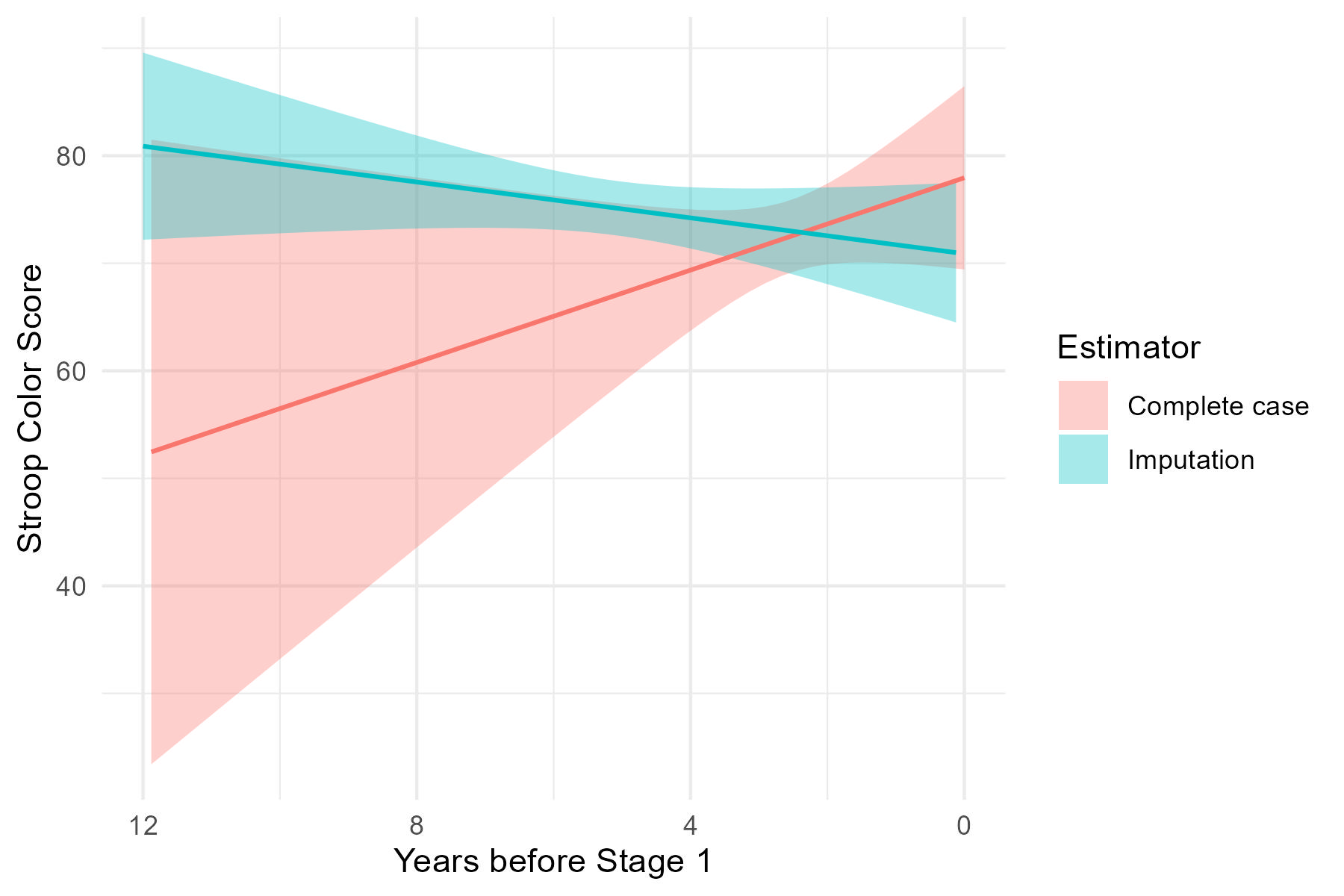}
    \caption{Contradictory estimates of how rapidly Stroop Color Word Test 
scores change as participants approach Stage 1, from two 
existing estimators. The $x$-axis shows years before Stage 1 
(0 = Stage 1 reached); the $y$-axis shows scores, where lower 
scores indicate worse cognition. The complete case estimator 
(using only data from participants who reached Stage 1) suggests 
implausible improvement, while the imputation estimator 
(replacing each unobserved time to Stage 1 with an estimated 
value) shows the expected decline. Shaded regions show 95\% 
confidence intervals.
    } 
    \label{fig:intro_plot}
\end{figure}

Treating neurodegenerative diseases early holds considerable promise, 
but clinical trials can only show treatment effectiveness if they 
track sensitive endpoints---outcomes such as cognitive scores, motor 
scores, and brain region volume ratios (a brain region's volume 
relative to total intracranial volume) that change rapidly enough that 
slowing this change through treatment produces detectable differences 
\citep{Aisenetal2022, Tabrizietal2022}. Huntington 
disease provides the ideal setting to identify such sensitive endpoints. 
Genetic testing reveals who will inevitably develop the disease---all 
participants carrying the Huntington disease gene mutation will eventually reach Stage 1, 
the earliest confirmed disease stage, and it is during this early period that intervention 
holds the most promise (see Section \ref{sec:stage1-details} for 
Stage 1 criteria). Estimating how rapidly each outcome changes as 
participants approach Stage 1 reveals which outcomes are sensitive 
endpoints. Yet when we apply existing estimators to data from PREDICT-HD, a 
long-running observational study of participants carrying the 
Huntington disease gene mutation, they produce contradictory results. One estimator suggests cognitive scores 
\emph{improve} as participants approach Stage 1, which is biologically 
implausible in a progressive disease \citep{paulsen2014prediction}, 
while another shows the expected \emph{decline} 
(Figure \ref{fig:intro_plot}). When estimators disagree this sharply, 
identifying sensitive endpoints becomes impossible. What causes 
existing estimators to give conflicting, sometimes clinically absurd 
results?

The culprit is that many participants exit PREDICT-HD before 
reaching Stage 1, so their time to Stage 1 is right-censored---we know only that they will reach Stage 1 after they exit the study, not exactly when. This creates a right-censored covariate 
problem \citep{dempsey2018survival, chu2020stochastic, 
kang2025dynamics}: estimating how rapidly each outcome changes 
as participants approach Stage 1 requires time to Stage 1 as a 
covariate, yet this covariate is right-censored for many 
participants. How existing estimators handle this right-censored 
covariate and what assumptions they make to do so determine 
whether they produce reliable or contradictory results.

Most existing estimators for right-censored covariate problems rely on
an assumption called \emph{outcome-independent censoring}
\citep{lotspeich2024making, lee2024robust,
zhang2025super}: among participants with the same time to Stage 1,
those with better and worse outcomes are equally likely to exit the
study at any given time. The assumption holds when participants exit
for reasons unrelated to their outcomes---the study ends or participants move away. Yet in reality, participants 
with worse outcomes often exit earlier. Worse cognitive scores 
reflect cognitive decline that makes keeping appointments difficult, 
and worse motor scores reflect motor deterioration that complicates 
travel to study sites. When participants with worse outcomes 
systematically exit earlier, how long they remain in the study 
depends on the very outcomes we are tracking---a violation that 
renders these existing estimators unreliable.

Estimators that allow for outcome-dependent censoring avoid 
this violation, yet each still falls short.  A closer look at the right-censored covariate problem reveals why. Let $Y$ denote the outcome, 
$X$ time to Stage 1, and $\Z$ baseline 
covariates such as age at study entry and gene mutation information. 
The goal in the right-censored covariate problem is to estimate 
$\bb$ in the outcome model $f_{Y|X,\Z}(y,x,\z;\bb)$, where the slope parameters in $\bb$ capture how rapidly each 
outcome changes as participants approach Stage 1.  Outcomes with steeper slopes are more sensitive endpoints, 
so reliable estimation of $\bb$ is essential for trial planning. 
Yet estimating $\bb$ is complicated by $X$ being right-censored. 
For participants who exit before reaching Stage 1, $X$ is not 
observed---we know only that $X$ exceeds their time to study exit 
$C$. From each participant we therefore observe $(Y, W, \Delta, 
\Z)$, where $W = \min(X, C)$ and $\Delta = I(X \leq C)$ indicates 
whether Stage 1 was reached ($\Delta = 1$) or the participant 
exited before reaching Stage 1 ($\Delta = 0$).

Under outcome-dependent censoring, existing estimators of $\bb$ 
are constructed from the likelihood function of $(Y, W, \Delta, \Z)$, 
which involves four density functions that we refer to as models 
throughout: the outcome model $f_{Y|X,\Z}$, the time to Stage 1 
model $f_{X|\Z}$, the censoring model $f_{C|Y,\Z}$, and the 
baseline covariate model $f_{\Z}$:
\be
\label{eq:model}
&&f_\Z(\z)\left\{f_{Y|X,\bZ}(y,w,\bz;\bb)f_{X|\bZ}(w,\bz)\int_{w}^{\infty}
f_{C|Y,\Z}(c,y,\z)
dc\right\}^{\delta}\nonumber\\
&&\times
\left\{f_{C|Y,\Z}(w,y,\z)\int_{w}^{\infty}f_{Y|X,\bZ}(y,x,\bz;\bb)f_{X|\bZ}(x,\bz)dx
\right\}^{1-\delta}.
\ee
The first line corresponds to participants who reached Stage 1 
($\delta = 1$); the second to participants who exited before 
reaching Stage 1 ($\delta = 0$). 

Existing estimators of $\bb$ differ primarily in what they 
assume about $f_{X|\Z}$ and $f_{C|Y,\Z}$, and these assumptions 
determine whether the resulting estimate $\wh{\bb}$ is reliable. 
The complete case estimator makes no assumptions about either 
model and cannot account for outcome-dependent censoring---a 
limitation that is easy to overlook. Instead it uses data only 
from participants who reached Stage 1, discarding data from all 
participants who exited earlier---those with the most rapid 
cognitive decline, greatest motor deterioration, and most severe 
brain atrophy. Their earlier exit is driven by the very outcomes 
we are tracking, so discarding their data biases $\wh{\bb}$, 
producing the implausible cognitive improvement shown in 
Figure~\ref{fig:intro_plot}.

Estimators adapted for outcome-dependent censoring model 
$f_{X|\Z}$, $f_{C|Y,\Z}$, or both to include data from 
participants who exited before reaching Stage 1, but each 
estimator requires at least one of these models to be correctly 
specified. The imputation (IMP) estimator and the maximum likelihood 
estimator (MLE) both require correct specification of $f_{X|\Z}$ 
\citep{atem2017linear}: the IMP estimator replaces each 
unobserved time to Stage 1 with a value estimated from 
$f_{X|\Z}$, while the MLE uses $f_{X|\Z}$ to integrate out 
the unobserved time to Stage 1 when maximizing the likelihood function
in \eqref{eq:model}. When $f_{X|\Z}$ is misspecified, the IMP 
estimator produces biased slope estimates, and the MLE compounds 
this bias: its slope estimates are biased \emph{and} precise, 
which is arguably more dangerous because a researcher might design 
a trial around them with false confidence. The weighted 
MLE \citep{lv2017maximum} adds inverse probability weighting to 
the likelihood function in \eqref{eq:model} but requires correct 
specification of both $f_{X|\Z}$ and $f_{C|Y,\Z}$, doubling
the risk of misspecification. The inverse probability weighting 
(IPW) estimator \citep{matsouaka2020regression} drops the 
$f_{X|\Z}$ requirement entirely, relying only on correct 
specification of $f_{C|Y,\Z}$, but with 58\% of participants 
exiting before reaching Stage 1, the resulting slope estimates 
are too imprecise for reliable trial planning. The estimator 
proposed by \citet{jhx2022nonparametric} takes a different 
approach, requiring correct specification of $f_{X|Y,\Z}$, the 
conditional density of time to Stage 1 given the outcome $Y$ and 
baseline covariates; however, these relationships that are not yet well understood biologically in Huntington disease.  Each estimator achieves 
consistency only when one particular model is correctly specified, 
but none provides any protection when that model is misspecified, 
making reliable identification of sensitive endpoints impossible.

What is needed is an estimator that remains consistent even when 
one of its required models is misspecified: a property called 
double robustness. Doubly robust, semiparametrically efficient 
estimators have been developed for right-censored covariate problems 
\citep{lee2024robust, zhang2025super}, but these estimators assume 
outcome-independent censoring, the assumption that fails in 
PREDICT-HD. We introduce SPYCE (the SemiParametric, $Y$-dependent 
right-Censored covariate Estimator) to address outcome-dependent 
censoring. SPYCE is doubly robust: consistent when either $f_{X|\Z}$ 
or $f_{C|Y,\Z}$ is correctly specified, so slope estimates remain reliable even when one model is misspecified. When both models are correctly 
specified, SPYCE achieves the semiparametric efficiency bound---the 
most precise slope estimates possible from the observed data---which translates directly to smaller sample sizes in trial 
planning. No existing estimator achieves both properties under 
outcome-dependent censoring.

When $f_{X|\Z}$ and $f_{C|Y,\Z}$ are difficult to specify 
correctly, as they often are in PREDICT-HD, both models can 
instead be estimated nonparametrically, avoiding the risk of 
misspecification altogether. Nonparametric estimation introduces 
a challenge, however: estimating $f_{X|\Z}$ and $f_{C|Y,\Z}$ 
nonparametrically requires knowing $\bb$, but $\bb$ is what we 
are trying to estimate.  We resolve this circular dependence 
through a novel reformulation of how the nuisance models 
enter the estimation, removing their dependence on $\bb$, 
and despite the additional 
complexity of nonparametric estimation, we show SPYCE still achieves 
the semiparametric efficiency bound, the same precision as when 
both models are correctly specified.  We establish these theoretical properties using semiparametric theory and demonstrate finite-sample performance through 
simulations designed to match the censoring rates observed in 
PREDICT-HD. Applying SPYCE to PREDICT-HD, where existing 
estimators produce contradictory or implausible slope estimates, 
SPYCE produces reliable ones, allowing researchers to identify 
sensitive endpoints and calculate the sample sizes needed to 
detect treatment effects in Huntington disease trials aimed at intervention before Stage 1.

%
%
%
%
%
%
%
%
%
%

\section{SPYCE: a Doubly Robust Estimator}
\label{sec:derivation-assumption2}

\subsection{Model Framework and Identifiability} \label{sec:model}
We observe $\bO = (Y, W, \Delta, \bZ)$, where $W = \min(X, C)$ and 
$\Delta = I(X \leq C)$ indicates whether a participant reached 
Stage 1. Under outcome-dependent censoring, time to study exit $C$ 
depends on the outcome $Y$, meaning $C \indep X \mid Y, \Z$, 
where $\indep$ denotes independence. This dependence shapes the 
likelihood function in \eqref{eq:model}, which involves four models: 
$f_{Y|X,\Z}$, which contains $\bb$, and $f_{X|\Z}$, $f_{C|Y,\Z}$, 
and $f_{\Z}$, none of which we want to constrain to a particular 
functional form, since misspecification can bias estimates of $\bb$. 
Before leaving the forms of $f_{X|\Z}$, $f_{C|Y,\Z}$, and $f_{\Z}$ 
unspecified, we must first confirm that $\bb$ and these models are 
identifiable from the observed data.
\begin{Pro}[Identifiability]\label{pro:iden}
  Let $S_{X|Y,\Z}$ and $S_{C|Y,\Z}$ denote the conditional survival
  functions of $X$ and $C$ given $(Y,\Z)$. If
  $S_{X|Y,\Z}(t,y,\z) > 0 \Longleftrightarrow S_{C|Y,\Z}(t,y,\z) > 0$
  for all $(t,y,\z)$, then $\bb$ and all models
  $f_{X|\Z}$, $f_{C|Y,\Z}$, and $f_{\Z}$ in \eqref{eq:model} are
  identifiable.
\end{Pro}
The proof is in Section~\ref{sec:ident}. 
Identifiability requires that time to Stage 1 and time to study 
exit are observable over the same range. Any estimator for a right-censored 
covariate problem requires this condition \citep{lee2024robust}, not 
just SPYCE. Without it, certain ranges of time to Stage 1 fall 
outside what the data can inform: if time to Stage 1 can extend 
to 15 years but time to study exit never exceeds 10 years, no 
participant remains enrolled long enough to reach Stage 1 beyond 
10 years, leaving $\bb$ impossible to estimate there. In PREDICT-HD, 
time to Stage 1 and time to study exit span the same range, so this 
condition is satisfied.

\subsection{Building Blocks of SPYCE} \label{sec:seff}
The models $f_{X|\Z}$, $f_{C|Y,\Z}$, and $f_{\Z}$ are necessary 
for estimation but are not our primary scientific target---we call 
them nuisance models, as their role is to support the estimation of 
$\bb$, which contains the slope parameters capturing how rapidly 
outcomes change as participants approach Stage 1. Misspecifying 
any one of these nuisance models may cause the resulting estimator 
of $\bb$ to be inconsistent.

Semiparametric theory \citep{bickel1993efficient, 
tsiatis2006semiparametric} addresses this risk through estimating 
functions, which are functions of the observed data and $\bb$ used 
to construct consistent estimators of $\bb$. This theory 
decomposes the space of all estimating functions into two key 
subspaces. The first is the nuisance tangent space $\Lambda$, 
spanned by the score functions of the nuisance models; estimating 
functions in $\Lambda$ are sensitive to nuisance model 
misspecification. The second is its orthogonal complement 
$\Lambda^{\perp}$, which contains estimating functions orthogonal 
to all score functions of the nuisance models; we show this 
orthogonality limits the influence of misspecified nuisance 
models on the resulting estimator of $\bb$.

The estimating function with the smallest asymptotic variance, 
$\S_{\text{eff}} \in \Lambda^{\perp}$, yields the most precise 
slope estimates possible. 
Deriving $\Lambda$, $\Lambda^{\perp}$, and $\S_{\text{eff}}$ 
under outcome-dependent censoring has not been done before 
and is not straightforward. Existing doubly robust, 
semiparametrically efficient estimators 
\citep{lee2024robust, zhang2025super} require 
outcome-independent censoring, under which the censoring 
model and outcome model appear separately in the estimating 
functions. Under outcome-dependent censoring, they are 
entangled throughout $\Lambda$, $\Lambda^{\perp}$, and 
$\S_{\text{eff}}$, so none of this machinery transfers 
directly and new technical arguments are required, which we 
provide in Proposition~\ref{pro:1} and 
Section~\ref{sec:pro1-3pf}.
\begin{Pro}[Tangent Spaces and Efficient Score] \label{pro:1}
For the semiparametric model \eqref{eq:model} with $\eta_1 = 
f_{X|\Z}$, $\eta_2 = f_{C|Y,\Z}$, and $\eta_3 = f_{\Z}$:
\begin{enumerate}[label=(\roman*),ref=(\roman*)]
\item The nuisance tangent space
$\Lambda = \Lambda_1 \oplus \Lambda_2 \oplus \Lambda_3$, where 
\bse
&\Lambda_1 &= \left[\delta \ba_1(w,\bz) + (1-\delta) \frac{E\{I(X> w)
    \ba_1 (X,\z)|y,\z\}}{E\{I(X > w)|y,\z\}} : E \{\ba_1 (X, \bz) |
  \bz\} = \mathbf{0} \right],\\ 
&\Lambda_2 &= \left[\delta \frac{E\{I(C \ge w) \ba_2
    (C,y,\z)|y,\z\}}{E\{I(C \ge w)|y,\z\}} + (1-\delta)\ba_2 (w,y,\z):
  E \left\{\ba_2 (C, y, \bz) | y,  \bz\right\} = \mathbf{0} \right],\\ 
&\Lambda_3 &= \left[\ba_3(\bz): E \{\ba_3 (\bZ)\} = \mathbf{0} \right].
\ese
\item The orthogonal complement $\Lambda^{\perp}$ 
  is
\bse
\Lambda^{\perp} &=& \left[\delta \bg_1(y,w,\z) + (1-\delta) \bg_2(y,w,\z): \right.\\
&& E\{I(x\le C) \bg_1 (Y,x,\z) + I(x>C) \bg_2 (Y,C,\z)| x,\z\} = \0,\\
&& E\{I(X\le c) \bg_1 (y,X,\z) + I(X>c) \bg_2 (y,c,\z)| c,y,\z\} \\
&&\left.= E\{I(X\le C) \bg_1 (y,X,\z) + I(X>C) \bg_2 (y,C,\z)| y,\z\} \right].
\ese
\item The efficient score function for $\bb$ is 
    \bse
    \S\eff (y,w,\delta, \z ; \bb) \equiv \S_\bb(y,w,\delta,\z;\bb) -
    \left[\delta \ba(w, \bz; \bb) + (1-\delta) \frac{ E\{I(X>
        w)\ba(X,\z;\bb) |y,\z\}}{ E\{I(X> w)|y,\z\}} \right], 
    \ese
where $\ba (x, \bz; \bb)$ satisfies
\bse
&&E\{I(x\le C)|x,\z\} \ba (x,\bz;\bb) +
E\left[\left.I(x>C)\frac{E\{I(X>C)\ba(X,\bz;\bb)
      |C,Y,\z\}}{E\{I(X>C)|C,Y,\z\}}\right|x,\bz\right] \\ 
&&= E\{I(x\le C)\S_\bb^F(Y,x,\bz;\bb)|x,\z\} +
E\left[\left.I(x>C)\frac{E\{I(X>C)\S_\bb^F(Y,X,\bz;\bb)
      |C,Y,\z\}}{E\{I(X>C)|C,Y,\z\}}\right|x,\bz\right], 
\ese
and $\S_\bb^F(y,x,\z;\bb) \equiv \partial \log
f_{Y|X,\Z}(y,x,\z;\bb)/ \partial \bb$.
\end{enumerate}
\end{Pro}

\subsection{SPYCE and Double Robustness} \label{sec:doublyrobust}
$\S\eff$ depends on the nuisance models $\eta_1 = f_{X|\Z}$ 
and $\eta_2 = f_{C|Y,\Z}$, both of which must be specified in 
practice and may be misspecified. Let $\eta_{10}$ and $\eta_{20}$ 
denote the true nuisance models and $\eta_1^*$ and $\eta_2^\star$ 
denote working models that may be misspecified, with $^*$ and 
$^\star$ indicating functions or expectations computed under 
each, respectively. Substituting $\eta_1^*$ and $\eta_2^\star$ 
into $\S\eff$ gives $\S\eff^{*\star}$, and SPYCE is the 
estimator $\wh{\bb}$ that solves $\sumi 
\S\eff^{*\star}(y_i, w_i, \delta_i, \z_i;\bb) = \0$.

Because $\S\eff^{*\star}$ is built from $\Lambda^{\perp}$, 
misspecification of either nuisance model need not compromise 
the consistency of $\wh\bb$; we now characterize exactly when 
consistency is guaranteed. The explicit form of $\S\eff^{*\star}$ is
\bse
    \S\eff^{*\star} (y,w,\delta, \z ; \bb) \equiv
    \S_\bb^*(y,w,\delta,\z;\bb) - \left[\delta \ba^{*\star}(w, \bz;
      \bb) + (1-\delta) \frac{ E^*\{I(X> w)\ba^{*\star}(X,\z;\bb)
        |y,\z\}}{ E^*\{I(X> w)|y,\z\}} \right],
\ese
    where $\ba^{*\star}(X,\z;\bb)$ satisfies    
    \be 
&& E^\star\{I(x\le C)|x,\z\} \ba^{*\star} (x,\bz;\bb) \label{eq:b1}\\
&&+
E^\star\left[\left. I(x>C) \frac{E^*\{I(X>C) \ba^{*\star}(X,\bz;\bb)
      |C,Y,\z\}}{E^*\{I(X>C)|C,Y,\z\}}\right|x,\bz\right] \n\\ 
&=& E^\star\{I(x\le C)\S_\bb^F(Y,x,\bz;\bb)|x,\z\} \n\\
&&+
E^\star\left[\left.I(x>C)\frac{E^*\{I(X>C)\S_\bb^F(Y,X,\bz;\bb)
      |C,Y,\z\}}{E^*\{I(X>C)|C,Y,\z\}}\right|x,\bz\right].\n
\ee

The density of $X|C,Y,\Z$ is $\eta_1^*(x,\z) f_{Y|X,\Z}(y,x,\z;\bb)\Big/\int \eta_1^*(x,\z)f_{Y|X,\Z}(y,x,\z;\bb)dx$,
which depends only on $\eta_1^*$ and not $\eta_2^\star$. The density
of $C,Y|X,\Z$ is $\eta_2^\star(c,y,\z) f_{Y|X,\Z}(y,x,\z;\bb)$,
which depends only on $\eta_2^\star$ and not $\eta_1^*$. Each 
expectation in $\S\eff^{*\star}$ therefore involves exactly 
one nuisance model, and this separation is what makes SPYCE 
doubly robust: consistent when either $\eta_1^*$ or $\eta_2^\star$ 
is correctly specified, but not necessarily both. The following 
theorem formalizes this guarantee under standard regularity 
conditions stated in Section~\ref{sec:pro4pf}.
\begin{Th}[Double Robustness] \label{th:doublyrobust}
Under Conditions \ref{con:bb}--\ref{con:conti}, SPYCE is doubly 
robust: $\wh\bb$ is consistent for $\bb_0$ if either $\eta_1^* = 
\eta_{10}$ or $\eta_2^\star = \eta_{20}$, but not necessarily both. 
\end{Th}
The proof is in Section~\ref{sec:pro4pf}. Double robustness is 
particularly valuable in PREDICT-HD because both nuisance models 
are difficult to specify correctly. The model for time to Stage 1, $f_{X|\Z}$, depends on baseline  covariates, such as age at study entry and gene mutation 
information, yet how these factors influence when participants 
reach Stage 1 remains biologically unclear. The model 
for time to study exit, $f_{C|Y,\Z}$, requires capturing the 
dependence between the outcome $Y$ and time to study exit $C$; 
for example, participants with a worse cognitive score may exit 
earlier, yet modeling the dependence between cognitive score 
and time to study exit precisely is difficult.  Because $f_{X|\Z}$ 
and $f_{C|Y,\Z}$ capture different biological processes, however, 
a researcher who is more confident in one than the other can 
rely on double robustness for protection: slope estimates remain 
reliable for identifying sensitive endpoints even when one 
nuisance model is misspecified.

\section{Various SPYCE Options} \label{sec:practical-spyce}

\subsection{Parametric Nuisance Models}\label{sec:par}

When parametric forms for $\eta_1$ and $\eta_2$ are available, 
setting $\eta_1^*(x,\z) = \eta_1(x,\z;\balpha_1^*)$ and 
$\eta_2^\star(c,y,\z) = \eta_2(c,y,\z;\balpha_2^\star)$ for 
finite-dimensional parameters $\balpha_1^*$ and $\balpha_2^\star$ 
allows the conditional expectations in \eqref{eq:b1} to be 
evaluated analytically or through standard numerical integration. 
Selecting flexible parametric forms reduces the risk of 
misspecification, with double robustness providing protection 
if one model is misspecified.

Estimating $\balpha_1^*$ and $\balpha_2^\star$ from the 
likelihood function in \eqref{eq:model} reveals an asymmetry that 
simplifies computation: since the censoring model $\eta_2^\star$ 
depends only on $(Y, \Z)$, which are fully observed, 
$\balpha_2^\star$ can be estimated from
\be\label{eq:ml2} 
\sumi \left[\delta_i \log \left\{\int_{w_i}^{\infty}
\eta_2(c, y_i, \z_i; \balpha_2^\star) dc\right\}+(1-\delta_i)\log \eta_2
(w_i,y_i,\z_i;\balpha_2^\star)\right]
\ee
independently of $\bb$.  The model for time to Stage 1, 
$\eta_1^*$, by contrast, cannot be separated from 
$f_{Y|X,\Z}(y,x,\z;\bb)$ in the likelihood function when participants 
exit before reaching Stage 1, entangling $\balpha_1^*$ 
with $\bb$, so $\balpha_1^*(\bb)$ must be estimated via 
the profile component
\be\label{eq:ml1}
\sumi \left[\delta_i \log\eta_1(w_i,\bz_i; \balpha_1^*)+ (1-\delta_i)
  \log \left\{\int_{w_i}^{\infty}
    f_{Y|X,\bZ}(y_i,x,\bz_i;\bb)\eta_1(x,\bz_i; \balpha_1^*)dx\right\}
\right]
\ee
for fixed $\bb$. The asymmetry gives a natural 
two-step workflow: obtain $\wh\balpha_2$ from 
\eqref{eq:ml2} first, for example by maximum 
likelihood or generalized method of moments. Then 
obtain $\wh\balpha_1(\bb)$ from \eqref{eq:ml1} 
by the same approach and $\wh\bb$ jointly by 
solving $\sumi \S\eff^{*\star}\{y_i, w_i, 
\delta_i, \z_i;\bb, \wh\balpha_1(\bb), \wh\balpha_2\} 
= \0$, yielding a consistent estimator of $\bb$ by 
Theorem~\ref{th:doublyrobust}.

A consistent estimator alone does not enable clinical 
trial planning: identifying sensitive endpoints requires 
reliable slope estimates, and standard errors are needed 
both to assess that reliability and to calculate how 
many participants a trial requires. Standard errors 
follow from the asymptotic normality of $\wh\bb$, which 
requires accounting for how estimation uncertainty in 
$\wh\balpha_1(\bb)$ and $\wh\balpha_2$ propagates to 
$\wh\bb$. This propagation is captured through the influence 
functions $\bphi_1$ and $\bphi_2$, defined by the 
asymptotic linearity conditions
\be
n^{1/2}\{\wh\balpha_1(\bb) - \balpha_1^*(\bb)\} &=& 
n^{-1/2}\sumi\bphi_1\{y_i, w_i, \delta_i, \z_i;
\balpha_1^*(\bb),\bb\} + o_p(1), \label{eqn:asymp-linearity}\\ 
n^{1/2}(\wh\balpha_2 - \balpha_2^\star) &=& 
n^{-1/2}\sumi\bphi_2(y_i, w_i, \delta_i, \z_i; 
\balpha_2^\star) + o_p(1). \nonumber
\ee
Maximum likelihood and generalized method of moments, 
introduced above as ways to obtain $\wh\balpha_1(\bb)$ 
and $\wh\balpha_2$, both satisfy these conditions, so 
the two-step workflow yields estimators with well-defined 
influence functions $\bphi_1$ and $\bphi_2$. The 
following theorem, proven in Section~\ref{sec:th1pf}, 
establishes asymptotic normality of $\wh\bb$.
\begin{Th}[Asymptotic Properties with Parametric Nuisance Models] 
\label{th:par}
Under the asymptotic linearity assumptions in 
\eqref{eqn:asymp-linearity} and Conditions 
\ref{con:pbb}--\ref{con:palpha1conv2} in Section~\ref{sec:th1pf},
\begin{enumerate}[label=(\roman*),ref=(\roman*)]
    \item\label{pcase1} if $\balpha_1^*(\bb_0) = \balpha_{10}$, then $\wh\bb$ is
        consistent and $n^{1/2}(\wh\bb - \bb_0) \stackrel{d}{\to} 
        \Normal\{\0, \bC^{-1}\bSigma_1(\bC^{-1})\trans\}$, where
        $\bSigma_1 = \var\{\S\eff^{\star}(Y,W,\Delta,\Z;\bb_0,
        \balpha_{10},\balpha_2^\star) + 
        \bA_1\bphi_1(Y,W,\Delta,\bZ;\balpha_{10},\bb_0)\}$.
    \item\label{pcase2} if $\balpha_2^\star = \balpha_{20}$, then $\wh\bb$ is
        consistent and $n^{1/2}(\wh\bb - \bb_0) \stackrel{d}{\to} 
        \Normal\{\0, \bB^{-1}\bSigma_2(\bB^{-1})\trans\}$, where
        $\bSigma_2 = \var[\S\eff^{*}\{Y,W,\Delta,\Z;\bb_0,
        \balpha_1^*(\bb_0),\balpha_{20}\} + 
        \bA_2\bphi_2(Y,W,\Delta,\bZ;\balpha_{20})]$.
    \item\label{pcase3} if both $\balpha_1^*(\bb_0) = \balpha_{10}$ and 
        $\balpha_2^\star = \balpha_{20}$, then $\wh\bb$ is
        consistent and semiparametrically efficient, specifically 
        $n^{1/2}(\wh\bb - \bb_0) \stackrel{d}{\to} \Normal(\0,
        [E\{\S\eff^{\otimes 2}(Y,W,\Delta,\Z;\bb_0)\}]^{-1})$.
\end{enumerate}
\end{Th}
Theorem~\ref{th:par} guarantees that the two-step workflow 
produces valid standard errors. When only one nuisance 
model is correctly specified, the variance accounts for 
estimation uncertainty in $\wh\balpha_1(\bb)$ and 
$\wh\balpha_2$ through $\bphi_1$ and $\bphi_2$. 
Ignoring the estimation uncertainty in $\wh\balpha_1(\bb)$ 
and $\wh\balpha_2$ would cause standard errors to be too 
small, leading researchers to design trials around slope 
estimates that appear more precise than they are.

When both nuisance models are correctly specified, the 
influence function contributions vanish asymptotically 
and $\wh\bb$ achieves the semiparametric efficiency 
bound---the smallest asymptotic variance any regular 
estimator can achieve from these data under 
outcome-dependent censoring. More precise slope estimates 
mean more confident identification of sensitive endpoints 
and smaller sample sizes for Huntington disease 
trials. These guarantees hold for any estimation method 
satisfying \eqref{eqn:asymp-linearity}; when 
$\wh\balpha_2$ and $\wh\balpha_1(\bb)$ are obtained 
by maximizing \eqref{eq:ml2} and \eqref{eq:ml1}, 
respectively, the influence functions take explicit 
forms that make computing standard errors straightforward.
\begin{Cor}[Asymptotic Properties under Maximum Likelihood 
Estimation] \label{cor:cor1}
Under the conditions of Theorem~\ref{th:par}, if 
$\wh\balpha_1(\bb)$ and $\wh\balpha_2$ are the maximum 
likelihood estimators $\wh\balpha_1^\ddagger(\bb)$ from 
\eqref{eq:ml1} and $\wh\balpha_2^\mathsection$ from 
\eqref{eq:ml2}, then the conclusions of 
Theorem~\ref{th:par}\ref{pcase1}--\ref{pcase3} hold with 
influence functions
$\bphi_1^\ddagger\{y,w,\delta,\bz;\balpha_1^\ddagger(\bb),\bb\} 
= \bI_1^{\ddagger-1}(\bb)\bS_1^\ddagger\{y,w,\delta,\bz;
\balpha_1^\ddagger(\bb),\bb\}$ and $\bphi_2^\mathsection(y,w,\delta,\bz;\balpha_2^\mathsection) 
= \bI_2^{\mathsection-1}\bS_2^\mathsection(y,w,\delta,\bz;
\balpha_2^\mathsection)$,
where $\bS_1^\ddagger$, $\bI_1^\ddagger$, $\bS_2^\mathsection$, 
and $\bI_2^\mathsection$ are the score functions and information 
matrices for \eqref{eq:ml2} and  \eqref{eq:ml1}, defined in 
Section~\ref{sec:th1pf}.
\end{Cor}
Parametric specification is computationally simpler and 
double robustness guarantees consistency even when one 
nuisance model is misspecified, but in PREDICT-HD both 
nuisance models are difficult to specify correctly, so 
misspecification of both remains a genuine risk. 
Section~\ref{sec:nonpar} avoids functional form assumptions 
entirely through nonparametric estimation, and as we show, 
without sacrificing precision. Researchers with strong 
prior knowledge about one or both nuisance models may 
still prefer parametric specification for its computational 
tractability; nonparametric estimation is the natural 
choice when such knowledge is unavailable.
%
%
%

\subsection{Nonparametric Nuisance Models} \label{sec:nonpar}

\subsubsection{Three Estimation Cases}

Nonparametric estimation of $\eta_1$ and $\eta_2$ requires 
no parametric assumptions, eliminating the risk of 
misspecification.  To make nonparametric estimation operational, we rewrite $\S\eff$ in terms of expectation operators $E_1$ and $E_2$:
\be \label{eq:n0}
&&\S\eff (y,w,\delta, \z ; \bb, E_1, \ba) \\
&&\equiv \S_\bb(y,w,\delta,\z;\bb, E_1) -
\left[\delta \ba(w, \bz; \bb) + (1-\delta) \frac{ E_1\{I(X>
    w)\ba(X,\z;\bb) |y,\z;\bb\}}{ E_1\{I(X> w)|y,\z;\bb\}} \right],\n
\ee
where $\ba (x, \bz; \bb, E_1, E_2)$ satisfies
\be \label{eq:n1}
&&E_2\{I(x\le C)|x,\z;\bb\} \ba (x,\bz;\bb) \\
&&+
E_2\left[\left.I(x>C)\frac{E_1\{I(X>C)\ba(X,\bz;\bb)
      |C,Y,\z;\bb\}}{E_1\{I(X>C)|C,Y,\z;\bb\}}\right|x,\bz;\bb\right] \n\\ 
&=& E_2\{I(x\le C)\S_\bb^F(Y,x,\bz;\bb)|x,\z;\bb\} \n\\
&&+
E_2\left[\left.I(x>C)\frac{E_1\{I(X>C)\S_\bb^F(Y,X,\bz;\bb)
      |C,Y,\z;\bb\}}{E_1\{I(X>C)|C,Y,\z;\bb\}}\right|x,\bz;\bb\right].\notag  
\ee
Double robustness carries over from 
Theorem~\ref{th:doublyrobust}: consistency requires only 
one expectation operator to be correctly estimated, so the 
other can be replaced by a working parametric model. A 
researcher who is more confident in one nuisance model than 
the other can therefore estimate that one parametrically 
while estimating the other nonparametrically, with 
consistency guaranteed as long as the parametric model is 
correctly specified. Let $\wh E_1 \equiv E_1(\cdot \mid \widehat{\eta}_1)$ and 
$\wh E_2 \equiv E_2(\cdot \mid \widehat{\eta}_2)$ denote 
nonparametric estimators of the expectation operators, and 
$E_1^* \equiv E_1(\cdot \mid \eta_1^*)$ and $E_2^\star 
\equiv E_2(\cdot \mid \eta_2^\star)$ denote expectation operators 
derived from working parametric models, which we refer to 
as working models for short. Three estimation cases arise depending 
on which expectation operators are estimated 
nonparametrically:
\begin{enumerate}[label=Case \arabic*., ref=Case \arabic*, 
start=1]
\item \label{case1} Estimate $E_1$ nonparametrically while 
using a working model for $E_2$. SPYCE is the estimator 
$\wh\bb$ that solves  $\sum_i
\S_{\text{eff}}^{\star}(y_i,w_i,\delta_i, \mathbf{z}_i ;
\boldsymbol{\beta}, \widehat{E}_1, 
\widehat{\mathbf{a}}^{\star}) = \mathbf{0}$, where 
$\widehat{\mathbf{a}}^{\star}(x, \mathbf{z};
\boldsymbol{\beta}) \equiv 
\mathbf{a}(x,\mathbf{z};\boldsymbol{\beta},
\widehat{E}_1, E_2^{\star})$. This case is appropriate when the researcher is confident 
in correctly specifying the censoring model but not the 
model for time to Stage 1.

\item \label{case2} Estimate $E_2$ nonparametrically while 
using a working model for $E_1$. SPYCE is the estimator 
$\wh\bb$ that solves $\sum_i
\S_{\text{eff}}^{*}(y_i,w_i,\delta_i, \mathbf{z}_i ;
\boldsymbol{\beta}, E_1^{*}, \widehat{\mathbf{a}}^{*}) = 
\mathbf{0}$, where $\widehat{\mathbf{a}}^{*}(x, 
\mathbf{z}; \boldsymbol{\beta}) \equiv 
\mathbf{a}(x,\mathbf{z};\boldsymbol{\beta}, E_1^{*},
\widehat{E}_2)$. This case is appropriate when the researcher is confident 
in correctly specifying the model for time to Stage 1 but 
not the censoring model.

\item \label{case3} Estimate both $E_1$ and $E_2$ 
nonparametrically. SPYCE is the estimator 
$\wh\bb$ that solves $\sum_i 
\S_{\text{eff}}(y_i,w_i,\delta_i, \mathbf{z}_i ;
\boldsymbol{\beta}, \widehat{E}_1, \widehat{\mathbf{a}}) 
= \mathbf{0}$, where $\widehat{\mathbf{a}}(x, \mathbf{z}; 
\boldsymbol{\beta}) \equiv 
\mathbf{a}(x,\mathbf{z};\boldsymbol{\beta}, \widehat{E}_1,
\widehat{E}_2)$. This case is most appropriate when the researcher has 
little confidence in correctly specifying either nuisance 
model.
\end{enumerate}

\subsubsection{Nonparametric Estimation of Expectation Operators}

In PREDICT-HD, where neither the model for time to Stage 1 nor the censoring model can be correctly specified, 
nonparametric estimation of $E_1$ and $E_2$ is the only path 
to reliable slope estimates. But nonparametric estimation is 
not straightforward: both $E_1$ and $E_2$ depend on $\bb$, 
whose slope parameters we want to estimate. So estimating 
$E_1$ and $E_2$ requires knowing $\bb$, but estimating $\bb$ 
requires $E_1$ and $E_2$, which creates a circular dependence 
that standard nonparametric tools cannot resolve.

The circular dependence is explicit in the forms of $E_1$ 
and $E_2$:
\bse
E_1\{g(y, X, c,\z)|y, c, \z;\bb\} &=& \frac{\int g(y,x,c,\z) 
f_{Y|X,\Z}(y,x,\z;\bb) \eta_1(x,\z)dx}{\int f_{Y|X,\Z}(y,x,\z;\bb) 
\eta_1(x,\z)dx},\\
E_2\{g(Y, x, C,\z)|x, \z;\bb\} &=& \iint g(y, x, c,\z)
f_{Y|X,\Z}(y,x,\z;\bb) \eta_2(c,y,\z) dydc.
\ese
Both $E_1$ and $E_2$ change as $\bb$ changes during estimation 
because $f_{Y|X,\Z}(y,x,\z;\bb)$ appears inside both operators. 
Ignoring this dependence, as standard nonparametric tools do, 
could bias estimates of $\bb$.

We resolve the circular dependence by exploiting a key 
structural feature: although $E_1$ and $E_2$ depend on $\bb$, 
this dependence enters only through $f_{Y|X,\Z}(y,x,\z;\bb)$ 
and $f_{Y|\Z}(y,\z;\bb)$, which can be evaluated at any 
candidate $\bb$ without estimation. We reformulate $E_1$ and 
$E_2$ so that the conditioning on $\bb$ is removed from the 
operators themselves:
\be
&&\label{eq:n2} E_1\{g(y,X,c,\z)\mid y, \z;\bb\} \\
&&= E_1\left\{\frac{ \Delta g(y, W, c, \z) f_{Y|X,\Z}(y,W,\z;\bb) }{ 
S_{C|Y,\Z}(W, Y, \z)}\mid \z\right\}\frac{1}{f_{Y|\Z}(y,\z;\bb)},\n
\ee
and
\be
\label{eq:n3}&&E_2\{g(Y,x, C,\z) \mid x,\z; \bb\} \\
&&= \int \frac{E_2\{(1-\Delta)
  g(y, x, W, \z)/S_{X|Y,\Z}(W, y, \z)\mid y, \z\}}{E_2\{(1-\Delta)/
  S_{X|Y,\Z}(W, y, \z)\mid y, \z\}} f_{Y|X,\Z}(y,x,\z;\bb)dy.\n
\ee
On the right-hand sides, $E_1$ and $E_2$ condition only on the 
nuisance models and observed data, so $\wh E_1$ and $\wh E_2$ 
can be constructed from the data and evaluated at any candidate 
$\bb$ during estimation, resolving the circular dependence. 
Two features of \eqref{eq:n2} and \eqref{eq:n3} are worth 
noting for correct implementation. First, in \eqref{eq:n2}, 
$Y$ appears as a random variable in the weight but $y$ is a 
fixed constant in the numerator. Second, in \eqref{eq:n3}, 
the denominator inside the integral equals 1, giving a 
Hájek-type normalization whose role in SPYCE's efficiency 
properties we return to in Section~\ref{sec:asymp-nonpar}.

With the reformulation in hand, we construct $\wh E_1$ and 
$\wh E_2$ using two ingredients. The first is conditional 
Kaplan-Meier estimators \citep{dabrowska1989uniform} with Nadaraya-Watson weights for 
$S_{C|Y,\Z}$ and $S_{X|Y,\Z}$, lower-bounded by $n^{-1}$ 
to prevent numerical instability when survival probabilities 
approach zero:
\be\label{eq:hatS}
&&\wh S_{C|Y,\Z}(t, y,\z)\\
&&= \max\left[ \prod_{j=1}^n \left\{1-
    \frac{K_{h_1}^{(m_1)}(y - y_j, \z - \z_j)}{\sum_{k=1}^n I(w_k \ge w_j)  
    K_{h_1}^{(m_1)}( y - y_k, \z - \z_k)}\right\}^{I(w_j \le t, \delta_j = 0)},   
    n^{-1}\right],\n\\
&&\wh S_{X|Y,\Z}(t, y,\z)\n\\
&&= \max\left[\prod_{j=1}^n \left\{1- \frac{K_{h_1}^{(m_1)}(y - y_j, \z - \z_j)}{
    \sum_{k=1}^n I(w_k \ge w_j) K_{h_1}^{(m_1)}(y - y_k, \z - \z_k)}
    \right\}^{I(w_j \le t, \delta_j = 1)}, n^{-1}\right],\n
\ee
where $K_h^{(m)}(\bt) = I(\bt_1 = \0)K^{(m)}(\bt_2/h)/h^a$ is a
kernel function with bandwidth $h$ and order $m$, with discrete
components $\bt_1$ and continuous components $\bt_2$ to accommodate
mixed $(Y,\Z)$. Product kernel functions are used for $K^{(m)}$, as
detailed in Condition~\ref{con:nker}.

The second ingredient is kernel smoothing, which turns the 
reformulated expectation operators in \eqref{eq:n2} and \eqref{eq:n3} 
into nonparametric estimators that can be evaluated from the observed data:
\be\label{eq:n15}
  &&\wh E_1\{g(y,X,c,\z)|y, c, \z,\bb\} \\
  &&=\sumi\frac{ \delta_i g(y, w_i, c, \z_i) f_{Y|X,\Z}(y,w_i,\z_i;\bb) 
  K_{h_2}^{(m_2)}(\z-\z_i)}{\wh S_{C|Y,\Z}(w_i,
  y_i, \z_i)\sum_{k=1}^n K_{h_2}^{(m_2)}(\z-\z_k)} 
  \frac{1}{f_{Y|\Z}(y,\z;\bb)},\n\\
&&\wh E_2\{g(Y,x, C,\z) | x,\z; \bb\}\label{eq:n16}\\
&=&\int  \frac{\sumi(1-\delta_i)
  g( y_i, x, w_i, \z_i)K_{h_3}^{(m_3)}(y-y_i,\z-\z_i)/\wh S_{X|Y,\Z}(w_i, y_i, \z_i)}{ 
  \sumi(1-\delta_i)K_{h_3}^{(m_3)}(y-y_i,\z-\z_i)/\wh S_{X|Y,\Z}(w_i, y_i, \z_i)}\n\\
  &&\times f_{Y|X,\Z} (y,x,\z;\bb)dy. \n
\ee
For 
$\wh E_1$, the factor $1/f_{Y|\Z}(y,\z;\bb)$ cancels within 
$\S\eff$, making $\wh E_1$ computationally tractable. For 
$\wh E_2$, the kernel smoothing weights and survival function 
estimates are constructed entirely from observed data, so 
$\wh E_2$ can be evaluated at any $\bb$ without re-estimation, 
and the Hájek-type normalization in \eqref{eq:n16} ensures 
$\wh E_2(1 \mid y, \z) = 1$, a requirement for $\wh E_2$ 
to be a valid expectation operator whose consequences for 
efficiency we detail in Remark~\ref{rem:r4}.

\begin{Rem}[The Need for Hájek Normalization]\label{rem:r4}
A natural but invalid alternative to $\wh E_2$ in 
\eqref{eq:n16} is to omit the denominator and use inverse 
probability weighting with kernel smoothing directly:
\be
\wt E_2\{g(y, x, C, \z)\mid y, \z\}= 
\sumi \frac{(1-\delta_i)
  g( y_i, x, w_i, \z_i)K_{h_3}^{(m_3)}(y-y_i,\z-\z_i)}{\wh
  S_{X|Y,\Z}(w_i, y_i, \z_i)\sum_{k=1}^n
  K_{h_3}^{(m_3)}(y-y_k,\z-\z_k)}. \label{eq:e5}
\ee
This alternative is invalid because $\wt E_2(1 \mid y, \z) 
\neq 1$ in general, violating the requirement that a valid 
expectation operator integrates to one, and when survival 
probabilities $\wh S_{X|Y,\Z}(w_i, y_i, \z_i)$ are small, 
the weights become large and $\wt E_2$ becomes unstable. 
Dividing by $\wt E_2(1 \mid y, \z)$ restores both properties:
\bse
\wh E_2\{g(y, x, C, \z)\mid y, \z\} = 
\frac{\wt E_2\{g(y, x, C, \z)\mid y, \z\}}
{\wt E_2(1\mid y, \z)},
\ese
the same normalization used in the Hájek estimator from 
survey sampling \citep{hajek1971comment}.
\end{Rem}
\begin{Rem}[Naive Approaches and Circular Dependence]\label{rem:r6}
A naive approach that appears to avoid the circular 
dependence is to construct $\wh E_1$ directly using inverse 
probability weighting and kernel smoothing:
\bse
\wc E_1\{g(y,X,c,\z)\mid y, c, \z\} = \sumi\frac{ \delta_i 
g(y_i, w_i, c, \z_i) K_{h_3}^{(m_3)}(y-y_i,\z-\z_i)}{\wh 
S_{C|Y,\Z}(w_i, y_i, \z_i) \sum_{k=1}^n 
K_{h_3}^{(m_3)}(y-y_k,\z-\z_k)}.
\ese
This alternative is invalid because $\wc E_1$ is constructed
entirely from observed data generated under $\bb_0$, so it is 
only valid at $\bb_0$ and does not correspond to any valid 
expectation operator $E_1(\cdot \mid \wc\eta_1, \bb)$ for 
general $\bb$. When $\wc E_1$ is substituted into $\S\eff$, 
the resulting estimating equation falls outside $\Lambda^\perp$, 
so $\wh\bb$ no longer achieves the semiparametric efficiency 
bound even when the other nuisance model is correctly 
specified---meaning a researcher who correctly specifies one 
nuisance model, and therefore believes double robustness 
guarantees reliable slope estimates, would still obtain 
unnecessarily imprecise estimates of how rapidly outcomes 
change as participants approach Stage 1.
\end{Rem}

\begin{table}[!t]
\centering
\caption{
How estimation errors from nonparametric estimation propagate 
to create additional variability in SPYCE's asymptotic 
variance. Each row shows how errors from a specific source 
flow through different components of the estimator to generate 
corresponding variability functions. $\wh f$ denotes quantities 
used in constructing the expectation operators rather than 
directly estimated density functions.
} 
\label{table:ta1}
\begin{tabular}{c|c|c|c|c}
Case & Error  & Affected & Individual Variability & Combined 
Variability\\
& Source & Component & Function & Function\\ \hline
\multirow{4}{*}{Case 1: $(\wh E_1, E_2^\star)$} & 
\multirow{2}{*}{$\wh S_{C|Y,\Z}$}  & $\wh E_1$ & 
$\bh_{1{\rm s}1}^\star(y_j,w_j, \delta_j, \z_j)$  & 
\multirow{2}{*}{$\bh_{1{\rm s}}^\star(y_j,w_j, \delta_j, 
\z_j)$} \\ \cline{3-4} 
&   &    $\wh \ba^\star$&  $\bh_{1{\rm s}2}^\star(y_j,w_j, 
\delta_j, \z_j)$ &\\ \cline{2-5}
& \multirow{2}{*}{$\wh f_{\Delta, W,Y|\Z}$} & $\wh E_1$& 
$\bh_{1{\rm k}1}^\star(y_j,w_j, \delta_j, \z_j)$ & 
\multirow{2}{*}{$\bh_{1{\rm k}}^\star(y_j,w_j, \delta_j, 
\z_j)$} \\ \cline{3-4} 
&   &    $\wh \ba^\star$& $\bh_{1{\rm k}2}^\star(y_j,w_j, 
\delta_j, \z_j)$ &\\ \hline
\multirow{2}{*}{Case 2: $(E_1^*, \wh E_2)$}     & 
$\wh S_{X|Y,\Z}$   & $\wh \ba^*$ & $\bh_{2{\rm s}}^*(y_j,
w_j, \delta_j, \z_j)$& $\bh_{2{\rm s}}^*(y_j,w_j, \delta_j, 
\z_j)$ \\ \cline{2-5} 
& $\wh f_{\Delta, W|Y,\Z}$   & $\wh \ba^*$ & 
$\bh_{2{\rm k}}^*(y_j,w_j, \delta_j, \z_j)$ & 
$\bh_{2{\rm k}}^*(y_j,w_j, \delta_j, \z_j)$
\end{tabular}
\end{table}

\subsubsection{Asymptotic Properties of SPYCE with Nonparametric 
Estimation}\label{sec:asymp-nonpar}

With parametric nuisance models, standard errors for $\wh\bb$ 
accounted for estimation uncertainty in $\wh\balpha_1(\bb)$ 
and $\wh\balpha_2$ through the influence functions $\bphi_1$ 
and $\bphi_2$. Nonparametric estimation introduces two 
additional sources of variability that must be accounted for 
in the same way: estimation error from the conditional 
Kaplan-Meier estimators $\wh S_{C|Y,\Z}$ and 
$\wh S_{X|Y,\Z}$, and approximation error from kernel 
smoothing. Ignoring either source would produce standard errors that 
are too small, giving researchers false confidence in how 
precisely $\bb$ is estimated.

How these errors propagate to $\wh\bb$ depends on which 
expectation operators are estimated nonparametrically, as 
summarized in Table~\ref{table:ta1} for Cases 1 and 2. Each 
source of error introduces a variability function---$\bh_{1{\rm s}}^\star$, 
$\bh_{1{\rm k}}^\star$, $\bh_{2{\rm s}}^*$, $\bh_{2{\rm k}}^*$---that 
captures the additional variability in the asymptotic 
variance of $\wh\bb$, with exact forms in 
Section~\ref{sec:nonpar-variability-functions}. Case 3, where both expectation operators are estimated 
nonparametrically, exhibits fundamentally different 
behavior: the penalty terms do not arise, and the asymptotic 
variance simplifies directly to the semiparametric efficiency 
bound, as Theorem~\ref{th:nonpar}\ref{th2case3} establishes.

A technical challenge arises across all three cases: the 
conditional Kaplan-Meier estimators $\wh S_{C|Y,\Z}$ and 
$\wh S_{X|Y,\Z}$ converge slower than $n^{-1/2}$, which 
could in principle compromise the $n^{1/2}$-consistency of 
$\wh\bb$. We therefore impose regularity conditions on the 
bandwidth, dimensionality, and smoothness of the 
nonparametric estimators of the nuisance models---detailed 
in Section~\ref{sec:nonpar-regularity-conditions}---to ensure their slower convergence does not 
affect estimation of $\bb$. The following theorem 
establishes the consistency, asymptotic normality, and 
semiparametric efficiency of SPYCE under nonparametric 
estimation.
\begin{Th} \label{th:nonpar}
In each case below, SPYCE is the estimator $\wh\bb$ that 
solves $\sumi \S\eff^{*\star}(y_i,w_i,\delta_i, \z_i ; \bb, 
E_1, \ba) = \0$, where $\ba = \ba(x,\z;\bb, E_1, E_2)$ as 
defined in \eqref{eq:n0} and \eqref{eq:n1}. The variability functions 
$\bh_{1{\rm s}}^\star$, $\bh_{1{\rm k}}^\star$, $\bh_{2{\rm s}}^*$, 
$\bh_{2{\rm k}}^*$, the matrices $\bB^\star$ and $\bB^*$, and 
the regularity conditions \ref{con:nker}, \ref{con:nkerbw'}, 
and \ref{con:npbb}--\ref{con:nBinv2} are detailed in 
Sections~\ref{sec:nonpar-variability-functions} and 
\ref{sec:nonpar-regularity-conditions}. 
Under the regularity conditions,
\begin{enumerate}[label=(\roman*),ref=(\roman*)]
    \item\label{th2case1} (\ref{case1}) if $E_1$ is estimated 
    nonparametrically as in \eqref{eq:n15} and $E_2$ is 
    replaced by a working model $E_2^\star$, then $\wh\bb$ 
    is consistent and $n^{1/2} (\wh\bb - \bb_0) 
    \stackrel{d}{\to} \Normal\{\0, \bB^{\star-1}\bSigma_1 
    (\bB^{\star-1})\trans\}$, where
    $\bSigma_1 = \var\{\S\eff^{\star}(Y,W,\Delta,\Z;\bb_0,
    E_{10},\ba_0^\star) + \bh_{1{\rm s}}^\star(Y,W,\Delta,
    \bZ) + \bh_{1{\rm k}}^\star(Y,W,\Delta,\bZ)\}$.
    \item\label{th2case2} (\ref{case2}) if $E_1$ is replaced 
    by a working model $E_1^*$ and $E_2$ is estimated 
    nonparametrically as in \eqref{eq:n16}, then $\wh\bb$ 
    is consistent and $n^{1/2}(\wh\bb - \bb_0) 
    \stackrel{d}{\to} \Normal\{\0, \bB^{*-1}\bSigma_2 
    (\bB^{*-1})\trans\}$, where
    $\bSigma_2 = \var\{\S\eff^*(Y,W,\Delta,\Z;\bb_0,E_1^*,
    \ba_0^*) + \bh_{2{\rm s}}^*(Y,W,\Delta,\bZ) + 
    \bh_{2{\rm k}}^*(Y,W,\Delta,\bZ)\}.$
    \item\label{th2case3} (\ref{case3}) if both $E_1$ and 
    $E_2$ are estimated nonparametrically as in 
    \eqref{eq:n15} and \eqref{eq:n16}, then $\wh\bb$ is 
    consistent and semiparametrically efficient, specifically 
    $n^{1/2}(\wh\bb - \bb_0) \stackrel{d}{\to} \Normal(\0, 
    [E\{\S\eff^{\otimes 2}(Y,W,\Delta,\Z;\bb_0)\}]^{-1})$.
\end{enumerate}
\end{Th}
The proof is in Section~\ref{sec:th2pf}. Like the parametric 
case, SPYCE achieves $n^{1/2}$-consistency, asymptotic 
normality, and double robustness across all three cases, so 
valid standard errors and confidence intervals for slope 
estimates are available regardless of which expectation 
operators are estimated nonparametrically. In Cases 1 and 2, 
the asymptotic variance includes the variability functions 
from Table~\ref{table:ta1}, which must be accounted for when 
computing standard errors; ignoring them would produce 
confidence intervals that are too narrow, giving researchers 
false confidence in the precision of their slope estimates.

Case 3 yields a result that is both surprising and 
practically important for PREDICT-HD. When both expectation 
operators are estimated nonparametrically---the case 
we adopt for our analysis of PREDICT-HD, where neither 
nuisance model is specified parametrically---the asymptotic variance 
simplifies to the semiparametric efficiency bound 
$[E\{\S_{\text{eff}}^{\otimes 2}(Y,W,\Delta,\Z;\bb_0)\}]^{-1}$ 
automatically. No working models are needed, yet SPYCE 
achieves the same precision as when both nuisance models are 
correctly specified parametrically---giving researchers the most precise slope estimates the data can 
support, without the risk of misspecification that parametric 
nuisance models carry.

Theorem~\ref{th:nonpar} also reveals a complementary result 
for Cases 1 and 2. When a working model is correct, the penalty terms it would otherwise contribute 
vanish entirely, as the following proposition establishes.
\begin{Pro}[Statistical Efficiency Preservation] \label{pro:5}
When working models are correct, the nonparametric 
penalty terms vanish. Specifically, if $E_2^\star = E_{20}$, 
then $\bh_{1{\rm s}} = \bh_{1{\rm k}} = \0$, and if $E_1^* = 
E_{10}$, then $\bh_{2{\rm s}} = \bh_{2{\rm k}} = \0$.
\end{Pro}
The proof is in Section~\ref{sec:th2pf}. A researcher who is more confident in one nuisance model 
can therefore estimate that model parametrically and the 
other nonparametrically, achieving the same precision as 
correctly specifying both parametrically.

\begin{Rem}[Hájek Normalization and Efficiency 
Preservation]\label{rem:r7}
Proposition~\ref{pro:5} depends on the Hájek-type 
normalization in $\wh E_2$. Using the unnormalized 
alternative $\wt E_2$ from \eqref{eq:e5} instead produces 
penalty functions $\wt\bh_{2{\rm s}}^*$ and $\wt\bh_{2{\rm k}}^*$ that 
do not vanish even when $E_1^* = E_{10}$. The consequence 
is concrete: in Case 2, where the model for time to Stage 1 
is correct, the penalty terms persist and the 
efficiency bound is not achieved, producing wider confidence 
intervals for slope estimates than necessary. In Case 3, 
where both operators are estimated nonparametrically, using 
$\wt E_2$ instead of $\wh E_2$ breaks the automatic 
efficiency guarantee of Theorem~\ref{th:nonpar}\ref{th2case3}, 
since the construction of $\wh E_2$ itself relies on the 
normalization. 
The Hájek normalization is therefore what makes it possible 
to achieve the semiparametric efficiency bound,
which is theoretically expected under correct specification of both nuisance models.
\end{Rem}

\section{Simulation Studies}\label{sec:simulation}

We designed simulations to match the outcome-dependent 
censoring and high censoring rates observed in PREDICT-HD. 
We simulated $1,000$ datasets of $n = 1,000$ observations each.
 We generated $X$ from 
$\TruncNormal(0,1; -1, 1)$, $Y$ from $\Normal(\beta_1 + 
\beta_2 X, 4^2)$ with $\bb_0 = (0,3)\trans$, and $C$ from 
$\TruncNormal(\alpha_{21} + 0.12Y, 1^2; -1, 1)$, where the 
dependence on $Y$ creates outcome-dependent censoring. 
Varying $\alpha_{21} = 3, 1, -1$ yielded low (10--20\%), 
moderate (30--40\%), and high (60--70\%) censoring rates, 
respectively; the high censoring setting approximates 
PREDICT-HD's 58\% rate.

We compared SPYCE against four benchmark estimators 
introduced in Section~\ref{sec:intro}: the complete case (CC) estimator, 
which requires no nuisance model but cannot account for 
outcome-dependent censoring; the IMP estimator and MLE, 
which both require $\eta_1$ to be correctly specified; and the 
IPW estimator, which requires $\eta_2$ to be correctly 
specified. SPYCE was tested under seven configurations 
varying $\eta_1$ and $\eta_2$ across three specifications---correctly specified, misspecified, or estimated 
nonparametrically; the IMP estimator and MLE each varied 
$\eta_1$ across the same three specifications; the IPW 
estimator varied $\eta_2$ across the same three 
specifications; and the CC estimator required neither 
nuisance model. Estimating equations for all four 
benchmark estimators and full configuration details are in 
Section~\ref{sec:benchmark-additional-sims}.

\begin{table}[!t]
\centering
\caption{Finite-sample performance of SPYCE and four benchmark estimators under high censoring rate (60--70\%). For each estimator, we report bias, standard deviation (SD), median estimated standard error (SE), and 95\% confidence interval coverage rate (CI). Par = parametric. Mis = misspecified. Non = nonparametric.   }
\label{table:ta6}
\small
\begin{tabular}{l|cc|rrrr|rrrr}
\multirow{2}{*}{Estimator} & \multirow{2}{*}{$\eta_1$} & \multirow{2}{*}{$\eta_2$} & \multicolumn{4}{c|}{$\beta_1$} & \multicolumn{4}{c}{$\beta_2$} \\
\cline{4-7}\cline{8-11}
 &  &  & Bias & SD & SE & CI & Bias & SD & SE & CI \\ \hline
\multirow{7}{*}{SPYCE}
 & Par & Par & -0.008 & 0.135 & 0.141 & 95.5\% & -0.051 & 0.286 & 0.304 & 96.0\% \\
 & Par & Mis & -0.012 & 0.133 & 0.134 & 94.6\% & -0.078 & 0.276 & 0.305 & 96.8\% \\
 & Mis & Par & -0.014 & 0.135 & 0.139 & 95.0\% & -0.064 & 0.286 & 0.302 & 95.7\% \\
 & Mis & Mis & -0.028 & 0.130 & 0.134 & 94.9\% & -0.117 & 0.274 & 0.293 & 95.0\% \\
 & Non & Non & -0.001 & 0.152 & 0.148 & 91.2\% &  0.060 & 0.318 & 0.336 & 92.1\% \\
 & Non & Mis &  0.038 & 0.138 & 0.138 & 94.1\% &  0.055 & 0.293 & 0.290 & 94.4\% \\
 & Mis & Non &  0.061 & 0.170 & 0.214 & 98.7\% &  0.191 & 0.361 & 0.402 & 97.5\% \\ \hline
CC   & --  & --  &  1.148 & 0.282 & 0.302 &  2.5\% &  1.098 & 0.486 & 0.519 & 43.2\% \\ \hline
\multirow{3}{*}{IMP}
 & Par & -- & 0.158 & 0.160 & 0.149 & 80.4\% & 2.336 & 0.348 & 0.308 & 0.0\% \\
 & Mis & -- & 0.033 & 0.136 & 0.147 & 95.8\% & 2.172 & 0.340 & 0.304 & 0.0\% \\
 & Non & -- & 0.204 & 0.180 & 0.145 & 67.8\% & 2.247 & 0.390 & 0.310 & 0.0\% \\ \hline
\multirow{3}{*}{IPW}
 & --  & Par &  0.045 & 0.365 & 0.379 & 96.1\% &  0.122 & 0.930 & 0.809 & 93.2\% \\
 & --  & Mis &  1.135 & 0.346 & 0.346 & 10.9\% &  1.064 & 0.710 & 0.669 & 65.3\% \\
 & --  & Non &  0.460 & 0.238 & 0.355 & 85.2\% &  0.490 & 0.554 & 0.747 & 94.7\% \\ \hline
\multirow{3}{*}{MLE}
 & Par & --  & -0.008 & 0.133 & 0.133 & 95.0\% & -0.090 & 0.270 & 0.291 & 96.3\% \\
 & Mis & --  & -0.050 & 0.126 & 0.132 & 94.4\% & -0.191 & 0.262 & 0.290 & 92.8\% \\
 & Non & --  &  0.075 & 0.146 & 0.133 & 89.6\% &  0.095 & 0.308 & 0.302 & 93.4\%
\end{tabular}
\end{table}

Table~\ref{table:ta6} presents results under high censoring 
(60--70\%), the setting that most closely approximates that of 
PREDICT-HD; results under low and moderate censoring appear 
in Section~\ref{sec:benchmark-additional-sims}. We organize 
the discussion around three questions a researcher would 
ask before applying SPYCE to PREDICT-HD: Do existing 
estimators produce misleading estimates of $\bb$ under 
PREDICT-HD's conditions? Does SPYCE remain reliable where 
existing estimators do not? Is SPYCE precise enough 
for trial planning?

\noindent\textbf{Existing estimators produce misleading 
estimates of $\bb$ under PREDICT-HD's conditions.}
The CC estimator has severe bias (1.148 for $\beta_1$, 
2.5\% coverage), as it ignores outcome-dependent censoring 
entirely. The IMP estimator has 0\% coverage for $\beta_2$ 
under high censoring regardless of how $\eta_1$ is 
specified---even when $\eta_1$ is correctly specified, 
bias of 2.336 renders it useless for identifying which 
outcomes change rapidly enough to serve as endpoints. 
Despite these biases, both have been used in Huntington 
disease research to select sensitive endpoints 
\citep{long2014tracking,langbehn2020clinical}; under 
outcome-dependent censoring, biases of this magnitude 
could lead a researcher to the wrong conclusions about 
which outcomes change rapidly enough to detect a 
treatment effect. The IPW estimator and MLE tell a 
different story. There is small bias and near-nominal 
coverage when their single required nuisance model is 
correctly specified, but either nuisance model is 
difficult to specify correctly: under misspecification, 
coverage drops to 10.9\% for the IPW estimator, and 
even the MLE's coverage of 92.8\% falls below the 
nominal 95\% with no fallback protection when the 
model is wrong.

\noindent\textbf{SPYCE remains reliable where existing 
estimators do not.} Across all settings, SPYCE has small 
bias and near-95\% coverage. The misspecification tested 
is severe: the misspecified models bear no resemblance to 
the true data-generating distributions. When at least one 
nuisance model is correctly specified, double robustness 
guarantees small bias and near-95\% coverage; in Case 3, 
where both expectation operators are estimated 
nonparametrically, SPYCE still achieves small bias and 
near-nominal coverage. 
Outside SPYCE's theoretical guarantees, when both 
nuisance models are misspecified, coverage drops to 
80.8\% and 84.0\% under low censoring 
(Table~\ref{table:ta4}) and is near-nominal under high 
censoring (Table~\ref{table:ta6}), but the latter is not 
because SPYCE performs well there. With 60--70\% of the data censored, confidence 
intervals widen enough to achieve nominal coverage 
despite the bias, a result of data scarcity rather 
than robustness.

The precision cost of nonparametric estimation is modest: 
Case 3's empirical standard deviation across simulations 
for $\beta_2$ is 0.318, compared to 0.286 under correct 
parametric specification. Standard errors track standard 
deviations closely across all configurations where at 
least one nuisance model is correctly specified, with modest 
undercoverage only under Case 3 and high censoring 
(91.2\% for $\beta_1$), where high censoring leaves fewer 
participants who reached Stage 1 to construct $\wh E_1$. 
When one nuisance model is estimated nonparametrically 
and the other is misspecified, an asymmetry emerges 
between Cases 1 and 2 as censoring increases: in Case 1, 
where $\eta_1$ is estimated nonparametrically and $\eta_2$ 
is misspecified, SPYCE's $\beta_2$ standard deviation 
increases modestly from 0.229 under low censoring to 
0.293 under high censoring; in Case 2, where $\eta_2$ is 
estimated nonparametrically and $\eta_1$ is misspecified, 
the increase is more pronounced, from 0.235 to 0.361. 
This pattern may reflect a difference in how the two 
nonparametric estimators are constructed: $\wh E_1$ 
relies on $S_{C|Y,\Z}$, the survival function for time 
to study exit, while $\wh E_2$ relies 
on $S_{X|Y,\Z}$, the survival function for time to 
Stage 1. Performance may be more 
influenced by accurate estimation of $S_{C|Y,\Z}$ than 
$S_{X|Y,\Z}$. If so, researchers who are uncertain about 
both nuisance models should prioritize nonparametric 
estimation of the model whose conditional survival 
function is more tractable to estimate accurately, 
as prioritizing the more tractable model is likely 
to yield more reliable slope estimates.

\noindent\textbf{SPYCE is precise enough for Huntington 
disease trial planning.} SPYCE achieves its smallest 
standard deviations when both nuisance models are 
correctly specified, consistent with semiparametric 
efficiency. The IPW estimator is the natural comparison 
since it is the only benchmark estimator that also 
achieves consistency through the censoring model alone: 
for $\beta_2$ under high censoring, SPYCE's standard 
deviation is 0.286 compared to the IPW estimator's 
0.930, meaning a study relying on the IPW estimator 
would need roughly ten times as many participants to 
achieve the same precision. Even under Case 3, where 
both nuisance models are estimated nonparametrically, 
SPYCE's standard deviation of 0.318 remains far below 
the IPW estimator's 0.930, so the precision advantage 
holds even when neither nuisance model can be correctly 
specified. This gap between SPYCE and the IPW estimator 
widens as censoring increases, because higher censoring 
rates leave fewer participants who reached Stage 1, and 
estimators that fall short of the efficiency bound use 
what remains less effectively. At PREDICT-HD's censoring 
rate, this efficiency gap is the difference between a 
trial that is feasible and one that is not.

\section{Resolving Contradictory Results in PREDICT-HD}
\label{sec:realdata}

We apply SPYCE to data from PREDICT-HD, a long-term observational 
study of 1,485 participants carrying the Huntington 
disease gene mutation from multiple sites and one of the 
few studies with the neuroimaging data needed to determine 
whether participants have reached Stage 1. Our goals are to 
identify which outcomes change rapidly enough before 
Stage 1 to serve as sensitive endpoints and to compute 
the sample sizes required for trials targeting those endpoints.

We take $X$ as time to Stage 1 and $C$ 
as time to study exit, both measured from study entry. We included only those 
participants who had not yet reached Stage 1 at study 
entry, were at least 18 years old, and who had at least one 
follow-up visit, which yielded $n = 448$ participants, with 
187 reaching Stage 1 during the study and a censoring rate 
of 58.3\%. We examine six outcomes that Huntington disease researchers 
have proposed as sensitive endpoint candidates: two 
cognitive scores (Color and Word) from the Stroop Color 
Word Test, where lower scores indicate worse cognition; 
the Total Motor Score (TMS, log-transformed as 
$\log(\mathrm{TMS}+1)$ to reduce skewness), where higher 
scores indicate greater motor deterioration; putamen and 
caudate volume ratios (a brain region's volume relative to 
total intracranial volume, measured in $10^{-2}\%$), 
where lower values indicate greater atrophy; and the 
composite Unified Huntington Disease Rating 
Scale score (cUHDRS), where lower scores indicate more 
advanced disease.

We model each outcome as $Y \mid X, \Z \sim 
\Normal(\beta_1 + \beta_2 X + \beta_3\Z + \beta_4 X\Z, 
\sigma^2)$, adjusting for a binary high-risk indicator 
$\Z = I(\mathrm{CAP} > 368)$, where CAP, the CAG-Age 
Product, measures cumulative disease risk, with larger 
values indicating higher risk \citep{zhang2011indexing}.
The slope parameters $\beta_2$ (low-risk) and $\beta_2 + 
\beta_4$ (high-risk) quantify how rapidly each outcome 
declines as participants approach Stage 1; a steeper 
negative slope indicates faster decline and requires 
fewer participants to detect a treatment effect. Since 
higher TMS scores indicate greater motor deterioration, 
the estimated slope for TMS is multiplied by $-1$ so 
that a negative slope indicates 
worsening for all six outcomes. To translate slope estimates into trial 
planning, we compute the required sample size per arm 
to detect a 50\% reduction in the slope, assuming 80\% 
power and two-sided $\alpha = 0.05$:
$n \geq (z_{\alpha/2} + z_{\beta})^2 \; 
\bd^\top\bSigma\,\bd / (d_1 - d_0)^2$,
where $d_0 = \bd^\top\wh\btheta$ is the estimated slope 
without treatment, $d_1 = 0.5 \times d_0$ is the slope 
if treatment slows decline by 50\%, and $\bd$ is a 
contrast vector selecting the slope of interest. We 
apply two versions of SPYCE and compare them to the CC, 
IMP, and IPW estimators and the MLE. SPYCE-Par (SPYCE 
with parametric specification of the nuisance models) 
specifies $X \mid \Z \sim \TruncNormal(\alpha_{11} + 
\alpha_{12}\Z, \tau_1^2;\, 0, 12)$ and $C \mid Y, \Z 
\sim \TruncNormal(\alpha_{21} + \alpha_{22}Y + 
\alpha_{23}\Z, \tau_2^2;\, 0, 12)$; SPYCE-Non (SPYCE 
with nonparametric estimation of the nuisance models) 
uses the expectation operators $\wh E_1$ in 
\eqref{eq:n15} and $\wh E_2$ in \eqref{eq:n16}.

We focus on the high-risk group because required sample 
sizes for the low-risk group exceed what any single 
rare-disease trial could feasibly enroll: low-risk 
participants are far from Stage 1, so slopes are not 
steep enough to detect a treatment effect over a 
realistic trial duration. Results for the low-risk group 
appear in Table~\ref{tab:stage1_low}.

\begin{figure}[!t]
    \centering
    \includegraphics[width=0.8\textwidth]{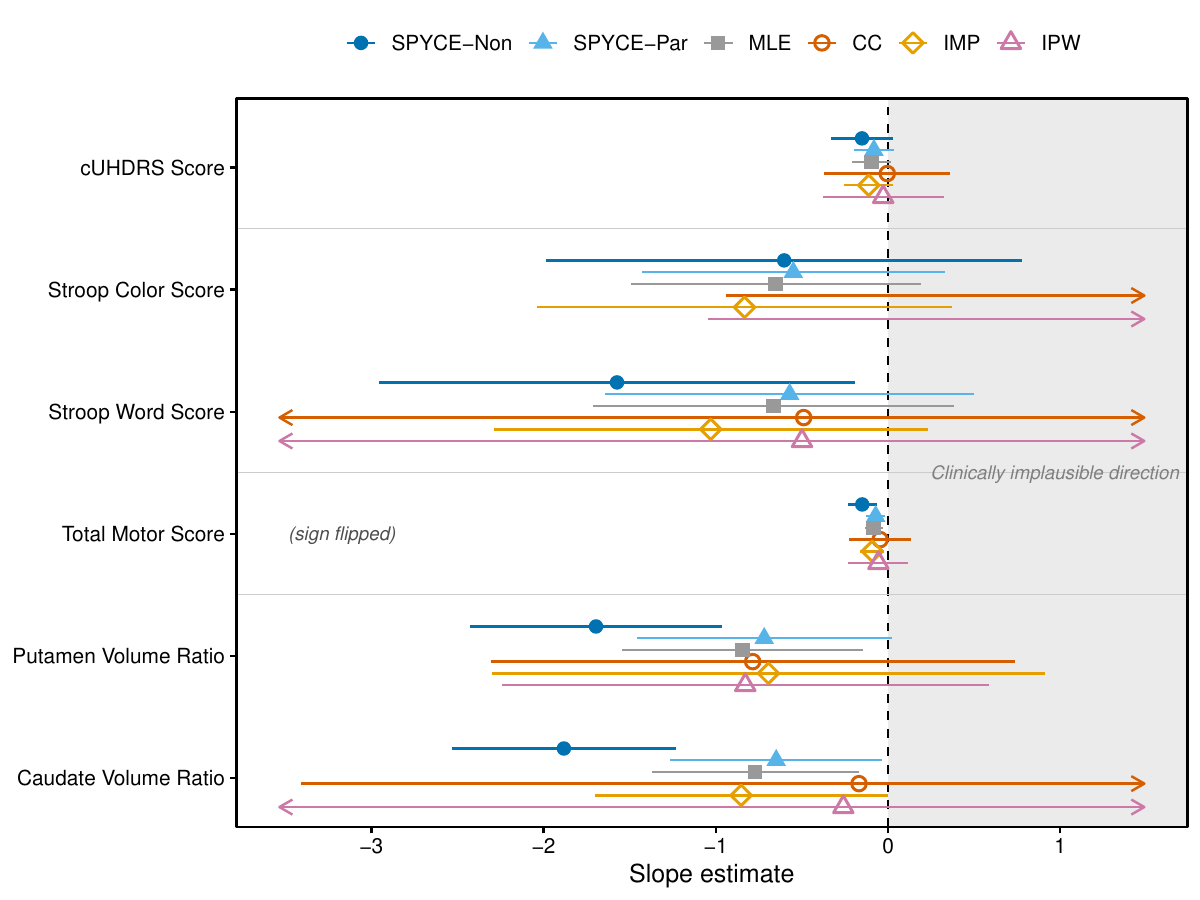}
    \label{Result}
    \caption{Slope estimates and 95\% confidence intervals 
for six outcomes in the high-risk group across six 
estimators. A negative slope indicates worsening as 
participants approach Stage 1; the gray shaded region 
indicates slopes in the clinically implausible 
direction. 
} \label{fig:forest_plot}
\end{figure}

\noindent\textbf{Existing estimators give contradictory 
or implausible slope estimates.} Prior analyses of 
PREDICT-HD data have used the CC estimator 
\citep{long2014tracking, langbehn2020clinical}, but 
under outcome-dependent censoring, it produces clinically 
implausible results. For the Stroop Color score, it 
returns a positive slope, suggesting cognitive 
improvement as participants approach Stage 1, as seen 
in Figures~\ref{fig:intro_plot} and 
\ref{fig:forest_plot}. For the cUHDRS score---now a 
primary endpoint in trials targeting early Huntington 
disease \citep{mccolgan2023tominersen}---the CC 
estimator returns $-0.004$, a slope so close to zero 
that it would lead a researcher to conclude the field's 
preferred endpoint is too slow-changing to detect a 
treatment effect (Table~\ref{tab:stage1_high}). 

\begin{table}[!ht]
\centering
\setlength{\tabcolsep}{6pt}
\renewcommand{\arraystretch}{1.25}
\small
\caption{Slope estimates, standard errors, and required 
sample sizes per arm for six outcomes in the high-risk 
group. Each cell shows slope (standard error) and 
[required sample size]. Sample sizes assume 50\% 
treatment slowing, 80\% power, and two-sided $\alpha 
= 0.05$. The smallest sample size in each row is 
bolded. $\dag$ indicates a slope in the clinically 
implausible direction (improvement as participants 
approach Stage 1); because the slope is in the wrong 
direction, a meaningful sample size cannot be computed. 
}
\label{tab:stage1_high}
\resizebox{\linewidth}{!}{%
\begin{tabular}{l|cc|cccc}
\toprule
Outcome & SPYCE-Non & SPYCE-Par & MLE & CC & IMP & IPW \\
\hline

\multirow{2}{*}{cUHDRS Score}
& $-0.151$ (0.092) & $-0.082$ (0.060) & $-0.096$ (0.059) & $-0.004$ (0.186) & $-0.112$ (0.073) & $-0.028$ (0.180) \\
& \textbf{[5,035]} & {[7,411]} & {[5,159]} & {[2.37e+07]} & {[5,848]} & {[558,506]} \\
\hline

\multirow{2}{*}{Stroop Color Score}
& $-0.603$ (0.706) & $-0.550$ (0.449) & $-0.652$ (0.430) & $2.146$ (1.575) & $-0.833$ (0.614) & $2.192$ (1.653) \\
& {[18,992]} & {[9,235]} & \textbf{[6,027]} & {$\dag$} & {[7,545]} & {$\dag$} \\

\multirow{2}{*}{Stroop Word Score}
& $-1.574$ (0.704) & $-0.571$ (0.547) & $-0.666$ (0.534) & $-0.490$ (1.962) & $-1.029$ (0.642) & $-0.500$ (2.040) \\
& \textbf{[2,786]} & {[12,771]} & {[8,943]} & {[222,968]} & {[5,411]} & {[231,708]} \\
\hline

Total Motor Score 
& $-0.150$ (0.043) & $-0.072$ (0.028) & $-0.084$ (0.027) & $-0.046$ (0.091) & $-0.092$ (0.036) & $-0.056$ (0.089) \\
\textit{(sign flipped)} & \textbf{[1,151]} & {[2,176]} & {[1,404]} & {[54,689]} & {[2,126]} & {[35,396]} \\
\hline

\multirow{2}{*}{Putamen Volume Ratio}
& $-1.696$ (0.372) & $-0.719$ (0.378) & $-0.846$ (0.357) & $-0.786$ (0.777) & $-0.695$ (0.818) & $-0.829$ (0.721) \\
& \textbf{[373]} & {[2,140]} & {[1,379]} & {[7,568]} & {[10,754]} & {[5,870]} \\

\multirow{2}{*}{Caudate Volume Ratio}
& $-1.882$ (0.332) & $-0.650$ (0.315) & $-0.772$ (0.307) & $-0.169$ (1.653) & $-0.853$ (0.434) & $-0.260$ (2.049) \\
& \textbf{[241]} & {[1,818]} & {[1,224]} & {[745,560]} & {[2,011]} & {[482,676]} \\
\bottomrule
\end{tabular}
}
\end{table}

\noindent\textbf{SPYCE gives consistent, clinically 
credible slope estimates.} SPYCE-Non returns a negative 
slope estimate for the Stroop Color score, whereas the 
CC and IPW estimators do not, consistent with expected 
cognitive decline, though the confidence interval 
includes zero so the direction cannot be confirmed. For the 
cUHDRS score, SPYCE-Non estimates a slope of $-0.151$, 
consistent with the field's decision to adopt it as a 
sensitive endpoint, and requires 5,035 participants per 
arm---a feasible target for a Huntington disease 
trial---compared to 558,506 for the IPW estimator and 
23.7 million for the CC estimator, numbers no 
rare-disease trial could realistically recruit.

Table~\ref{tab:bump-chart} ranks all six outcomes by 
estimated effect size---the estimated slope difference 
divided by its standard error, $|d_1 - d_0| / 
\sqrt{\bd^\top\bSigma\,\bd}$, which is equivalent to 
ranking by required sample size since larger effect 
sizes require fewer participants to detect a treatment 
effect. SPYCE-Par and the IPW estimator are omitted from this table because their estimates closely track those of the MLE and CC estimator, respectively. 
Caudate and putamen volume ratios change most 
rapidly in the earliest stages of Huntington disease 
\citep{paulsen2014prediction}, making them the most 
promising sensitive endpoints for trials targeting 
early Huntington disease, and SPYCE-Non ranks them 
first and second, consistent with the established 
literature. The other estimators diverge from SPYCE-Non's rankings 
in ways that align with their known vulnerabilities: the 
MLE agrees with SPYCE-Non on the top three outcomes 
but diverges in the bottom half, where its rankings 
may reflect its dependence on the specification of 
$\eta_1$; the IMP estimator ranks putamen volume ratio 
last; and the CC estimator drops caudate volume ratio 
to fourth.

\begin{table}[!t]
\centering
\caption{Outcome rankings by estimated effect size 
across four estimators, high-risk group. Outcomes are 
ranked from 1 (largest estimated effect size, fewest 
participants required) to 6 (smallest). For the CC 
estimator, the Stroop Color score produced a slope in 
the clinically implausible direction, so the effect 
size cannot be computed. 
}
\footnotesize
\setlength{\tabcolsep}{6pt}
\resizebox{\textwidth}{!}{
\begin{tabular}{
    c
    >{\centering\arraybackslash}m{4.1cm}
    >{\centering\arraybackslash}m{4.1cm}
    >{\centering\arraybackslash}m{4.1cm}
    >{\centering\arraybackslash}m{4.1cm}
}
\toprule
{Rank} & {SPYCE-Non} & {MLE} & {IMP} & {CC} \\
\midrule
1 & \caudatecell{Caudate Volume Ratio}
  & \caudatecell{Caudate Volume Ratio}
  & \caudatecell{Caudate Volume Ratio}
  & \putamencell{Putamen Volume Ratio} \\

2 & \putamencell{Putamen Volume Ratio}
  & \putamencell{Putamen Volume Ratio}
  & \tmscell{Total Motor Score}
  & \tmscell{Total Motor Score} \\

3 & \tmscell{Total Motor Score}
  & \tmscell{Total Motor Score}
  & \stroopwordcell{Stroop Word Score}
  & \stroopwordcell{Stroop Word Score} \\

4 & \stroopwordcell{Stroop Word Score}
  & \cuhdrcell{cUHDRS Score}
  & \cuhdrcell{cUHDRS Score}
  & \caudatecell{Caudate Volume Ratio} \\

5 & \cuhdrcell{cUHDRS Score}
  & \stroopcolorcell{Stroop Color Score}
  & \stroopcolorcell{Stroop Color Score}
  & \cuhdrcell{cUHDRS Score} \\

6 & \stroopcolorcell{Stroop Color Score}
  & \stroopwordcell{Stroop Word Score}
  & \putamencell{Putamen Volume Ratio}
  & \multicolumn{1}{c}{\textit{---}} \\


\bottomrule
\end{tabular}
}
\label{tab:bump-chart}
\end{table}

\begin{figure}[!t]
    \centering
    \includegraphics[width=0.82\textwidth]{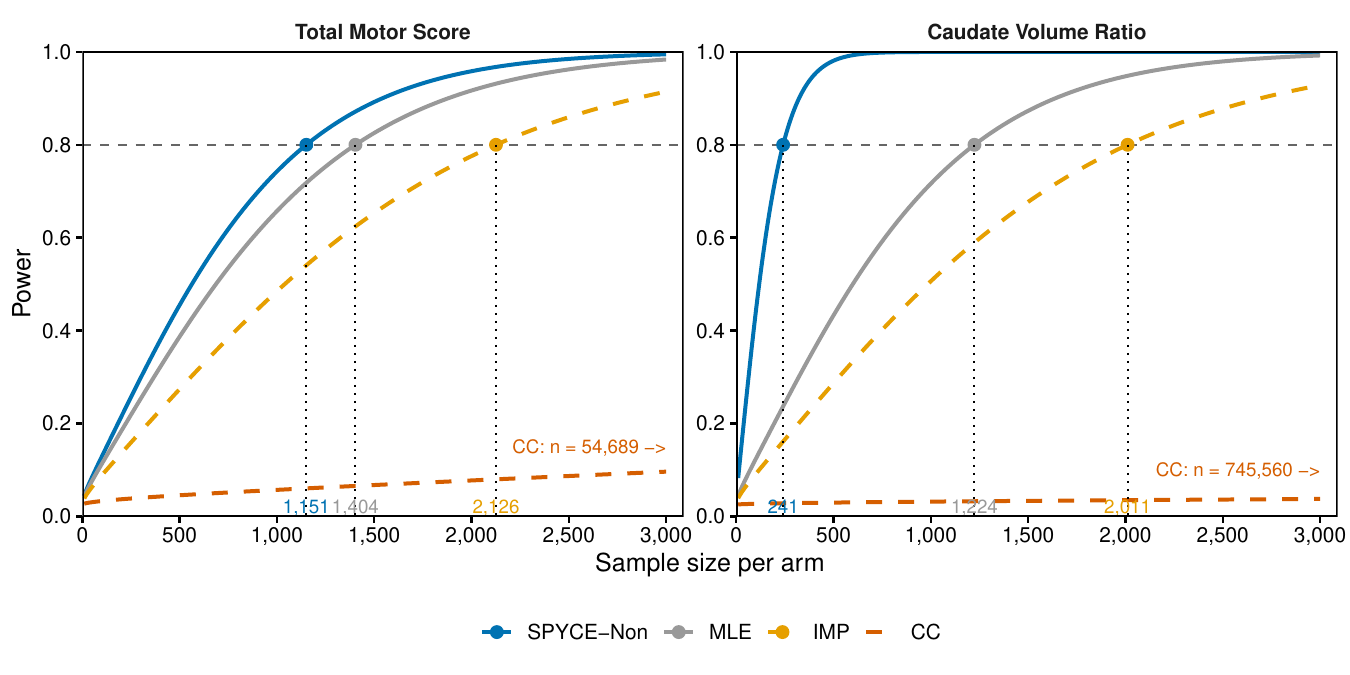}
    \caption{Power curves for detecting 50\% treatment 
slowing in the high-risk group, assuming two-sided 
$\alpha = 0.05$. Total Motor Score (left panel) and 
caudate volume ratio (right panel) are the 
highest-ranked outcomes in the motor and neuroimaging 
domains, respectively. The horizontal dashed line 
indicates 80\% power; vertical dotted lines mark 
where each estimator reaches 80\% power. 
}\label{fig:power_curves}
\end{figure}

A researcher relying on the IMP estimator or the CC estimator would deprioritize exactly 
these outcomes. The 
practical consequence is visible in the power curves 
(Figure~\ref{fig:power_curves}): for the Total Motor Score, the highest-ranked 
non-neuroimaging outcome, achieving 80\% power 
requires 1,151 participants per arm under SPYCE-Non---a 
feasible target for a Huntington disease trial---compared 
to 2,126 for the IMP estimator and 1,404 for the MLE; 
none of the remaining estimators yield feasible sample 
sizes, with the CC estimator requiring 54,689 
participants per arm and the IPW estimator 35,396. 
For caudate volume ratio, ranked first 
overall, achieving 80\% power requires 241 
participants per arm under SPYCE-Non compared to 
2,011 for the IMP estimator and 1,224 for the MLE; 
the CC and IPW estimators require 745,560 and 482,676 
per arm, respectively---far beyond what any 
trial could recruit for a rare disease 
affecting fewer than 5 per 100,000 people worldwide 
(Table~\ref{tab:stage1_high}). A trial requiring 
hundreds of thousands of participants per arm cannot 
be run. One requiring 241 can.

\section{Discussion}\label{sec:discussion}

Huntington disease researchers developed the staging 
system to identify when early intervention holds the 
most promise, but how rapidly outcomes change as 
participants approach Stage 1 under outcome-dependent 
censoring remained an open statistical problem. 
The sample size differences in Table~\ref{tab:stage1_high} 
show what that costs: an estimator that cannot handle 
outcome-dependent censoring deprioritizes the field's 
preferred sensitive endpoints and requires enrolling 
more participants than any rare-disease trial could 
feasibly recruit.

 Before this work, researchers facing outcome-dependent censoring 
had no doubly robust estimator, and the only alternative was 
accepting bias from singly robust estimators. SPYCE removes 
that constraint: it remains consistent when either nuisance 
model is correctly specified, and when neither can be correctly 
specified, nonparametric estimation of both expectation operators 
still achieves the semiparametric efficiency bound. The most precise 
slope estimates the data can support are therefore 
available without requiring correct specification of 
either nuisance model.

Identifying sensitive endpoints and computing required 
sample sizes, however, are only two of three statistical 
problems that must be solved before a trial targeting 
early Huntington disease can be run. Careful trial 
planning requires solving a third problem: identifying which 
participants will reach Stage 1 during the trial. 
This problem remains open. Solving it will require a 
model for time to Stage 1 that may be difficult to specify 
correctly, and no current estimator fully addresses this 
challenge. SPYCE-Non does offer partial protection: 
because it estimates $\eta_1$ nonparametrically rather 
than relying on a parametric form, it produces valid slope 
estimates and required sample sizes even when the 
time to Stage 1 model is misspecified---but predicting 
which specific participants will reach Stage 1 during 
a trial remains a separate, unsolved problem.

The problem SPYCE addresses is not unique to Huntington 
disease. Studies in which the participants declining most 
rapidly drop out earliest are common in diseases where 
early intervention holds the most promise. In 
Parkinson disease, participants identified as at high 
risk through genetic markers or REM sleep behavior 
disorder exit studies before the onset of motor symptoms; 
in spinocerebellar ataxia, participants identified as 
at high risk through genetic testing exit before the 
onset of ataxia. In both settings, as in Huntington 
disease, the participants lost to follow-up are those 
whose data matter most for understanding early disease 
progression and for planning trials that aim to 
intervene before it is too late. Ignoring outcome-dependent censoring in such studies 
carries a scientific cost: endpoints are misidentified 
and trials are underpowered to detect the treatments 
that might help. In diseases where early intervention 
holds the most promise, SPYCE brings researchers a 
step closer to trials that can actually be run.


\section*{Disclosure Statement}\label{disclosure-statement}

No potential conflict of interest is reported by the authors.

\section*{Data Availability Statement}

PREDICT-HD data are available upon request from the CHDI Foundation, Inc.
  
\phantomsection\label{supplementary-material}

\begin{center}

{\large\bf SUPPLEMENTARY MATERIALS}

\end{center}

\begin{description}
\item Supplementary materials include implementation details, additional numerical results, and derivations for all theoretical claims. 
\end{description}

\bibliography{censcov_odc_ref}

@article{Aisenetal2022,
author={Aisen, P.S. and Jimenez-Maggiora, G.A. and Rafii, M.S. and others},
title={Early-stage {A}lzheimer disease: getting trial-ready},
journal={Nature Reviews Neurology},
volume={18},
pages={389–399},
year={2022}
}

@article{Tabrizietal2022,
author={Tabrizi, S.J. and others},
title={Potential disease-modifying therapies for {H}untington's disease: lessons learned and future opportunities},
journal={Lancet Neurology},
year={2022},
volume={21},
pages={645-658}
}

@article{newey1994large,
  title={Large sample estimation and hypothesis testing},
  author={Newey, Whitney K and McFadden, Daniel},
  journal={Handbook of Econometrics},
  volume={4},
  pages={2111--2245},
  year={1994},
  publisher={Elsevier}
}

@article{jhx2022nonparametric,
author = {Jiang, Hui and Huang, Lei and Xia, Yingcun},
title = {Nonparametric regression with right-censored covariate via conditional density function},
journal = {Statistics in Medicine},
volume = {41},
number = {11},
pages = {2025-2051},
year = {2022}
}

@article{matsouaka2020regression,
  title={Regression with a right-censored predictor using inverse probability weighting methods},
  author={Matsouaka, Roland A and Atem, Folefac D},
  journal={Statistics in Medicine},
  volume={39},
  number={27},
  pages={4001--4015},
  year={2020},
  publisher={Wiley Online Library}
}

@article{long2014tracking,
  title={Tracking motor impairments in the progression of {Huntington}'s disease},
  author={Long, Jeffery D and Paulsen, Jane S and Marder and others},
  journal={Movement Disorders},
  volume={29},
  number={3},
  pages={311--319},
  year={2014},
  publisher={Wiley Online Library}
}

@article{zhang2011indexing,
  title={Indexing disease progression at study entry with individuals at-risk for {Huntington} disease},
  author={Zhang, Ying and Long, Jeffrey D and Mills, James A and Warner, John H and Lu, Wenjing and others},
  journal={American Journal of Medical Genetics Part B: Neuropsychiatric Genetics},
  volume={156},
  number={7},
  pages={751--763},
  year={2011},
  publisher={Wiley Online Library}
}

@article{lee2024robust,
  author = {Lee, S. and Richardson, B. D. and Ma, Y. and Marder, K. S. and Garcia, T. P.},
title = {{SPARCC}: Semi-Parametric Robust Estimation in a Right-Censored Covariate Model},
journal = {Journal of the American Statistical Association},
volume = {0},
number = {0},
pages = {1--12},
year = {2026},
publisher = {Taylor \& Francis},
doi = {10.1080/01621459.2025.2562645},
}

@article{zhang2025super,
  title={Super doubly robust and efficient estimator for informative covariate censoring},
  author={Zhang, Zhewei and Ma, Yanyuan and Marder, Karen and Garcia, Tanya P},
  journal={arXiv preprint arXiv:2511.02187},
  year={2025}
}

@article{lv2017maximum,
  title={Maximum weighted likelihood for discrete choice models with a dependently censored covariate},
  author={Lv, Xiaofeng and Zhang, Gupeng and Li, Qinghai and Li, Rui},
  journal={Journal of the Korean Statistical Society},
  volume={46},
  number={1},
  pages={15--27},
  year={2017},
  publisher={Elsevier}
}

@article{atem2017linear,
journal={Biostatistics and Biometrics Open Access Journal},
author={Folefac D Atem and Roland A Matsouaka},
title={Linear Regression Model with a Randomly Censored Predictor: Estimation Procedures},
year={2017},
month={March},
pages={21-32},
volume={1},
number={2},
doi={10.19080/BBOAJ.2017.01.555556},
}

@article{lotspeich2024making,
  title={Making sense of censored covariates: statistical methods for studies of {Huntington}'s disease},
  author={Lotspeich, Sarah C and Ashner, Marissa C and Vazquez, Jesus E and Richardson, Brian D and Grosser, Kyle F and Bodek, Benjamin E and Garcia, Tanya P},
  journal={Annual Review of Statistics and its Application},
  volume={11},
  year={2024},
  publisher={Annual Reviews}
}

@article{dempsey2018survival,
  title={Survival models and health sequences},
  author={Dempsey, Walter and McCullagh, Peter},
  journal={Lifetime Data Analysis},
  volume={24},
  pages={550--584},
  year={2018},
  publisher={Springer}
}

@article{chu2020stochastic,
  title={Stochastic functional estimates in longitudinal models with interval-censored anchoring events},
  author={Chu, Chenghao and Zhang, Ying and Tu, Wanzhu},
  journal={Scandinavian Journal of Statistics},
  volume={47},
  number={3},
  pages={638--661},
  year={2020},
  publisher={Wiley Online Library}
}

@article{kang2025dynamics,
  title={The dynamics of cognitive decline toward {Alzheimer}'s disease progression: Results from {ADSP-PHC}'s harmonized cognitive composites},
  author={Kang, Kaidi and Zhang, Panpan and Dumitrescu, Logan and Mukherjee, Shubhabrata and Lee, Michael L and others},
  journal={Alzheimer's \& Dementia},
  volume={21},
  number={6},
  pages={e70335},
  year={2025},
  publisher={Wiley Online Library}
}

@article{paulsen2014prediction,
  title={Prediction of manifest {Huntington}'s disease with clinical and imaging measures: a prospective observational study},
  author={Paulsen, Jane S and Long, Jeffrey D and Ross, Christopher A and Harrington, Deborah L and others},
  journal={The Lancet Neurology},
  volume={13},
  number={12},
  pages={1193--1201},
  year={2014},
  publisher={Elsevier}
}

@book{bickel1993efficient,
  title={Efficient and adaptive estimation for semiparametric models},
  author={Bickel, Peter J and Klaassen, Chris AJ and Ritov, Ya’acov and Wellner, Jon A},
  volume={4},
  year={1993},
  publisher={Springer}
}

@book{tsiatis2006semiparametric,
  title={Semiparametric theory and missing data},
  author={Tsiatis, Anastasios A},
  volume={4},
  year={2006},
  publisher={Springer},
 address = {New York}
}

@article{langbehn2020clinical,
  title={Clinical outcomes and selection criteria for prodromal Huntington's disease trials},
  author={Langbehn, Douglas R and Hersch, Steven},
  journal={Movement Disorders},
  volume={35},
  number={12},
  pages={2193--2200},
  year={2020},
  publisher={Wiley Online Library}
}

@article{mccolgan2023tominersen,
  title={Tominersen in adults with manifest {H}untington’s disease},
  author={McColgan, Peter and Thobhani, Alpa and Boak, Lauren and others},
  journal={New England Journal of Medicine},
  volume={389},
  number={23},
  pages={2203--2205},
  year={2023},
  publisher={Mass Medical Soc}
}

@article{macdonald1993novel,
  title={A novel gene containing a trinucleotide repeat that is expanded and unstable on {Huntington}'s disease chromosomes},
  author={MacDonald, Marcy E and Ambrose, Christine M and Duyao, Mabel P and Myers, Richard H and Lin, Carol and Srinidhi, Lakshmi and Barnes, Glenn and Taylor, Sherryl A and James, Marianne and Groot, Nicolet and others},
  journal={Cell},
  volume={72},
  number={6},
  pages={971--983},
  year={1993},
  publisher={Elsevier}
}

@article{tabrizi2022biological,
  title={A biological classification of {Huntington}'s disease: the {Integrated Staging System}},
  author={Tabrizi, Sarah J and Schobel, Scott and Gantman, Emily C and Mansbach, Alexandra and Borowsky, Beth and Konstantinova, Pavlina and Mestre, Tiago A and Panagoulias, Jennifer and Ross, Christopher A and Zauderer, Maurice and others},
  journal={The Lancet Neurology},
  volume={21},
  number={7},
  pages={632--644},
  year={2022},
  publisher={Elsevier}
}

@article{hajek1971comment,
  title={Comment on “An essay on the logical foundations of survey sampling, part one”},
  author={H{\'a}jek, Jaroslav},
  journal={Foundations of Statistical Inference},
  volume={236},
  year={1971},
  publisher={Holt, Rinehart, and Winston Toronto}
}

@article{gm1994asymptotic,
author = {W. Gonzalez-Manteiga and C. Cadarso-Suarez},
title = {Asymptotic properties of a generalized {Kaplan-Meier} estimator with some applications},
journal = {Journal of Nonparametric Statistics},
volume = {4},
number = {1},
pages = {65-78},
year = {1994},
publisher = {Taylor \& Francis}
}

@article{dabrowska1989uniform,
  title={Uniform consistency of the kernel conditional {Kaplan-Meier} estimate},
  author={Dabrowska, Dorota M},
   journal = {Annals of Statistics},
 number = {3},
 pages = {1157--1167},
 publisher = {Institute of Mathematical Statistics},
 volume = {17},
 year = {1989}
}

\clearpage
\newpage

{\small

\begin{center}
    {\LARGE\bf Supplementary Materials}
\end{center}

\setcounter{section}{0}\renewcommand{\thesection}{\Alph{section}}
\setcounter{subsection}{0}\renewcommand{\thesubsection}{A.\arabic{subsection}}
\setcounter{equation}{0}\renewcommand{\theequation}{A.\arabic{equation}}
\setcounter{Pro}{0}\renewcommand{\thePro}{A.\arabic{Pro}}
\setcounter{Th}{0}\renewcommand{\theTh}{A.\arabic{Th}}
\setcounter{Lem}{0}\renewcommand{\theLem}{A.\arabic{Lem}}
\setcounter{Rem}{0}\renewcommand{\theRem}{A.\arabic{Rem}}
\setcounter{Cor}{0}\renewcommand{\theCor}{A.\arabic{Cor}}
\setcounter{table}{0}\renewcommand{\thetable}{A.\arabic{table}}

\section{Implementation Details and Additional Numerical Results}

\subsection{HD-ISS Classification Details}
\label{sec:stage1-details}
Huntington disease is caused by an unstable expansion of CAG (cytosine-adenine-guanine) repeats in the huntingtin gene \citep{macdonald1993novel}. Participants with 40 or more CAG repeats are certain to develop the disease, and the Huntington Disease Integrated Staging System (HD-ISS) provides a standardized means of staging their disease progression from birth onward \citep{tabrizi2022biological}.

The HD-ISS is designed to reflect the monotonic progression of Huntington disease, so that participants in higher stages are considered more advanced than those in lower stages, similar to cancer staging. By sharing a common staging system, researchers can consistently compare participants across different studies and design clinical trials that target interventions at clearly defined milestones.

In the HD-ISS, Stage 0 begins at birth for participants whose CAG-repeat length is $\geq 40$. Stage 1 is reached when the putamen or caudate---brain structures that support motor and cognitive function---shows atrophy relative to the intracranial volume. Specifically, a participant reaches Stage 1 when either the putamen volume ratio (putamen volume divided by intracranial volume) or caudate volume ratio (caudate volume divided by intracranial volume) falls below predetermined thresholds.

These thresholds for Stage 1 are selected as the extreme 5\% of a reference distribution from healthy participants---people with fewer than 35 CAG repeats who are not expected to develop Huntington disease. The thresholds for putamen volume ratio and caudate volume ratio are set as the lower 5\% of values observed in healthy controls, representing the point below which brain atrophy is considered abnormal and Stage 1 is declared.

\subsection{Estimation of the Asymptotic Variance} \label{sec:estimation}

We estimate the asymptotic variances of $\wh \bb$ from Theorem \ref{th:par} and Theorem \ref{th:nonpar}  to investigate the statistical inference of those estimators. For parametric nuisance models, for a function $g(\bO; \bb, \balpha_1, \balpha_2)$ with 
$E\{g(\bO; \bb, \balpha_1, \balpha_2)\} = \0$, we have 
\bse
E[\partial g(\bO; \bb, \balpha_1, \balpha_2)/\partial \balpha_1 \trans]
&=& - E[g(\bO; \bb, \balpha_1, \balpha_2) \S_1\trans(\bO; \bb, \balpha_1, \balpha_2) ],\\
E[\partial g(\bO; \bb, \balpha_1, \balpha_2)/\partial \balpha_2 \trans]
&=& - E[g(\bO; \bb, \balpha_1, \balpha_2) \S_2\trans(\bO; \bb, \balpha_1, \balpha_2) ],\\
E[\partial g(\bO; \bb, \balpha_1, \balpha_2)/\partial \bb \trans]
&=& - E[g(\bO; \bb, \balpha_1, \balpha_2) \S_\bb\trans(\bO; \bb, \balpha_1, \balpha_2)].
\ese
When the forms of the parametric nuisance models are specified, the score functions $\bS_1$ and $\bS_2$ are readily computable from the model structures. For  nonparametric estimation, the expectation of the following form can be calculated as
\bse
E[\partial g(\bO; \bb)/\partial \bb \trans]
&=& - E[g(\bO; \bb) \S_\bb\trans(\bO; \bb)].
\ese
In Theorem \ref{th:nonpar}, the variance estimation requires evaluating the functions  $\bh_{1{\rm s}1}^\star$, $\bh_{1{\rm s}2}^\star$, $\bh_{1{\rm k}1}^\star$, 
$\bh_{1{\rm k}2}^\star$, $\bh_{2{\rm s}}^*$, and $\bh_{2{\rm k}}^*$ at each observation $(y_j,w_j, \delta_j, \z_j)$.
For \ref{ncase1} of Theorem \ref{th:nonpar}, we have
\bse
&&\bh_{1{\rm s}1}^\star(y_j,w_j, \delta_j, \z_j)\n\\
&=&- E\left[ I(X>C) \{\S\eff^\star(Y,
    X, 1, \z_j;\bb_0) -\S\eff^\star(Y, C, 0, \z_j;\bb_0)\} \right.\n\\
&& \left.\times\frac{\xi_C (w_j, \delta_j, X,y_j,\z_j) f_{Y|X,\Z}(y_j,
    X, \z_j)}{S_{C|Y,\Z} (X,y_j,\z_j)f_{Y|\Z}(y_j, \z_j)} \mid \Z = \z_j,\bo_j\right]\n\\
&=&- E\left( E\left[ \{1-S_{C|Y,\Z}(X,Y,\z_j)\} \S\eff^\star(Y,
    X, 1, \z_j;\bb_0) -E\{I(X>C)\S\eff^\star(Y, C, 0, \z_j;\bb_0)\mid X,Y,\Z\} \mid X,\Z\right]\right.\n\\
&& \left.\times\frac{\xi_C (w_j, \delta_j, X,y_j,\z_j)}{S_{C|Y,\Z} (X,y_j,\z_j)} \mid Y = y_j, \Z = \z_j,\bo_j\right).
\ese
From \eqref{eq:n0} and \eqref{eq:n1}, we obtain
\be\label{eq:n31}
&&E[E_2^\star\{ I(X\le C) \S\eff^\star(Y,X,1,\Z;\bb_0)\\
&&+ I(X>C) \S\eff^\star(Y,C,0,\Z;\bb_0) \mid X,Y,\Z \}|X,\Z]
= \0,\n
\ee
which leads to 
\bse
&&\bh_{1{\rm s}2}^\star(y_j,w_j, \delta_j, \z_j)\\
&=& E_2^\star\left[ I(X>C)\{\S\eff^\star(Y,X,1,\z_j;\bb_0) -\S\eff^\star(Y,C,0,\z_j;\bb_0) \}
\right.\n\\
&&\times \left.\frac{ \xi_C (w_j, \delta_j, X,y_j,\z_j)f_{Y|X,\Z}(y_j,X,\z_j)}{S_{C|Y,\Z} (X,y_j,\z_j)f_{Y|\Z}(y_j,\z_j)} \mid \Z = \z_j,\bo_j;\bb_0\right]\\
&=& E\left[E\{\S\eff^\star(Y,X,1,\z_j;\bb_0)\mid X,\Z\} \frac{ \xi_C (w_j, \delta_j, X,y_j,\z_j)}{S_{C|Y,\Z} (X,y_j,\z_j)} \mid Y = y_j, \Z = \z_j,\bo_j\right].
\ese
Then $\bh_{1{\rm s}}^\star(y_j,w_j, \delta_j, \z_j)$ becomes
\bse
&&\bh_{1{\rm s}}^\star(y_j,w_j, \delta_j, \z_j)\\
&=&\bh_{1{\rm s}1}^\star(y_j,w_j, \delta_j, \z_j) + \bh_{1{\rm s}2}^\star(y_j,w_j, \delta_j, \z_j)\\
&=& E\left( E\left[ S_{C|Y,\Z}(X,Y,\z_j) \S\eff^\star(Y,
    X, 1, \z_j;\bb_0) + E\{I(X>C)\S\eff^\star(Y, C, 0, \z_j;\bb_0)\mid X,Y,\Z\} \mid X,\Z\right]\right.\n\\
&& \left.\times\frac{\xi_C (w_j, \delta_j, X,y_j,\z_j)}{S_{C|Y,\Z} (X,y_j,\z_j)} \mid Y = y_j, \Z = \z_j,\bo_j\right),
\ese
which can be estimated as
\bse
&&\wh \bh_{1{\rm s}}^\star(y_j,w_j, \delta_j, \z_j)\\
    &=& \wh E_1\left( E\left[ \wh S_{C|Y,\Z}(X,Y,\z_j) \S\eff^\star(Y,
    X, 1, \z_j;\wh\bb)\right.\right.\\
    &&\left.\left.+ \wh E_2\{I(X>C)\S\eff^\star(Y, C, 0, \z_j;\wh\bb)\mid X,Y,\Z\} \mid X,\Z;\wh\bb\right]\right.\n\\
&& \left.\times\frac{\xi_C (w_j, \delta_j, X,y_j,\z_j)}{S_{C|Y,\Z} (X,y_j,\z_j)} \mid Y = y_j, \Z = \z_j,\bo_j\right).
\ese
From \eqref{eq:n31}, we obtain
\bse
&&\bh_{1{\rm k}2}^\star(y_j,w_j, \delta_j, \z_j)\\
&=& -\frac{\delta_j }{S_{C|Y,\Z} (w_j,y_j,\z_j)}E_2^\star\left[I(w_j>C) \{\S\eff^\star(Y,w_j,1,\z_j;\bb_0)- \S\eff^\star(Y,C,0,\z_j;\bb_0)\}\right.\n\\
&&\left.\mid X=w_j, \Z = \z_j,\bo_j;\bb_0\right]\n\\
&=& -\frac{\delta_j }{S_{C|Y,\Z} (w_j,y_j,\z_j)}E\{\S\eff^\star(Y,w_j,1,\z_j;\bb_0) \mid X=w_j, \Z = \z_j,\bo_j\}.
\ese
Then $\bh_{1{\rm k}}^\star(y_j,w_j, \delta_j, \z_j)$ becomes
\bse
&&\bh_{1{\rm k}}^\star(y_j,w_j, \delta_j, \z_j)\\
&=&\bh_{1{\rm k}1}^\star(y_j,w_j, \delta_j, \z_j) + \bh_{1{\rm k}2}^\star(y_j,w_j, \delta_j, \z_j)\\
    &=&\frac{-\delta_j }{S_{C|Y,\Z} (w_j,y_j,\z_j)}E\left[S_{C|Y,\Z}(w_j,Y,\z_j)\S\eff^\star(Y,
    w_j, 1, \z_j;\bb_0)\right.\\
    &&\left.+E\{I(w_j>C)\S\eff^\star(Y, C, 0, \z_j;\bb_0)\mid Y,\Z\}\mid X = w_j, \Z = \z_j,\bo_j \right],
\ese
which can be estimated as
\bse
\wh \bh_{1{\rm k}}^\star(y_j,w_j, \delta_j, \z_j)&=& \frac{-\delta_j }{\wh S_{C|Y,\Z} (w_j,y_j,\z_j)}E\left[\wh S_{C|Y,\Z}(w_j,Y,\z_j)\S\eff^\star(Y,
    w_j, 1, \z_j;\wh\bb)\right.\\
    &&\left.+\wh E_2\{I(w_j>C)\S\eff^\star(Y, C, 0, \z_j;\wh\bb)\mid Y,\Z\}\mid X = w_j, \Z = \z_j,\bo_j; \wh\bb \right].
\ese
For \ref{ncase2} of Theorem \ref{th:nonpar}, we have
\bse
&&\bh_{2{\rm s}}^*(y_j,w_j, \delta_j, \z_j)\n\\
&=& E\left( \frac{ \xi_X (w_j, \delta_j, C,y_j,\z_j)}{S_{X|Y,\Z}(C,y_j,\z_j)} [I(X\le C)\S\eff^* (y_j,X,1,\z_j;\bb_0) + I(X>C)\S\eff^* (y_j,C,0,\z_j;\bb_0)\right.\n\\
&& - E\{I(X\le C)\S\eff^* (y_j,X,1,\z_j;\bb_0) + I(X>C)\S\eff^* (y_j,C,0,\z_j;\bb_0)\mid X, Y = y_j,\Z = \z_j\}]\n\\
&&\left.\mid Y = y_j,\Z = \z_j,\bo_j\right)\\
&=& \cov\left[E\{I(X\le C)\S\eff^* (y_j,X,1,\z_j;\bb_0)\mid C, Y = y_j,\Z = \z_j\} \right.\\
&&\left.+ S_{X|Y,\Z}(C,y_j,\z_j)\S\eff^* (y_j,C,0,\z_j;\bb_0), \frac{ \xi_X (w_j, \delta_j, C,y_j,\z_j)}{S_{X|Y,\Z}(C,y_j,\z_j)}\mid Y = y_j,\Z = \z_j,\bo_j\right],
\ese
which can be estimated as
\bse
&&\wh \bh_{2{\rm s}}^*(y_j,w_j, \delta_j, \z_j)\n\\
&=& \wh \cov_2\left[ \wh E_1 \{I(X\le C)\S\eff^* (y_j,X,1,\z_j;\wh\bb)\mid C, Y = y_j,\Z = \z_j\} \right.\\
&&\left.+ \wh S_{X|Y,\Z}(C,y_j,\z_j)\S\eff^* (y_j,C,0,\z_j;\wh\bb), \frac{ \xi_X (w_j, \delta_j, C,y_j,\z_j)}{S_{X|Y,\Z}(C,y_j,\z_j)}\mid Y = y_j,\Z = \z_j,\bo_j\right],
\ese
where $\wh\cov_2$ indicates the conditional covariance under $\wh E_2$. Lastly, in a similar manner,
\bse
&&\bh_{2{\rm k}}^*(y_j,w_j, \delta_j, \z_j)\n\\
&=& -\frac{1-\delta_j}{S_{X|Y,\Z} (w_j,y_j,\z_j)} E\left[ I(X \le w_j)\S\eff^*(y_j,X,1,\z_j;\bb_0) + I(X > w_j) \bS\eff^*(y_j,w_j,0,\z_j;\bb_0)\right.\n\\
&& - E\{I(X\le C)\S\eff^* (y_j,X,1,\z_j;\bb_0) + I(X>C)\S\eff^* (y_j,C,0,\z_j;\bb_0)\mid X, Y = y_j,\Z = \z_j\}\n\\
&& \left.\mid Y=y_j,\Z = \z_j,\bo_j\right]\\
&=& -\frac{1-\delta_j}{S_{X|Y,\Z} (w_j,y_j,\z_j)} (E\{I(X\le w_j)\S\eff^* (y_j,X,1,\z_j;\bb_0)\mid Y = y_j,\Z = \z_j\} \\
&&+ S_{X|Y,\Z}(w_j,y_j,\z_j)\S\eff^* (y_j,w_j,0,\z_j;\bb_0)\n\\
&& - E[E\{I(X\le C)\S\eff^* (y_j,X,1,\z_j;\bb_0)\mid C, Y = y_j,\Z = \z_j\} \\
&&+ S_{X|Y,\Z}(C,y_j,\z_j)\S\eff^* (y_j,C,0,\z_j;\bb_0)\mid Y = y_j,\Z = \z_j]),
\ese
which can be estimated as
\bse
&&\wh \bh_{2{\rm k}}^*(y_j,w_j, \delta_j, \z_j)\n\\
&=& -\frac{1-\delta_j}{\wh S_{X|Y,\Z} (w_j,y_j,\z_j)} (\wh E_1\{I(X\le w_j)\S\eff^* (y_j,X,1,\z_j;\wh\bb)\mid Y = y_j,\Z = \z_j\} \\
&&+ \wh S_{X|Y,\Z}(w_j,y_j,\z_j)\S\eff^* (y_j,w_j,0,\z_j;\wh\bb)\n\\
&& - \wh E_2[\wh E_1\{I(X\le C)\S\eff^* (y_j,X,1,\z_j;\wh\bb)\mid C, Y = y_j,\Z = \z_j\} \\
&&+ \wh S_{X|Y,\Z}(C,y_j,\z_j)\S\eff^* (y_j,C,0,\z_j;\wh\bb)\mid Y = y_j,\Z = \z_j]).
\ese

\subsection{Benchmark Estimators and Additional Simulation Results}
\label{sec:benchmark-additional-sims}

Using the data structure and likelihood defined in 
Section~\ref{sec:intro}, we present the estimating equations for 
four benchmark estimators used as comparators in our simulations 
and data analysis. As in the main text, we refer to 
$\eta_1 = f_{X|\Z}$ (the time to Stage 1 model) and 
$\eta_2 = f_{C|Y,\Z}$ (the censoring model) as nuisance models, 
since their role is to support estimation of $\bb$ rather than 
serve as scientific targets.

The complete case (CC) estimator uses only uncensored observations 
($\Delta_i = 1$), solving
\be
\sumi \delta_i \S_\bb^F(y_i, w_i, \z_i; \bb) = \0,
\ee
where $\S_\bb^F(y, x, \z; \bb) \equiv \partial \log 
f_{Y|X,\Z}(y, x, \z; \bb) / \partial \bb$ denotes the score 
function for the outcome model. No nuisance model is required.

The imputation (IMP) estimator \citep{atem2017linear} replaces each unobserved time to Stage 1 with its conditional expectation $\wc x_i \equiv E\{I(X_i > w_i) X_i \mid y_i, \z_i\} / E\{I(X_i > w_i) \mid y_i, \z_i\}$, solving
\be
\sumi \left\{ \delta_i \S_\bb^F(y_i, w_i, \z_i; \bb) + (1 - \delta_i) \S_\bb^F(y_i, \wc x_i, \z_i; \bb) \right\} = \0.
\ee
Consistency requires the score function $\S_\bb^F(y, x, \z; \bb)$ to be linear in $x$ and $\eta_1=f_{X|\Z}$ to be correctly specified.

The IPW estimator \citep{matsouaka2020regression} upweights uncensored participants by the inverse of their conditional survival function, solving
\be
\sumi \frac{\delta_i \S_\bb^F(y_i, w_i, \z_i; \bb)}{\wh{S}_{C|Y,\Z}(w_i, y_i, \z_i)} = \0,
\ee
where $\wh{S}_{C|Y,\Z}(w_i, y_i, \z_i)$ estimates the probability that a participant with outcome $y_i$ and baseline characteristics $\z_i$ remains in the study beyond time $w_i$. Consistency requires $\eta_2=f_{C|Y,\Z}$ to be correctly specified.

The MLE \citep{atem2017linear} maximizes the log-likelihood of \eqref{eq:model}, solving
\be
\sumi \left[ \delta_i \S_\bb^F(y_i, w_i, \z_i; \bb) + (1 - \delta_i) \frac{E\left\{ I(X_i > w_i) \S_\bb^F(y_i, X_i, \z_i; \bb) \mid y_i, \z_i \right\}}{E\left\{ I(X_i > w_i) \mid y_i, \z_i \right\}} \right] = \0.
\ee
Consistency requires $\eta_1=f_{X|\Z}$ to be correctly specified.

We evaluate these benchmark estimators and SPYCE across simulation settings 
that vary which nuisance models are correctly specified, 
misspecified, or estimated nonparametrically. Throughout, correct 
specification means $\eta_1$ fit as 
$\TruncNormal\{\wh\alpha_1^\ddagger(\bb), 1^2; -1,1\}$ via 
\eqref{eq:ml1} and $\eta_2$ as 
$\TruncNormal(\wh\alpha_{21}^\mathsection + 
\wh\alpha_{22}^\mathsection Y, 1^2; -1,1)$ via \eqref{eq:ml2}; 
misspecified means replaced by ${\rm Unif}(-1,1)$; and 
nonparametric means $\wh{E}_1$ via \eqref{eq:n15} or $\wh{E}_2$ 
via \eqref{eq:n16}. The configurations for each estimator are 
as follows.

\noindent\textit{SPYCE} (seven configurations):
\begin{enumerate}[label=(\roman*),ref=(\roman*)]
  \item Both $\eta_1$ and $\eta_2$ correctly specified
  \item $\eta_1$ correctly specified; $\eta_2$ misspecified
  \item $\eta_1$ misspecified; $\eta_2$ correctly specified
  \item Both $\eta_1$ and $\eta_2$ misspecified
\item $\wh{E}_1$ estimated nonparametrically via \eqref{eq:n15}; 
        $\wh{E}_2$ estimated nonparametrically via \eqref{eq:n16}
  \item $\wh{E}_1$ estimated nonparametrically via \eqref{eq:n15}; 
        $\eta_2$ misspecified
  \item $\wh{E}_2$ estimated nonparametrically via \eqref{eq:n16}; 
        $\eta_1$ misspecified.
\end{enumerate}

\noindent\textit{IMP estimator and MLE} (three $\eta_1$ configurations each): 
$\eta_1$ correctly specified, misspecified, or estimated 
nonparametrically via \eqref{eq:n15}; $\eta_2$ not required.

\noindent\textit{IPW estimator} (three $\eta_2$ configurations): 
$\eta_2$ correctly specified, misspecified, or estimated 
nonparametrically via \eqref{eq:n16}; $\eta_1$ not required.

\noindent\textit{CC estimator}: no nuisance model required.

The configurations above are evaluated under low (10--20\%), 
moderate (30--40\%), and high (60--70\%) censoring rates. 
Section~\ref{sec:simulation} focused on the high censoring 
setting, which most closely approximates PREDICT-HD's 58\% 
censoring rate; the low and moderate censoring results, 
reported below, show consistent patterns.

\begin{table}[htbp]
\centering
\caption{Finite-sample performance of SPYCE and four benchmark estimators under low censoring rate (10--20\%). For each estimator, we report bias, standard deviation (SD), median estimated standard error (SE), and 95\% confidence interval coverage rate (CI). Par = parametric. Mis = misspecified. Non = nonparametric.}  
\label{table:ta4}
\small
\begin{tabular}{l|c|c|rrrr|rrrr}
\multirow{2}{*}{Estimator} & \multirow{2}{*}{$\eta_1$} & \multirow{2}{*}{$\eta_2$} & \multicolumn{4}{c|}{$\beta_1$} & \multicolumn{4}{c}{$\beta_2$} \\
\cline{4-11}
 &  &  & Bias & SD & SE & CI & Bias & SD & SE & CI \\ \hline
\multirow{7}{*}{SPYCE}
 & Par & Par & -0.004 & 0.122 & 0.129 & 95.2\% & -0.026 & 0.230 & 0.244 & 96.1\% \\
 & Par & Mis &  0.032 & 0.124 & 0.142 & 96.2\% &  0.048 & 0.243 & 0.253 & 95.0\% \\
 & Mis & Par & -0.005 & 0.122 & 0.129 & 95.2\% & -0.029 & 0.230 & 0.243 & 96.2\% \\
 & Mis & Mis &  0.153 & 0.128 & 0.134 & 80.8\% &  0.234 & 0.261 & 0.250 & 84.0\% \\
 & Non & Non &  0.004 & 0.122 & 0.130 & 95.1\% &  0.017 & 0.234 & 0.246 & 95.3\% \\
 & Non & Mis & -0.040 & 0.125 & 0.136 & 95.2\% & -0.051 & 0.229 & 0.242 & 95.8\% \\
 & Mis & Non &  0.011 & 0.123 & 0.127 & 94.7\% &  0.040 & 0.235 & 0.236 & 93.7\% \\ \hline
CC   & --  & --  &  0.151 & 0.137 & 0.146 & 84.3\% &  0.220 & 0.272 & 0.291 & 89.1\% \\ \hline
\multirow{3}{*}{IMP}
 & Par & -- &  0.001 & 0.122 & 0.129 & 95.1\% & 0.082 & 0.235 & 0.244 & 93.9\% \\
 & Mis & -- & -0.010 & 0.122 & 0.129 & 95.3\% & 0.055 & 0.235 & 0.244 & 94.3\% \\
 & Non & -- &  0.006 & 0.122 & 0.129 & 95.4\% & 0.085 & 0.236 & 0.244 & 93.7\% \\ \hline
\multirow{3}{*}{IPW}
 & --  & Par &  0.000 & 0.154 & 0.157 & 96.5\% & -0.014 & 0.372 & 0.342 & 96.2\% \\
 & --  & Mis &  0.159 & 0.150 & 0.153 & 82.4\% &  0.285 & 0.538 & 0.464 & 91.1\% \\
 & --  & Non &  0.061 & 0.130 & 0.153 & 96.2\% &  0.089 & 0.265 & 0.329 & 98.0\% \\ \hline
\multirow{3}{*}{MLE}
 & Par & --  &  0.000 & 0.122 & 0.129 & 95.1\% & -0.015 & 0.230 & 0.243 & 96.4\% \\
 & Mis & --  & -0.010 & 0.122 & 0.129 & 95.3\% & -0.047 & 0.229 & 0.243 & 95.7\% \\
 & Non & --  &  0.005 & 0.122 & 0.129 & 95.3\% &  0.001 & 0.231 & 0.244 & 96.2\%
\end{tabular}
\end{table}

\begin{table}[htbp]
\centering
\caption{
Finite-sample performance of SPYCE and four benchmark estimators under moderate censoring rate (30--40\%). For each estimator, we report bias, standard deviation (SD), median estimated standard error (SE), and 95\% confidence interval coverage rate (CI). Par = parametric. Mis = misspecified. Non = nonparametric.}
\label{table:ta5}
\small
\begin{tabular}{l|c|c|rrrr|rrrr}
\multirow{2}{*}{Estimator} & \multirow{2}{*}{$\eta_1$} & \multirow{2}{*}{$\eta_2$} & \multicolumn{4}{c|}{$\beta_1$} & \multicolumn{4}{c}{$\beta_2$} \\
\cline{4-7}\cline{8-11}
 &  &  & Bias & SD & SE & CI & Bias & SD & SE & CI \\ \hline
\multirow{7}{*}{SPYCE}
 & Par & Par & -0.006 & 0.124 & 0.131 & 95.4\% & -0.034 & 0.243 & 0.259 & 96.2\% \\
 & Par & Mis &  0.015 & 0.127 & 0.137 & 96.2\% &  0.012 & 0.252 & 0.266 & 96.1\% \\
 & Mis & Par & -0.009 & 0.124 & 0.131 & 95.4\% & -0.042 & 0.243 & 0.259 & 96.2\% \\
 & Mis & Mis &  0.076 & 0.130 & 0.134 & 91.8\% &  0.089 & 0.265 & 0.263 & 93.2\% \\
 & Non & Non &  0.008 & 0.134 & 0.135 & 94.1\% &  0.040 & 0.271 & 0.270 & 93.5\% \\
 & Non & Mis & -0.025 & 0.126 & 0.134 & 95.6\% & -0.028 & 0.242 & 0.251 & 95.9\% \\
 & Mis & Non &  0.024 & 0.129 & 0.150 & 96.4\% &  0.079 & 0.260 & 0.263 & 94.1\% \\ \hline
CC   & --  & --  &  0.510 & 0.162 & 0.178 & 15.9\% &  0.549 & 0.318 & 0.346 & 66.7\% \\ \hline
\multirow{3}{*}{IMP}
 & Par & -- &  0.011 & 0.125 & 0.131 & 94.9\% & 0.491 & 0.270 & 0.260 & 53.6\% \\
 & Mis & -- & -0.021 & 0.124 & 0.131 & 95.8\% & 0.434 & 0.271 & 0.260 & 61.5\% \\
 & Non & -- &  0.025 & 0.126 & 0.131 & 94.8\% & 0.475 & 0.275 & 0.262 & 56.8\% \\
\hline
\multirow{3}{*}{IPW}
 & --  & Par &  0.013 & 0.183 & 0.202 & 97.4\% &  0.030 & 0.501 & 0.453 & 96.2\% \\
 & --  & Mis &  0.520 & 0.188 & 0.197 & 25.1\% &  0.596 & 0.577 & 0.514 & 79.8\% \\
 & --  & Non &  0.204 & 0.146 & 0.195 & 88.0\% &  0.234 & 0.322 & 0.422 & 96.5\% \\ \hline
\multirow{3}{*}{MLE}
 & Par & --  &  0.001 & 0.124 & 0.130 & 95.0\% & -0.027 & 0.239 & 0.257 & 96.3\% \\
 & Mis & --  & -0.026 & 0.123 & 0.130 & 95.5\% & -0.104 & 0.238 & 0.257 & 94.5\% \\
 & Non & --  &  0.017 & 0.125 & 0.130 & 95.1\% &  0.020 & 0.247 & 0.259 & 95.8\% 
\end{tabular}
\end{table}

\clearpage\newpage

\subsection{Additional PREDICT-HD Results}
\label{sec:additional-real-data}

\begin{table}[!htbp]
\centering
\setlength{\tabcolsep}{6pt}
\renewcommand{\arraystretch}{1.25}
\small
\caption{Slope estimates, standard errors, and required 
sample sizes per arm for six outcomes in the low-risk 
group. Each cell shows slope (standard error) and 
[required sample size]. Sample sizes assume 50\% 
treatment slowing, 80\% power, and two-sided $\alpha 
= 0.05$. The smallest sample size in each row is 
bolded. $\dag$ indicates a slope in the clinically 
implausible direction (improvement as participants 
approach Stage 1); sample size cannot be computed. 
Putamen and caudate volume ratios are measured in 
$10^{-2}\%$.}
\label{tab:stage1_low}
\resizebox{\linewidth}{!}{%
    \begin{tabular}{l|cc|cccc}
\toprule
Outcome & SPYCE-Non & SPYCE-Par & MLE & CC & IMP & IPW \\
\hline

\multirow{2}{*}{cUHDRS Score}
& $-0.132$ (0.046) & $-0.065$ (0.033) & $-0.077$ (0.033) & $-0.128$ (0.102) & $-0.027$ (0.052) & $-0.134$ (0.107) \\
& \textbf{[1,647]} & {[3,594]} & {[2,556]} & {[8,626]} & {[51,239]} & {[8,713]} \\
\hline

\multirow{2}{*}{Stroop Color Score}
& $-0.844$ (0.340) & $-0.218$ (0.265) & $-0.257$ (0.261) & $-0.952$ (0.604) & $0.042$ (0.424) & $-0.969$ (0.692) \\
& \textbf{[2,255]} & {[20,549]} & {[14,270]} & {[5,593]} & {$\dag$} & {[7,077]} \\

\multirow{2}{*}{Stroop Word Score}
& $-1.199$ (0.382) & $-0.449$ (0.318) & $-0.538$ (0.314) & $-1.161$ (0.772) & $-0.132$ (0.512) & $-1.294$ (0.905) \\
& \textbf{[1,409]} & {[6,965]} & {[4,738]} & {[6,152]} & {[208,015]} & {[6,794]} \\
\hline

Total Motor Score
& $-0.051$ (0.021) & $-0.021$ (0.017) & $-0.025$ (0.016) & $-0.076$ (0.049) & $-0.010$ (0.025) & $-0.058$ (0.059) \\
 \textit{(sign flipped)} & \textbf{[2,464]} & {[8,387]} & {[6,118]} & {[5,869]} & {[85,481]} & {[14,361]} \\
\hline
\multirow{2}{*}{Putamen Volume Ratio}
& $-0.989$ (0.124) & $-0.560$ (0.145) & $-0.633$ (0.142) & $-0.276$ (0.249) & $-0.017$ (0.250) & $-0.384$ (0.279) \\
& \textbf{[122]} & {[520]} & {[390]} & {[6,302]} & {[1.77e+06]} & {[4,097]} \\

\multirow{2}{*}{Caudate Volume Ratio}
& $-0.877$ (0.119) & $-0.454$ (0.114) & $-0.508$ (0.112) & $-0.615$ (0.195) & $-0.091$ (0.191) & $-0.874$ (0.322) \\
& \textbf{[144]} & {[493]} & {[376]} & {[777]} & {[33,945]} & {[1,055]} \\
\bottomrule
\end{tabular}
}
\end{table}

\clearpage\newpage

\section{Technical Proofs}
\setcounter{subsection}{0}\renewcommand{\thesubsection}{B.\arabic{subsection}}
\setcounter{equation}{0}\renewcommand{\theequation}{B.\arabic{equation}}
\setcounter{Pro}{0}\renewcommand{\thePro}{B.\arabic{Pro}}
\setcounter{Th}{0}\renewcommand{\theTh}{B.\arabic{Th}}
\setcounter{Lem}{0}\renewcommand{\theLem}{B.\arabic{Lem}}
\setcounter{Rem}{0}\renewcommand{\theRem}{B.\arabic{Rem}}
\setcounter{Cor}{0}\renewcommand{\theCor}{B.\arabic{Cor}}
\setcounter{table}{0}\renewcommand{\thetable}{B.\arabic{table}}
\subsection{Proof of Proposition \ref{pro:iden}} \label{sec:ident}


We can write the probability density function of an observation as 
\bse
\{
f_{Y,\Z}(y,\z)f_{X|Y,\bZ}(x,y,\bz)\int_{x}^{\infty}
f_{C|Y,\Z}(c,y,\z)
dc\}^\delta
\{f_{Y,\Z}(y,\z)f_{C|Y,\bZ}(c,y,\bz)\int_{c}^{\infty}
f_{X|Y,\Z}(x,y,\z)
dx\}^{1-\delta}.
\ese
If the model is not identifiable, then we have
\bse
&&\{f_{Y,\Z}(y,\z)f_{X|Y,\bZ}(x,y,\bz)\int_{x}^{\infty}
f_{C|Y,\Z}(c,y,\z)
dc\}^\delta
\{f_{Y,\Z}(y,\z)f_{C|Y,\bZ}(c,y,\bz)\int_{c}^{\infty}
f_{X|Y,\Z}(x,y,\z)
dx\}^{1-\delta}\\
&=&\{\wt f_{Y,\Z}(y,\z)\wt f_{X|Y,\bZ}(x,y,\bz)\int_{x}^{\infty}
\wt f_{C|Y,\Z}(c,y,\z)
dc\}^\delta
\{\wt f_{Y,\Z}(y,\z)\wt f_{C|Y,\bZ}(c,y,\bz)\int_{c}^{\infty}
\wt f_{X|Y,\Z}(x,y,\z)
dx\}^{1-\delta}.
\ese

Then $\delta=1$ and $\delta=0$, respectively, lead to
\bse 
&&f_{Y,\Z}(y,\z)f_{X|Y,\bZ}(x,y,\bz)\int_{x}^{\infty}
f_{C|Y,\Z}(c,y,\z)
dc =\wt f_{Y,\Z}(y,\z) \wt f_{X|Y,\bZ}(x,y,\bz)\int_{x}^{\infty}
\wt f_{C|Y,\Z}(c,y,\z)
dc,\\
&&f_{Y,\Z}(y,\z)f_{C|Y,\bZ}(c,y,\bz)\int_{c}^{\infty}
f_{X|Y,\Z}(x,y,\z)
dx= \wt f_{Y,\Z}(y,\z)\wt f_{C|Y,\bZ}(c,y,\bz)\int_{c}^{\infty}
\wt f_{X|Y,\Z}(x,y,\z)
dx.\\
\ese
%
Integrating the first equality over all values of $x$
and integrating the second equality over all values of $c$ gives
\bse
&&f_{Y,\Z}(y,\z)\iint_{x<c}
f_{C|Y,\Z}(c,y,\z)
f_{X|Y,\Z}(x,y,\z)
dcdx\\ 
&&=\wt f_{Y,\Z}(y,\z)\iint_{x<c}
\wt f_{C|Y,\Z}(c,y,\z)
\wt f_{X|Y,\Z}(x,y,\z)
dcdx,\\
&&f_{Y,\Z}(y,\z)\iint_{c<x}
f_{C|Y,\bZ}(c,y,\bz)
f_{X|Y,\bZ}(x,y,\bz)
dxdc\\
&&= \wt f_{Y,\Z}(y,\z)\iint_{c<x}
\wt f_{C|Y,\bZ}(c,y,\bz)
\wt f_{X|Y,\bZ}(x,y,\bz)
dxdc.
\ese
Adding the two equalities yields $f_{Y,\Z}(y,\z) = \wt f_{Y,\Z}(y,\z)$. 
Then we have
\bse 
&&f_{X|Y,\bZ}(x,y,\bz)\int_{x}^{\infty}
f_{C|Y,\Z}(c,y,\z)
dc = \wt f_{X|Y,\bZ}(x,y,\bz)\int_{x}^{\infty}
\wt f_{C|Y,\Z}(c,y,\z)
dc,\\
&&f_{C|Y,\bZ}(c,y,\bz)\int_{c}^{\infty}
f_{X|Y,\Z}(x,y,\z)
dx= \wt f_{C|Y,\bZ}(c,y,\bz)\int_{c}^{\infty}
\wt f_{X|Y,\Z}(x,y,\z)
dx.\\
\ese
Replacing $c$ and $x$ with $s$ and using the conditional survival function of
$C|Y,\Z$ and $X|Y,\Z$, we have 
\be\label{eq:e3}
&&f_{X|Y,\bZ}(s,y,\bz) S_{C|Y,\Z}(s,y,\z)
= \wt f_{X|Y,\bZ}(s,y,\bz) \wt S_{C|Y,\Z}(s,y,\z), \\
&&f_{C|Y,\bZ}(s,y,\bz) S_{X|Y,\Z}(s,y,\z)
= \wt f_{C|Y,\bZ}(s,y,\bz)\wt S_{X|Y,\Z}(s,y,\z).\n
\ee
Adding these two equalities yields 
\bse
&&\frac{\partial}{\partial s} \{S_{X|Y,\bZ}(s,y,\bz) S_{C|Y,\Z}(s,y,\z)\}
= \frac{\partial}{\partial s} \{\wt S_{X|Y,\bZ}(s,y,\bz) \wt S_{C|Y,\Z}(s,y,\z)\}.
\ese
Then 
\bse
&&S_{X|Y,\bZ}(s,y,\bz) S_{C|Y,\Z}(s,y,\z)
= \wt S_{X|Y,\bZ}(s,y,\bz) \wt S_{C|Y,\Z}(s,y,\z) + K(y,\z),
\ese
for some function $K(y,\z)$.
Taking $s \to \infty$, we have that $K(y,\z)=0$.
Since $K(y,\z)=0$, \eqref{eq:e3} becomes
\bse 
&&\frac{f_{X|Y,\bZ}(s,y,\bz)}{S_{X|Y,\Z}(s,y,\z)}
= \frac{\wt f_{X|Y,\bZ}(s,y,\bz)} {\wt S_{X|Y,\Z}(s,y,\z)},\\
&&\frac{f_{C|Y,\bZ}(s,y,\bz)}{S_{C|Y,\Z}(s,y,\z)}
= \frac{\wt f_{C|Y,\bZ}(s,y,\bz)} {\wt S_{C|Y,\Z}(s,y,\z)},\n
\ese
where $(s, y, \bz)$ lies on the common support of $X|Y,\bZ$ and
$C|Y,\bZ$. Hence $f_{X|Y,\bZ}(x,y,\bz)=\wt f_{X|Y,\bZ}(x,y,\bz)$ and
$f_{C|Y,\bZ}(c,y,\bz)=\wt f_{C|Y,\bZ}(c,y,\bz)$.

Since $f_{X|Y,\bZ}(x,y,\bz)$, $f_{C|Y,\bZ}(c,y,\bz)$, and $f_{Y,\Z}(y,\z)$ are each uniquely determined, the joint density
\bse
f_{Y,X,C,\bZ}(y,x,c,\bz;\bb) &=&
f_{C|Y,\Z}(c,y,\z)f_{X|Y,\bZ}(x,y,\bz)f_{Y,\bZ}(y,\bz)\\
&=&f_\Z(\z)f_{X|\Z}(x,\z)f_{Y\mid X,\Z}(y,x,\z,\bb)f_{C\mid Y,\Z}(c,y,\z)
\ese
%
is also uniquely determined. This uniqueness of the joint density implies the uniqueness of each component: $f_\Z(\z)$, $f_{X|\bZ}(x,\bz)$, $f_{C|Y,\Z}(c,y,\z)$, and $\bb$. Therefore, the model is identifiable.

\subsection{Proofs of Proposition \ref{pro:1}}\label{sec:pro1-3pf}
\subsubsection{Proof of (i) of Proposition \ref{pro:1}}\label{sec:pro1pf}
Consider a parametric submodel with parameters $\bxi = (\bxi_1\trans, \bxi_2\trans, \bxi_3\trans)\trans$.
The log-likelihood of one observation is
\bse
l_{Y,W,\Delta, \bZ}(y, w, \delta, \bz; \bb, \bxi) &=&
\log \eta_3(\z;\bxi_3)\\
&&+ \delta \left[\log f_{Y|X,\bZ}(y,w,\bz;\bb) + \log\eta_1(w,\bz;
  \bxi_1) + \log \left\{\int_{w}^{\infty} 
\eta_2(c,y,\z; \bxi_2)
dc\right\} \right]\\
&&+  (1-\delta) \left[\log \eta_2 (w,y,\z;\bxi_2) + \log
  \left\{\int_{w}^{\infty} f_{Y|X,\bZ}(y,x,\bz;\bb)\eta_1(x,\bz;
    \bxi_1)dx\right\} 
\right].
\ese
Then the nuisance score functions of $\eta_1$, $\eta_2$, and $\eta_3$ are, respectively
\bse
\bS_1 (y,w,\delta, \bz; \bxi) &=& \delta \ba_1 (w,\z;\bxi_1) +
(1-\delta) \frac{\int_w^\infty f_{Y|X,\Z}(y,x,\z,\beta) \ba_1
  (x,\z;\bxi_1) \eta_1 (x,\z; \bxi_1) dx}{\int_w^\infty
  f_{Y|X,\Z}(y,x,\z,\beta) \eta_1 (x,\z; \bxi_1) dx}\\ 
&=& \delta \ba_1(w,\bz;\bxi_1) + (1-\delta) \frac{E\{I(X > w) \ba_1
  (X,\z,\bxi_1)|y,\z\}}{E\{I(X> w)|y,\z\}},\\ 
\bS_2 (y,w,\delta, \bz; \bxi)&=& 
\delta \frac{\int_w^\infty \ba_2 (c,y,\z;\bxi_2) \eta_2 (c,y, \z;
  \bxi_2) dc}{\int_w^\infty \eta_2 (c,y, \z; \bxi_2) dc} + (1-\delta)
\ba_2 (w,y,\z;\bxi_2)\\ 
&=& \delta \frac{E\left\{\ba_2 (C,y,\z;\bxi_2) I(C\ge
    w)\right|y,\z\}}{E\left\{I(C\ge w)\right|y,\z\}} + (1-\delta)
\ba_2 (w,y,\z;\bxi_2), \\ 
 \bS_3 (\z; \bxi) &=& \ba_3(\bz; \bxi_3),
\ese
where $\ba_1(x,\z;\bxi_1) = \partial\log
  \eta_1(x,\z;\bxi_1)/\partial \bxi_1$, $\ba_2(c,y,\z;\bxi_2) = \partial
\log \eta_2(c,y,\z;\bxi_2)/\partial \bxi_2$, $\ba_3(\z;\bxi_3) =
\partial \log \eta_3(\z;\bxi_3)/\partial \bxi_3$. Note that
$E\{\ba_1(X,\z;\bxi_1)|\z\} = 0$,
$E\{\ba_2(C,y,\z;\bxi_2)|y,\z\} = 0$, and
$E\{\ba_3(\Z;\bxi_3)\} = 0$. In other words, the score
functions $\bS_1$, $\bS_2$, $\bS_3$ are elements of $\Lambda_1$,
$\Lambda_2$, $\Lambda_3$, respectively. 


Conversely, for any functions $\ba_1(x,\z), \ba_2(c,y,\z), \ba_3(\z)$ satisfying $E\{\ba_1(X,\z)|\z\} = 0$, $E\{\ba_2(C,y,\z)|y,\z\} = 0$, and $E\{\ba_3(\Z)\} = 0$, we can always construct a parametric submodel with $\bxi_1,\bxi_2,\bxi_3$ whose score functions are $\ba_1(x,\z), \ba_2(c,y,\z), \ba_3(\z)$, respectively. Thus, the nuisance tangent spaces $\Lambda_1,\Lambda_2,\Lambda_3$ are indeed as described.

It remains to prove the orthogonality between $\Lambda_1$, $\Lambda_2$, 
and $\Lambda_3$. First, $\Lambda_1 \perp \Lambda_2$ because for some
$\ba_1(x,\bz)$ and  $\ba_2(c,y,\bz)$ satisfying 
$E \{\ba_1 (X, \bz) |  \bz\} = 0$ and $E
\left\{\ba_2 (C, y, \bz) | y,  \bz\right\} = 0$,

\bse
&&E\{\bS_1 (Y,W,\Delta, \bZ) \bS_2\trans (Y,W,\Delta, \bZ)\}   \\
&=&E\left(\left[\Delta \ba_1(X,\bZ) + (1-\Delta) \frac{E\{I(X> C)
      \ba_1 (X,\Z)|C,Y,\Z\}}{E\{I(X> C)|C,Y,\Z\}}\right]\right.\\ 
&&\times\left.\left[\Delta \frac{E\{I(C\ge X) \ba_2
      (C,Y,\Z)|X,Y,\Z\}}{E\{I(C\ge X)|X,Y,\Z\}} + (1-\Delta)\ba_2 (C,Y,\Z)\right]\trans \right)\\
&=& E\left[ \Delta \ba_1(X,\bZ)  \frac{E\{I(C\ge X) \ba_2\trans
      (C,Y,\Z)|X,Y,\Z\}}{E\{I(C\ge X)|X,Y,\Z\}} \right.\\
&&\left.+ (1-\Delta) \frac{E\{I(X> C) \ba_1 (X,\Z)|C,Y,\Z\}}{E\{I(X> C)|C,Y,\Z\}}\ba_2\trans
    (C,Y,\Z)\right]\\
&=&E\left[E\{ I(C\ge X) \ba_1(X,\bZ)\ba_2\trans
      (C,Y,\Z)|X,Y,\Z\}\right]\\
&&+E\left[  E\{I(X> C) \ba_1 (X,\Z) \ba_2\trans
    (C,Y,\Z)|C,Y,\Z\}\right]\\
&=& E\{I(C\ge X) \ba_1(X,\bZ)\ba_2\trans
      (C,Y,\Z)\}+E\{I(X > C) \ba_1 (X,\Z) \ba_2\trans
    (C,Y,\Z)\}\\
&=& E[\ba_1(X,\bZ)E\{\ba_2\trans
      (C,Y,\Z)\mid X,Y,\bZ\}]\\
&=& E[\ba_1(X,\bZ)E\{\ba_2\trans
      (C,Y,\Z)\mid Y,\bZ\}]\\
&=&\0.
\ese

Moreover, $\Lambda_1 \perp \Lambda_3$ because for any $\ba_1 (x, \bz)$
with $E \{\ba_1 (X, \bz) |  \bz\} = 0$, 
\bse
&&E \{\bS_1 (Y,W,\Delta, \bz)|\z\}   \\
&=& E \left[\left.\Delta \ba_1(X,\bz) + (1-\Delta) \frac{E\{I(X>C)
      \ba_1 (X,\z)|C,Y,\z\}}{E\{I(X > C)|C,Y,\z\}}\right|\z \right]\\ 
&=& E \left\{\left. I(C \ge X) \ba_1(X,\bz)\right|\z \right\} + E
\left[\left.E\{I(X> C) \ba_1 (X,\z)|C, Y,\z\}\right|\z \right]\\ 
&=& E \left\{\left. I(C \ge X) \ba_1(X,\bz)\right|\z \right\} +
E\{\left.I(X> C) \ba_1 (X,\z)\right|\z\}\\ 
&=& E \left\{\left. \ba_1(X,\bz)\right|\z \right\}\\
&=& \0,
\ese
which implies that $E \{\bS_1 (Y,W,\Delta, \bZ) \bS_3\trans
(Y,W,\Delta, \bZ)\} = E\left[E \left\{\bS_1 (Y,W,\Delta,
    \bZ)|\Z\right\} \ba_3\trans(\Z) \right] = \mathbf{0}$ for any
$\ba_3(\z)$ with $E\{\ba_3(\Z)\} = 0$.

Furthermore, $\Lambda_2 \perp \Lambda_3$ follows since for
any $\ba_2(c,y,\z)$ with
$E\left\{\ba_2(C,y,\z)|y,\z\right\} =\0,$ 
\bse
&&E \{\bS_2 (Y,W,\Delta, \bz)|\z\} \\
&=& E \left[\left.\Delta \frac{E\{I(C\geq X) \ba_2
      (C,Y,\z)|X,Y,\z\}}{E\{I(C\ge X)|X,Y,\z\}} + (1-\Delta)\ba_2
    (C,Y,\z) \right| \z \right]\\ 
&=& E \left[E\{I(C \ge X) \ba_2 (C,Y,\z)|X,Y,\z\} + I(X>C)\ba_2 (C,Y,\z) | \z \right]\\
&=& E\{I(C \ge X) \ba_2 (C,Y,\z)|\z\} + E\left\{I(X>C)\ba_2 (C,Y,\z) | \z \right\}\\
&=& E \left\{\ba_2(C,Y,\z)|\z \right\}\\
&=& E \left[E \left\{\ba_2(C,Y,\z)|Y, \z \right\}|\z\right]\\
&=& \mathbf{0},
\ese
and thus $E \{\bS_2 (Y,W,\Delta, \bZ) \bS_3\trans (Y,W,\Delta, \bZ)\}
= E\left[E \left\{\bS_2 (Y,W,\Delta, \bZ)|\Z\right\} \ba_3\trans(\Z)
\right] =  \0$  for any $\ba_3(\z)$ satisfying $E\{\ba_3(\Z)\} = \0$.  
\qed

\subsubsection{Proof of (ii) of Proposition \ref{pro:1}}\label{sec:pro2pf}
Let $\bg(y,w,\delta,\bz) \in \H$ be written as $\delta
\bg_1(y,w,\bz) + (1-\delta) \bg_2(y,w,\bz)$. Define 
\bse
\bh_1(x,\z) \equiv
E\{\bg(Y,W,\Delta,\bz)\mid x,\z\}=
E\left\{I(x \le C) \bg_1(Y,x,\bz) + I(x> C)
  \bg_2(Y,C,\bz)|x,\bz\right\}.
\ese
Then $\bg(y,w,\delta,\z) \perp
(\Lambda_1\oplus \Lambda_3)$ is equivalent to 
\bse
\0 &=& E\left(\bg(Y,W,\Delta,\bZ)\left[\Delta \ba_1(X,\bZ) +
    (1-\Delta) \frac{E\{I(X> C) \ba_1 (X,\Z)|C,Y,\Z\}}{E\{I(X> C)
      |C,Y,\Z\}} + \ba_3(\Z)\right]\trans\right)\\ 
&=& E\left[\Delta \bg_1(Y,X,\bZ) \{\ba_1(X,\bZ)+\ba_3 (\Z)\}\trans \right]\\
&&+ E\left((1-\Delta) \bg_2(Y,C,\bZ) \frac{E\left[I(X> C)
      \{\ba_1(X,\bZ)+\ba_3 (\Z)\}\trans|C,Y,\Z\right]}{E\{I(X> C)
    |C,Y,\Z\}}\right)\\ 
&=& E\left[I(X\le C)\bg_1(Y,X,\bZ) \{\ba_1(X,\bZ)+\ba_3 (\Z)\}\trans\right]\\
&&+ E\left( E[\bg_2(Y,C,\bZ)I(X> C) \{\ba_1(X,\bZ)+\ba_3 (\Z)\}\trans|C,Y,\Z]\right)\\
&=& E\left[I(X\le C)\bg_1(Y,X,\bZ) \{\ba_1(X,\bZ)+\ba_3 (\Z)\}\trans
  +I(X> C) \bg_2(Y,C,\bZ)\{\ba_1(X,\bZ)+\ba_3 (\Z)\}\trans\right]\\ 
&=& E\left[E\left\{I(X \le C) \bg_1(Y,X,\bZ) + I(X> C)
    \bg_2(Y,C,\bZ)|X,\bZ\right\} \{\ba_1(X,\bZ)+\ba_3
  (\Z)\}\trans\right]\\ 
&=& E[\bh_1 (X,\Z) \{\ba_1(X,\bZ)+\ba_3 (\Z)\}\trans] 
\ese
for any $\ba_1(x,\z)$ and $\ba_3(\z)$ with $E\{ \ba_1(X,\z)|\z\} =
\0$, $E\{ \ba_3(\Z)\} = \0$. This condition is equivalent to $\bh_1(x,\z)=\0$
because otherwise we can set $\ba_3(\z)=E\{\bh_1(X,\z)\mid\z\}$ and
$\ba_1(x,\z)=\bh_1(x,\z)-\ba_3(\z)$ and obtain a contradiction.

In a similar fashion, define 
\bse
\bh_2(c,y,\z) \equiv E\{I(X\le c) \bg_1
(y,X,\z) + I(X>c) \bg_2 (y    ,c,\z)| c,y,\z\}.
\ese
Then
$\bg(y,w,\delta,\z) \perp \Lambda_2$ is equivalent to 
\bse
\0 &=& E\left(\bg(Y,W,\Delta,\bZ)\left[\Delta \frac{E\{I(C\ge X)
      \ba_2(C,Y,\Z)|X,Y,\Z\}}{E\{I(C\ge X)|X,Y,\Z\}} + (1-\Delta)\ba_2
    (C,Y,\Z) \right]\trans \right)\\ 
&=&E\left[\Delta \bg_1(Y,X,\bZ) \frac{E\{I(X\le C)
    \ba_2\trans(C,Y,\Z)|X,Y,\Z\}} {E\{I(X \le C)|X,Y,\Z\}}\right] +
E\left\{(1-\Delta) \bg_2(Y, C, \bZ) \ba_2\trans(C,Y,\Z)\right\}\\ 
&=& E\left[E\left\{I(X\le C)\bg_1(Y,X,\bZ) \ba_2\trans(C,Y,\Z)|X,Y,\Z
  \right\}\right]+ E\{I(X> C)\bg_2(Y,C,\bZ) \ba_2\trans(C,Y,\Z)\}\\ 
&=& E\left\{I(X\le C)\bg_1(Y,X,\bZ) \ba_2\trans(C,Y,\Z) + I(X> C)
  \bg_2(Y,C,\bZ)\ba_2\trans(C,Y,\Z)\right\}\\ 
&=& E\left[E\left\{I(X \le C) \bg_1(Y,X,\bZ) + I(X> C)
    \bg_2(Y,C,\bZ)|C,Y,\bZ\right\}\ba_2\trans(C,Y,\Z)\right]\\ 
&=& E\left\{\bh_2(C,Y,\Z) \ba_2\trans(C,Y,\Z) \right\}
\ese
for any $\ba_2(c,y,\z)$ with $E\{ \ba_2(C,y,\z)|y,\z\}
= \0$. The condition $E\left\{\bh_2(C,Y,\Z) \ba_2\trans(C,Y,\Z) \right\} = 0$ holds if and only if $\bh_2(c,y,\bz) =
E\{\bh_2(C,y,\bz)|y,\bz\}$ because otherwise taking $\ba_2(c,y,\z) =
\bh_2(c,y,\bz) - E\{\bh_2(C,y,\bz)|y,\bz\}$ yields a contradiction.

Since $\Lambda_1,\Lambda_2,\Lambda_3$ are orthogonal to each other,
$\bg(y,w,\delta,\z) \perp \Lambda$ if and only if $\bg(y,w,\delta,\z)
\perp (\Lambda_1 \oplus \Lambda_3)$ and $\bg(y,w,\delta,\z) \perp
\Lambda_2$, which is equivalent to $\h_1(x,\z)= \0$ and $\bh_2(c,y,\z)
= E\{\bh_2(C,y,\z)|y,\z\}$. \qed

\subsubsection{Proof of (iii) of Proposition \ref{pro:1}}\label{sec:pro3pf}
It suffices to prove that $\S_\bb - \S\eff \in \Lambda$ and
$\S\eff \perp \Lambda$. We note that
\bse 
&&E\{\ba (X,\bz;\bb)|\bz\}\\
&=&E\{I(X\le C) \ba (X,\bz;\bb)|\z\} +
E\left[\left.E\{I(X>C)\ba(X,\bz;\bb) |C,Y,\z\}\right|\bz\right] \\ 
&=&E\{I(X\le C) \ba (X,\bz;\bb)|\z\} +
E\left[\left.I(X>C)\frac{E\{I(X>C)\ba(X,\bz;\bb)
      |C,Y,\z\}}{E\{I(X>C)|C,Y,\z\}}\right|\bz\right] \\ 
&=& E\{I(X\le C)\S_\bb^F(Y,X,\bz;\bb)|\z\} +
E\left[\left.I(X>C)\frac{E\{I(X>C)\S_\bb^F(Y,X,\bz;\bb)
      |C,Y,\z\}}{E\{I(X>C)|C,Y,\z\}}\right|\bz\right]\\ 
&=& E\{I(X\le C)\S_\bb^F(Y,X,\bz;\bb)|\z\} +
E\left[\left.E\{I(X>C)\S_\bb^F(Y,X,\bz;\bb)
    |C,Y,\z\}\right|\bz\right]\\ 
&=& E\{\S_\bb^F(Y,X,\bz;\bb)|\z\}\\
&=& E[E\{\S_\bb^F(Y,X,\bz;\bb)|X,\z\}\z]\\
&=& \0.
\ese
Thus $\S_\bb - \S\eff \in \Lambda_1 \subset \Lambda$ by Proposition \ref{pro:1}.
Now, let
\bse
\h_1(y,w,\z;\bb) &\equiv& \S_\bb^F(y,w,\z;\bb) - \ba(w, \bz; \bb) ,\\
\h_0(y,w,\z;\bb) &\equiv&\frac{ E[I(X> w)\{\S_\bb^F(y,X,\z;\bb) -
  \ba(X,\z;\bb)\} |y,\z]}{ E\{I(X> w)|y,\z\}}.
\ese
Then we obtain
\bse
\S\eff (y,w,\delta, \z ; \bb) &=& \S_\bb(y,w,\delta,\z;\bb) -
\left[\delta \ba(w, \bz; \bb) + (1-\delta) \frac{ E\{I(X>
    w)\ba(X,\z;\bb) |y,\z\}}{ E\{I(X> w)|y,\z\}} \right]\\ 
&=& \delta \S_\bb^F(y,w,\z;\bb) + (1-\delta)\frac{ E\{I(X>
  w)\S_\bb^F(y,X,\z;\bb) |y,\z\}}{ E\{I(X> w)|y,\z\}}\\ 
&&-\left[\delta \ba(w, \bz; \bb) + (1-\delta) \frac{ E\{I(X>
    w)\ba(X,\z;\bb) |y,\z\}}{ E\{I(X> w)|y,\z\}} \right]\\ 
&=& \delta \{\S_\bb^F(y,w,\z;\bb) - \ba(w, \bz; \bb) \}\\
&&+ (1-\delta)\frac{ E[I(X> w)\{\S_\bb^F(y,X,\z;\bb) -
  \ba(X,\z;\bb)\} |y,\z]}{ E\{I(X> w)|y,\z\}}\\
&=& \delta \h_1(y,w,\z;\bb)  + (1-\delta)
\h_0(y,w,\z;\bb).
\ese

Then the condition of $\ba (x,\bz;\bb)$ leads to
\bse
&&E\{I(x\le C) {\h_1 (Y,x,\z) + I(x>C) \h_0 (Y,C,\z)}| x,\z\} \\
&=&E\left[I(x\le C) \{\S_\bb^F(Y,x,\z;\bb) - \ba(x, \bz; \bb)\}| x,\z\right]\\
&&+ E\left(\left.I(x>C) \frac{ E[I(X> C)\{\S_\bb^F(Y,X,\z;\bb) -
      \ba(X,\z;\bb)\} |C,Y,\z]}{ E\{I(X> C)|C,Y,\z\}}\right|
  x,\z\right)\\ 
&=& E\{I(x\le C)\S_\bb^F(Y,x,\bz;\bb)|x,\z\} +
E\left[\left.I(x>C)\frac{E\{I(X>C)\S_\bb^F(Y,X,\bz;\bb)
      |C,Y,\z\}}{E\{I(X>C)|C,Y,\z\}}\right|x,\bz\right]\\ 
&& - E\{I(x\le C)|x,\z\} \ba (x,\bz;\bb) -
E\left[\left.I(x>C)\frac{E\{I(X>C)\ba(X,\bz;\bb)
      |C,Y,\z\}}{E\{I(X>C)|C,Y,\z\}}\right|x,\bz\right]\\ 
&=& \0,
\ese
and the outcome-dependence assumption $X\indep C|Y,\Z$ gives 
\bse
&& E\{I(X\le c) \h_1 (y,X,\z) + I(X>c) \h_0 (y,c,\z)| c,y,\z\} \\
&=&E[I(X\le c) \{\S_\bb^F(y,X,\z;\bb) - \ba(X, \bz; \bb)\} | c,y,\z]\\
&&+ E\left( \left.I(X>c) \frac{ E[I(X> c)\{\S_\bb^F(y,X,\z;\bb) -
      \ba(X,\z;\bb)\} |c,y,\z]}{ E\{I(X> c)|c,y,\z\}}\right| c,y,\z
\right) \\ 
&=&E\left[\left.I(X\le c) \{\S_\bb^F(y,X,\z;\bb) - \ba(X, \bz; \bb)\} \right| c,y,\z \right] \\
&&+ E[I(X> c)\{\S_\bb^F(y,X,\z;\bb) - \ba(X,\z;\bb)\} |c,y,\z] \\
&=&E\left\{\S_\bb^F(y,X,\z;\bb) - \ba(X, \bz; \bb)| c,y,\z \right\}  \\
&=&E\left\{\S_\bb^F(y,X,\z;\bb) - \ba(X, \bz; \bb)| y,\z \right\}  \\
&=& E\{I(X\le C) \h_1 (y,X,\z) + I(X>C) \h_0 (y,C,\z)| y,\z\}.
\ese
Hence, combining these results with (ii) of Proposition \ref{pro:1}, we have
that $\S\eff \in \Lambda^\perp$. \qed

\subsection{Proof of Theorem \ref{th:doublyrobust}}\label{sec:pro4pf}

The proof assumes the following standard regularity conditions for
estimating equation estimators \citep{newey1994large}, where
$\|\cdot\|_2$ denotes the vector $L_2$ norm:
\begin{enumerate}[label=(C\arabic*),ref=(C\arabic*),start=1]
    \item\label{con:bb}
    The true parameter $\bb_0$ is
contained in a compact set $\Omega$;
    \item\label{con:seffub}
    $E\{\sup_{\bb \in \Omega}\|\S\eff^{*\star} (Y,W,\Delta, \Z ;
\bb)\|_2\}< \infty$;
    \item\label{con:partinv}
    $E\{\partial \S\eff^{*\star} (Y,W,\Delta, \Z ; \bb_0)/ \partial \bb\trans\}$ is invertible;
    \item\label{con:conti}
    The mapping $\bb \mapsto E\{\S\eff^{*\star} (Y,W,\Delta, \Z ; \bb)\}$ is continuous on $\Omega$.
\end{enumerate}

We now prove consistency under each condition separately.
First, assume that $\eta_2^\star (c,y,\z) = \eta_2 (c,y,\z)$. Then
\bse
    && E\{\S\eff^* (Y,W,\Delta, \Z ; \bb_0)\}\\
    &=& E\{\S_\bb^*(Y,W,\Delta,\Z;\bb_0)\}\\
    &&- E\left[I(X\le C) \ba^{*}(X, \bZ; \bb_0) + I(X> C) \frac{
        E^*\{I(X> C)\ba^{*}(X,\Z;\bb_0) |C,Y,\Z\}}{E^*\{I(X>
        C)|C,Y,\Z\}} \right]\\ 
    &=& E\left[I(X\le C) \S_\bb^F(Y, X, \bZ; \bb_0) + I(X> C) \frac{
        E^*\{I(X> C)\S_\bb^F(Y, X,\Z;\bb_0) |C,Y,\Z\}}{E^*\{I(X>
        C)|C,Y,\Z\}} \right]\\ 
    &&- E\left[I(X\le C) \ba^{*}(X, \bZ; \bb_0) + I(X> C) \frac{
        E^*\{I(X> C)\ba^{*}(X,\Z;\bb_0) |C,Y,\Z\}}{E^*\{I(X>
        C)|C,Y,\Z\}} \right]\\ 
    &=&\0.
\ese
Next, assume that $\eta_1^* (x,\z) = \eta_1 (x,\z)$. Then \eqref{eq:b1} implies that
\bse
\0 &=& E^\star\{I(X\le C)\S_\bb^F(Y,X,\bZ;\bb_0)\} +
E^\star\left[I(X>C)\frac{E\{ I(X>C) \S_\bb^F(Y,X,\bz;\bb_0)
    |C,Y,\Z\}}{ E\{I(X>C)|C,Y,\Z\}}\right]\\ 
&&- E^\star \{I(X\le C) \ba^\star (X,\Z;\bb_0) \} -
E^\star\left[I(X>C)\frac{E\{ I(X>C)\ba^{\star}(X,\Z;\bb_0)
    |C,Y,\Z\}}{E\{I(X>C)|C,Y,\Z\}}\right]\\ 
&=& E^\star\{I(X\le C)\S_\bb^F(Y,X,\bZ;\bb_0)\} + E^\star\left[E\{
  I(X>C) \S_\bb^F(Y,X,\bz;\bb_0) |C,Y,\Z\}\right]\\ 
&&- E^\star \{I(X\le C) \ba^\star (X,\Z;\bb_0) \} - E^\star\left[E\{
  I(X>C)\ba^{\star}(X,\Z;\bb_0) |C,Y,\Z\}\right]\\ 
&=& E^\star\{\S_\bb^F(Y,X,\bZ;\bb_0)\} - E^\star \{\ba^\star (X,\Z;\bb_0) \}\\
&=& - E \{\ba^\star (X,\Z;\bb_0) \}
\ese
since the models for $X|C,Y,\Z$ and $Y,X,\Z$ are correctly specified. Thus,
 \bse
    && E\{\S\eff^{\star} (Y,W,\Delta, \Z ; \bb_0)\}\\
    &=& E\{\S_\bb(Y,W,\Delta,\Z;\bb_0)\}\\
    &&- E\left[I(X\le C) \ba^{\star}(X, \bZ; \bb_0) + I(X> C) \frac{
        E\{I(X> C)\ba^{\star}(X,\Z;\bb_0) |C,Y,\Z\}}{E\{I(X>
        C)|C,Y,\Z\}} \right]\\ 
    &=& - E\left\{I(X\le C) \ba^{\star}(X, \bZ; \bb_0)\right\} -
    E\left[ E\{I(X> C)\ba^{\star}(X,\Z;\bb_0) |C,Y,\Z\} \right]\\ 
    &=& - E\left\{\ba^{\star}(X, \bZ; \bb_0)\right\}\\
    &=&\0.
\ese
We have shown that $E\{\S\eff^{*\star}(Y,W,\Delta, \Z ; \bb_0)\}=\0$ as long as either $\eta_1^* (x,\z) = \eta_1 (x,\z)$ or $\eta_2^\star(c,y,\z) = \eta_2(c,y,\z)$. By the inverse function theorem and Condition \ref{con:partinv}, the solution of $E\{\S\eff^{*\star} (Y,W,\Delta, \Z ; \bb)\}=\0$ is unique in a neighborhood of $\bb_0$. Conditions \ref{con:bb}, \ref{con:seffub}, and \ref{con:conti} guarantee that $n^{-1}\sumi \S\eff^{*\star}(y_i, w_i, \delta_i, \z_i;\bb)$ converges uniformly in probability to $E\{\S\eff^{*\star} (Y,W,\Delta, \Z ; \bb)\}$ by Lemma 2.4 of \cite{newey1994large}.

Let $Q_0(\bb) = -\|E\{\S\eff^{*\star} (Y,W,\Delta, \Z ; \bb)\}\|_2^2$ and $\wh Q_n(\bb) = -\|n^{-1}\sumi \S\eff^{*\star}(y_i, w_i, \delta_i, \z_i;\bb)\|_2^2$. Since the function $t \mapsto -\|t\|_2^2$ is continuous and $n^{-1}\sumi \S\eff^{*\star}(y_i, w_i, \delta_i, \z_i;\bb)$ converges uniformly in probability to $E\{\S\eff^{*\star} (Y,W,\Delta, \Z ; \bb)\}$, it follows from Condition \ref{con:bb} that $\wh Q_n(\bb)$ converges uniformly in probability to $Q_0(\bb)$.

Since $Q_0(\bb)$ is uniquely maximized at $\bb = \bb_0$, and given Conditions \ref{con:bb} and \ref{con:conti}, all conditions in Theorem 2.1 of \cite{newey1994large} are satisfied. Therefore, $\wh \bb$, which maximizes $\wh Q_n(\bb)$, is consistent for $\bb_0$.
\qed

\subsection{Proof of Theorem \ref{th:par}}\label{sec:th1pf}

The proof assumes the following conditions, which extend those 
in Theorem~\ref{th:doublyrobust} to cover uncertainty from 
estimating $\balpha_1^*(\bb)$ and $\balpha_2^\star$, and require 
$\wh\balpha_1(\bb)$ and $\wh\balpha_2$ to be asymptotically 
linear, satisfying
\be
n^{1/2} \{\wh \balpha_1(\bb) - \balpha_1^*(\bb)\} &=& n^{-1/2} \sumi \bphi_1\{y_i,
w_i, \delta_i, \z_i;\balpha_1^*(\bb), \bb\} + o_p(1), \\ 
n^{1/2} (\wh \balpha_2 - \balpha_2^\star) &=& n^{-1/2} \sumi
\bphi_2(y_i, w_i, \delta_i, \z_i; \balpha_2^\star) + o_p(1), \nonumber
\ee
where $\bphi_1$ and $\bphi_2$ are influence functions. Because
$\balpha_1^*$ depends on $\bb$, we also require $\bphi_1\{y, w,
\delta, \z;\balpha_1^*(\bb), \bb\}$ to be continuously differentiable
in $\bb$.

\begin{enumerate}[label=(P\arabic*),ref=(P\arabic*),start=1]
    \item\label{con:pbb}
    The true parameter $\bb_0$ is
contained in a compact set $\Omega$.
\item \label{con:pseffconti}
$\S\eff^{*\star} (y,w,\delta, \z ;
    \bb, \balpha_1, \balpha_2)$ is a 
    continuous function of
    $(\bb\trans, \balpha_1\trans, \balpha_2\trans)\trans$.
    \item\label{con:pseffub} 
    $(\balpha_{10}\trans, \balpha_{20}\trans)\trans$ is in a compact set $D$, and
 \bse
    E\left\{\sup_{\bb, \balpha_1, \balpha_2}  \|\S\eff^{*\star}
      (Y,W,\Delta, \Z ;\bb, \balpha_1, \balpha_2)\|_2\right\} 
    \infty.
    \ese
   
\item \label{con:pderivub}
For all $(\balpha_1\trans, \balpha_2\trans)\trans \in D$ and $\bb\in \Omega$,
\bse
E\{\sup_{\bb, \balpha_1, \balpha_2}  \|\partial \S\eff^{*\star} (Y,W,\Delta, \Z ;
\bb, \balpha_1, \balpha_2)/\partial (\bb\trans, \balpha_1\trans, \balpha_2\trans)\|_2\} < \infty.
\ese

\item\label{con:pBinv2}
 The following matrices exist and are well-defined:
\bse
\bA_1 &\equiv& E[\partial \S\eff^{*\star} \{Y,W,\Delta, \Z ;
\bb_0, \balpha_1^*(\bb_0), \balpha_2^\star\}/\partial \balpha_1 \trans],\\ 
\bA_2 &\equiv& E[\partial \S\eff^{*\star} \{Y,W,\Delta, \Z ;
\bb_0, \balpha_1^*(\bb_0), \balpha_2^\star\}/\partial \balpha_2 \trans],\\ 
\bB &\equiv& E[d \S\eff^{*\star} \{Y,W,\Delta, \Z ;
\bb_0, \balpha_1^*(\bb_0), \balpha_2^\star\}/d \bb \trans]\\ 
&=& E[\partial \S\eff^{*\star} \{Y,W,\Delta, \Z ;
\bb_0, \balpha_1^*(\bb_0), \balpha_2^\star\}/\partial \bb \trans]\\
&&+ E[\partial \S\eff^{*\star} \{Y,W,\Delta, \Z ;
\bb_0, \balpha_1^*(\bb_0), \balpha_2^\star\}/\partial \balpha_1 \trans] \partial
\balpha_1^*(\bb_0)/\partial \bb \trans,
\ese
where $\bB$ is the total derivative with respect to $\bb$ (accounting for the dependence $\balpha_1^*(\bb)$), and
\bse
\bC &\equiv& \bB + \bA_1 E\left\{d \bphi_1(Y,W,\Delta,\Z; \balpha_{10}, \bb_0)/d \bb \trans\right\},
\ese
where $\bphi_1$ is the influence function of the estimator for $\balpha_1$. Moreover, $\bB$ and $\bC$ are invertible.
\item\label {con:palpha1conv2}
$\wh \balpha_1(\bb)$ converges uniformly in probability to $\balpha_1^*(\bb)$ for $\bb \in \Omega$, i.e.,
\bse
\sup_{\bb \in \Omega} \|\wh \balpha_1(\bb) - \balpha_1^*(\bb)\|_2 = o_p(1).
\ese
\end{enumerate}
Conditions \ref{con:pbb}--\ref{con:pBinv2} are standard smoothness
and boundedness requirements. Condition \ref{con:palpha1conv2}
requires uniform convergence of $\wh \balpha_1(\bb)$ over $\bb \in
\Omega$, which typically holds when $\wh \balpha_1(\bb)$ and
$\balpha_1^*(\bb)$ are continuous in $\bb$ and $\Omega$ is compact.

The proof proceeds via a lemma establishing uniform convergence,
followed by the main argument.

\subsubsection{Lemma for Theorem \ref{th:par}}\label{sec:l1pf}
\begin{Lem} \label{lem:l1}
    Under Conditions \ref{con:pbb}--\ref{con:pseffub} and \ref{con:palpha1conv2}, if
    $\wh\balpha_1(\bb)$ and $\wh\balpha_2$ converge in probability to
    $\balpha_1^*(\bb)$ and $\balpha_2^\star$, respectively,
    then $n^{-1} \sumi \S\eff^{*\star}\{y_i, w_i, \delta_i, \z_i; \bb, \wh\balpha_1(\bb),
    \wh\balpha_2\}$ converges uniformly to $E[\S\eff^{*\star}
    \{Y,W,\Delta, \Z ; \bb, \balpha_1^*(\bb), 
    \balpha_2^\star\}]$ in
    probability for $\bb \in \Omega$.
\end{Lem} 

\textit{Proof of Lemma \ref{lem:l1}.}
Let $\epsilon>0$ and $\nu>0$ be arbitrary. Let
\bse
&&R(Y,W,\Delta,\Z; \bb, d) \\
&&\equiv  \sup_{\| \bzeta - \bb\|_2<d} \sup_{\| (\balpha_1^{\rm T},
  \balpha_2^{ \rm T})\trans - \{\balpha_1^{*\rm T}(\bb), \balpha_2^{\star
    \rm T}\}\trans \|_2<d} \|\S\eff^{*\star} (Y,W,\Delta, \Z
; \bzeta, \balpha_1, \balpha_2) - \S\eff^{*\star} \{Y,W,\Delta, \Z
; \bb, \balpha_1^*(\bb), \balpha_2^\star\}\|_2.
\ese
Then Conditions \ref{con:pseffconti}, \ref{con:pseffub}, and the
dominated convergence theorem lead to $\lim_{d \to 0+}
    R(Y,W,\Delta,\Z; \bb, d) = 0$ almost surely. Let $a(\bb)>0$ be such that 
\be \label{eq:t4} 
E\left[R\{Y,W,\Delta,\Z; \bb, a(\bb)\}\right] < \nu.
\ee
Meanwhile, Condition \ref{con:pseffconti} implies the existence of $b(\bb)>0$ such that 
$\sup_{\|\bzeta - \bb\|_2 < b(\bb)} \|\balpha_1^*(\bzeta) - \balpha_1^*(\bb)\|_2  < a(\bb)/2.$
Define $d(\bb) = \min\{a(\bb), b(\bb)\}$. Since $\Omega$ is compact by Condition \ref{con:pbb}, the neighborhoods $[N\{\bb, d(\bb)\}]_{\bb \in \Omega}$ form an open cover of $\Omega$. Therefore, we can extract a finite subcover $[N\{\bb_k, d(\bb_k)\}]_{k = 1,\ldots,K}$.
Let $N_1(k) = N\{\bb_k, d(\bb_k)\}$ and $N_2(k) = N[\{\balpha_1^{*\rm T}(\bb_k), \balpha_2^{\star \rm T}\}\trans, d(\bb_k)]$.

By Condition \ref{con:palpha1conv2}, there exists a constant $M$ such that for $n>M$, with probability at least $1-\epsilon/(3K)$,
\bse
\sup_{\bb \in \Omega}\|\wh\balpha_1(\bb) - \balpha_1^*(\bb)\|_2  < \min_{j=1,\ldots,K} d(\bb_j)/2,
\ese
which implies that
\bse
\sup_{\bb \in N_1(k)} \|\wh\balpha_1(\bb) - \balpha_1^*(\bb_k)\|_2 &\le& \sup_{\bb \in \Omega}\|\wh\balpha_1(\bb) - \balpha_1^*(\bb)\|_2 +  \sup_{\|\bb - \bb_k\|_2 < b(\bb_k)} \|\balpha_1^*(\bb) - \balpha_1^*(\bb_k)\|_2  \\
&<& \min_{j=1,...,K} d(\bb_j)/2 + a(\bb_k)/2 \\
&\le& a(\bb_k).
\ese
Then $\Pr\{\sup_{\bb \in N_1(k)} \|\wh\balpha_1(\bb) -
\balpha_1^*(\bb_k)\|_2 \le a(\bb_k) \mbox{ for all } k=1,...,K\} \ge
1-\epsilon/3$. Thus, 
\bse 
&&\limsup_{n \to \infty}\Pr \left(\max_{k=1,...,K}\sup_{\bb \in N_1(k)} \| n^{-1} \sumi [\S\eff^{*\star}\{y_i,
w_i, \delta_i, \z_i;\bb, \wh \balpha_1(\bb), \wh\balpha_2\}  \right.\\
&&\left.-
\S\eff^{*\star}\{y_i, w_i, \delta_i, \z_i; \bb_k, \balpha_1^*(\bb_k),
\balpha_2^\star\}]\|_2 > \nu \right) \n\\
&\le& \limsup_{n \to \infty}\Pr \left(\max_{k=1,...,K}\sup_{\bb \in N_1(k)} \| n^{-1} \sumi [\S\eff^{*\star}\{y_i,
w_i, \delta_i, \z_i;\bb, \wh \balpha_1(\bb), \wh\balpha_2\}  \right.\\
&&-
\S\eff^{*\star}\{y_i, w_i, \delta_i, \z_i; \bb_k, \balpha_1^*(\bb_k),
\balpha_2^\star\}]\|_2 > \nu \\
&&\left.\mbox{ and } \sup_{\bb \in N_1(k)} \|\wh\balpha_1(\bb) - \balpha_1^*(\bb_k)\|_2  \le  a(\bb_k) \mbox{ for all } k=1,...,K \right) \n \\
&&+ \limsup_{n \to \infty}\Pr \left\{\sup_{\bb \in N_1(k)} \|\wh\balpha_1(\bb) - \balpha_1^*(\bb_k)\|_2  > a(\bb_k)\mbox{ for some } k=1,...,K \right\} \n \\
&\le& \limsup_{n \to \infty} \Pr\left[\max_{k=1,...,K} n^{-1} \sumi R\{y_i,w_i,\delta_i,\z_i; \bb_k, a(\bb_k)\} > \nu \right] + \frac{\epsilon}{3} \n \\
&=& \Pr\left(\max_{k=1,...,K} E [R\{Y,W,\Delta,\Z; \bb_k, a(\bb_k)\}] > \nu \right) + \frac{\epsilon}{3} \n \\
&=& \frac{\epsilon}{3}, 
\ese 
where the first equality follows by the strong law of large
numbers. 
Hence, we may find $M'$ such that for $n>M'$,
\be \label{eq:t5} 
&&\Pr \left(\max_{k=1,...,K}\sup_{\bb \in N_1(k)} \| n^{-1} \sumi [\S\eff^{*\star}\{y_i,
w_i, \delta_i, \z_i;\bb, \wh \balpha_1(\bb), \wh\balpha_2\} \right.\\
&&\left.-
\S\eff^{*\star}\{y_i, w_i, \delta_i, \z_i; \bb_k, \balpha_1^*(\bb_k),
\balpha_2^\star\}]\|_2 > \nu \right)
\le \frac{\epsilon}{2}.\n
\ee 
From \eqref{eq:t4}, we obtain
\be \label{eq:t6}
&&\max_{k=1,...,K} \sup_{\bb \in N_1(k)} \left\| E[\S\eff^{*\star} \{Y,W,\Delta, \Z ;
  \bb, \balpha_1^*(\bb), \balpha_2^\star\}\right. \\
  &&\left.- \S\eff^{*\star} \{Y,W,\Delta, \Z
  ; \bb_k, \balpha_1^*(\bb_k), \balpha_2^\star\}]\right\|_2 \n \\
  &\le& \max_{k=1,...,K} E [R\{Y,W,\Delta,\Z; \bb_k, a(\bb_k)\}] \n \\
  &<& \nu.\n
\ee 
Moreover, by the weak law of large numbers, for each $k = 1,\ldots,K$, there exists a constant $M_k$ such that for any $n>M_k$,
\bse
\|n^{-1} \sumi \S\eff^{*\star}\{y_i, w_i,
\delta_i, \z_i; \bb_k, \balpha_1^*(\bb_k),
\balpha_2^\star\}
-E[\S\eff^{*\star} \{Y,W,\Delta, \Z ; \bb_k, \balpha_1^*(\bb_k), \balpha_2^\star\}] \|_2 \le \nu
\ese
holds with probability at least $1-\epsilon/(2K)$. Therefore, for $n>\max\{M_1,\ldots,M_K\}$,
\be \label{eq:t7}
&&\Pr \left(\max_{k=1,...,K}\|\frac{1}{n} \sumi \S\eff^{*\star}\{y_i, w_i,
\delta_i, \z_i; \bb_k, \balpha_1^*(\bb_k),
\balpha_2^\star\}\right.\\
&&\left.-E[\S\eff^{*\star} \{Y,W,\Delta, \Z ; \bb_k, \balpha_1^*(\bb_k), \balpha_2^\star\}] \|_2 > \nu\right) < \frac{\epsilon}{2}.\n
\ee 
By combining   \eqref{eq:t5}, \eqref{eq:t6}, and \eqref{eq:t7}, it follows that for any $n>\max\{M',M_1,\ldots,M_K\}$, with probability at least $1-\epsilon$,
\bse
&& \sup_{\bb \in \Omega}\|n^{-1} \sumi \S\eff^{*\star}\{y_i, w_i,
\delta_i, \z_i;\bb, \wh\balpha_1(\bb), \wh\balpha_2\}  - E[\S\eff^{*\star}
\{Y,W,\Delta, \Z ; \bb, \balpha_1^*(\bb), \balpha_2^\star\}]\|_2\\ 
&\le& \max_{k=1,...,K}\sup_{\bb \in N_1(k)} \| n^{-1} \sumi [\S\eff^{*\star}\{y_i,
w_i, \delta_i, \z_i;\bb, \wh \balpha_1(\bb), \wh\balpha_2\}  -
\S\eff^{*\star}\{y_i, w_i, \delta_i, \z_i; \bb_k, \balpha_1^*(\bb_k),
\balpha_2^\star\}]\|_2 \\ 
&& + \max_{k=1,...,K}\|n^{-1} \sumi \S\eff^{*\star}\{y_i, w_i,
\delta_i, \z_i; \bb_k, \balpha_1^*(\bb_k),
\balpha_2^\star\}
-E[\S\eff^{*\star} \{Y,W,\Delta, \Z ; \bb_k, \balpha_1^*(\bb_k), \balpha_2^\star\}] \|_2\\ 
&& + \max_{k=1,...,K}\sup_{\bb \in N_1(k)}\left\| E\{\S\eff^{*\star} (Y,W,\Delta, \Z ;
  \bb_k, \balpha_1^*(\bb_k), \balpha_2^\star\}  -E[\S\eff^{*\star}
  \{Y,W,\Delta, \Z ; \bb, \balpha_1^*(\bb), \balpha_2^\star\}]\right\|_2\\ 
&\le& \nu + \nu + \nu  = 3\nu.
\ese
Hence, we have shown that $n^{-1} \sumi \S\eff^{*\star}\{y_i, w_i,
\delta_i, \z_i;\bb, \wh\balpha_1(\bb), \wh\balpha_2\}$ converges in
probability to $E[\S\eff^{*\star} 
\{Y,W,\Delta, \Z ; \bb, \balpha_1^*(\bb), \balpha_2^\star\}]$.
\qed

\subsubsection{Main Proof of Theorem \ref{th:par}}\label{sec:mainth1pf}
 Assume that either $\balpha_1^*(\bb_0) = \balpha_{10}$ or $\balpha_2^\star = \balpha_{20}$. Then by the proof of Theorem \ref{th:doublyrobust}, we have $E[\S\eff^{*\star}\{Y,W,\Delta, \Z ; \bb_0, \balpha_1^*(\bb_0), \balpha_2^\star\}]=\0.$
Let 
\bse 
Q_0(\bb) &=& -\|E[\S\eff^{*\star} \{Y,W,\Delta, \Z ; \bb, \balpha_1^*(\bb),
\balpha_2^\star\}]\|_2^2,\\ 
\wh Q_n(\bb) &=& -\|n^{-1}\sumi
\S\eff^{*\star}\{y_i, w_i, \delta_i, \z_i;\bb, \wh\balpha_1(\bb), \wh\balpha_2\}\|_2^2.
\ese 
First, by the inverse function theorem and Condition \ref{con:pBinv2}, the solution to $$E[\S\eff^{*\star} (Y,W,\Delta, \Z ; \bb, \balpha_1^*(\bb), \balpha_2^\star)] = \0$$ is unique in a neighborhood of $\bb = \bb_0$. Therefore, $Q_0(\bb)$ is uniquely maximized at $\bb = \bb_0$ in this neighborhood. We restrict our analysis to a compact subset of this neighborhood.

By Conditions \ref{con:pseffconti} and \ref{con:pseffub}, the function $E[\S\eff^{*\star} \{Y,W,\Delta, \Z ; \bb, \balpha_1^*(\bb), \balpha_2^\star\}]$ is continuous. Under Conditions \ref{con:pbb}--\ref{con:pseffub}, Lemma \ref{lem:l1} and the continuity of $-\|t\|_2^2$ as a function of $t$ imply that $\wh Q_n(\bb)$ converges uniformly in probability to $Q_0(\bb)$. Since $\wh \bb$ maximizes $\wh Q_n(\bb)$, Theorem 2.1 of \cite{newey1994large} establishes that $\wh \bb$ is consistent for $\bb_0$.

Note that Condition \ref{con:palpha1conv2} and the continuity of $\balpha_1^*(\bb)$ imply that
\bse
\|\wh\balpha_1(\wh \bb) - \balpha_1^*(\bb_0))\|_2 &=& \|\wh\balpha_1(\wh \bb) - \balpha_1^*(\wh \bb)\|_2 + \|\balpha_1^*(\wh \bb) - \balpha_1^*(\bb_0)\|_2 \\
&\le& \sup_{\bb \in \Omega} \|\wh\balpha_1(\bb) - \balpha_1^*(\bb)\|_2 + o_p(1)\\
&=& o_p(1).
\ese
Also, by Taylor's theorem, we have that $\limsup_{d\to0+}
R(Y,W,\Delta,\Z;d)=0$ almost surely, where 
\bse
 &&R(Y,W,\Delta,\Z;d)\\
 &\equiv& \sup_{\| (\bb\trans, \balpha_1^{\rm T}, \balpha_2^{ \rm T})\trans - \{\bb_0\trans, \balpha_1^{*\rm T}(\bb_0), \balpha_2^{\star \rm T}\}\trans \|_2<d}   \left\| \S\eff^{*\star}(Y,W,\Delta, \Z
   ; \bb, \balpha_1, \balpha_2) - \S\eff^{*\star}\{Y,W,\Delta, \Z
   ;\bb_0, \balpha_1^*(\bb_0), \balpha_2^\star\}\right.\\ 
 && - \frac{\partial \S\eff^{*\star}\{Y,W,\Delta, \Z ; \bb_0, \balpha_1^*(\bb_0),
   \balpha_2^\star\}}{\partial \balpha_1\trans}  \{\balpha_1-
 \balpha_1^*(\bb_0)\} -   \frac{\partial \S\eff^{*\star}\{Y,W,\Delta, \Z ;
   \bb_0, \balpha_1^*(\bb_0), \balpha_2^\star\}}{\partial \balpha_2\trans}
 (\balpha_2 - \balpha_2^\star )\\ 
 &&\left. -\frac{\partial \S\eff^{*\star}\{Y,W,\Delta, \Z ;
     \bb_0, \balpha_1^*(\bb_0), \balpha_2^\star\}}{\partial \bb\trans} (\bb
   - \bb_0) \right\|_2  /{ \| (\bb\trans, \balpha_1^{\rm T}, \balpha_2^{ \rm T})\trans - \{\bb_0\trans, \balpha_1^{*\rm T}(\bb_0), \balpha_2^{\star \rm T}\}\trans \|_2}. 
\ese
By applying Condition \ref{con:pderivub} and the dominated convergence theorem, we obtain $$\lim_{d\to 0+} E \{R(Y,W,\Delta,\Z;d)\}=0.$$ Therefore, for any $\epsilon>0$, there exists $d>0$ such that $E \{R(Y,W,\Delta,\Z;d)\}<\epsilon/2$, which gives

\bse
\Pr\left\{n^{-1} \sumi R(y_i, w_i, \delta_i, \bz_i; d)>\epsilon\right\}
\le \Pr\left[n^{-1} \sumi R(y_i, w_i, \delta_i, \bz_i; d) - E
  \{R(Y,W,\Delta,\Z;d)\}>\epsilon/2\right] \to 0 
 \ese
as $n\to \infty$, by the weak law of large numbers.
Therefore the consistency of
$\wh \balpha_1(\wh \bb)$, $\wh \balpha_2$, and  $\wh \bb$ further
yields that for any $\epsilon>0$,
\bse
&&\left\|n^{-1} \sumi \left\{\S\eff^{*\star}\{y_i, w_i, \delta_i,
    \z_i;\wh \bb, \wh\balpha_1(\wh \bb), \wh\balpha_2\} - \S\eff^{*\star}\{y_i,
    w_i, \delta_i, \z_i;\bb_0, \balpha_1^*(\bb_0), \balpha_2^\star\}\right.\right.\\ 
&& -\frac{\partial \S\eff^{*\star}\{y_i, w_i, \delta_i,
  \z_i;\bb_0, \balpha_1^*(\bb_0), \balpha_2^\star\}}{\partial \balpha_1\trans}
\{\wh \balpha_1(\wh \bb) - \balpha_1^*(\bb_0)\} - \frac{\partial \S\eff^{*\star}\{y_i,
  w_i, \delta_i, \z_i;\bb_0, \balpha_1^*(\bb_0), \balpha_2^\star\}}{\partial
  \balpha_2\trans} (\wh \balpha_2 - \balpha_2^\star)\\ 
&& -\left.\left.\frac{\partial \S\eff^{*\star}\{y_i, w_i, \delta_i,
    \z_i;\bb_0, \balpha_1^*(\bb_0), \balpha_2^\star\}}{\partial \bb\trans}
  (\wh \bb - \bb_0)\right\}\right\|_2/{ \| \{\wh \bb\trans, \wh \balpha_1\trans(\wh \bb), \wh
  \balpha_2\trans\}\trans - \{\bb_0\trans, \balpha_1^{*\rm T}(\bb_0),
  \balpha_2^{\star \rm T})\}\trans \|_2}\\ 
&&\le n^{-1} \sumi R(y_i, w_i, \delta_i, \bz_i; d) + o_p(1) \le \epsilon + o_p(1),
\ese
which gives that the left-hand side is $o_p(1)$. In addition,
\bse
&&n^{1/2}\{\wh\balpha_1(\wh \bb) - \balpha_1^*(\bb_0)\}\\
&=& n^{1/2}\{\wh\balpha_1(\wh \bb) - \balpha_1^*(\wh \bb)\}+n^{1/2}\{\balpha_1^*(\wh \bb) - \balpha_1^*(\bb_0)\}\\
&=&  n^{-1/2} \sumi \bphi_1\{y_i,
w_i, \delta_i, \z_i; \balpha_1^*(\wh \bb), \wh \bb\} + n^{1/2}\frac{\partial
\balpha_1^*(\bb_0)}{\partial \bb \trans}(\wh \bb - \bb_0) + o_p(1)\\
&=&  n^{-1/2} \sumi \left[\bphi_1\{y_i,
w_i, \delta_i, \z_i; \balpha_1^*(\bb_0), \bb_0\} + \frac{d \bphi_1\{y_i,
w_i, \delta_i, \z_i; \balpha_1^*(\bb_0), \bb_0\}}{d \bb \trans}(\wh \bb - \bb_0)\right]  \\
&&+ n^{1/2}\frac{\partial
\balpha_1^*(\bb_0)}{\partial \bb \trans}(\wh \bb - \bb_0) + o_p(1)\\
&=&  n^{-1/2} \sumi \bphi_1\{y_i,
w_i, \delta_i, \z_i; \balpha_1^*(\bb_0), \bb_0\} \\
&&+ \left(E\left[\frac{d \bphi_1\{Y,W,\Delta,\Z; \balpha_1^*(\bb_0), \bb_0\}}{d \bb \trans}\right] + \frac{\partial \balpha_1^*(\bb_0)}{\partial \bb \trans} + o_p(1)\right) n^{1/2}(\wh \bb - \bb_0) + o_p(1),
\ese
where the last equality holds by the weak law of large numbers.
Then 
\bse
\0 &=& n^{-1/2} \sumi \S\eff^{*\star}\{y_i, w_i, \delta_i,
\z_i;\wh \bb, \wh\balpha_1(\wh \bb), \wh\balpha_2\}  \\ 
&=& n^{-1/2} \sumi \frac{\partial \S\eff^{*\star}\{y_i, w_i, \delta_i,
  \z_i;\bb_0, \balpha_1^*(\bb_0), \balpha_2^\star\}}{\partial \balpha_1\trans}
\{\wh \balpha_1(\wh \bb) - \balpha_1^*(\bb_0)\}\\
&&+ n^{-1/2} \sumi \frac{\partial
  \S\eff^{*\star}\{y_i, w_i, \delta_i, \z_i;\bb_0, \balpha_1^*(\bb_0),
  \balpha_2^\star\}}{\partial \balpha_2\trans} (\wh \balpha_2 -
\balpha_2^\star)\\ 
&& + n^{-1/2} \sumi \frac{\partial \S\eff^{*\star}\{y_i, w_i, \delta_i,
  \z_i;\bb_0, \balpha_1^*(\bb_0), \balpha_2^\star\}}{\partial \bb\trans} (\wh
\bb - \bb_0)\\
&&+ n^{1/2}\| \{\wh \bb\trans, \wh \balpha_1\trans(\wh \bb), \wh \balpha_2\trans\}\trans - \{\bb_0\trans, \balpha_1^{*\rm T}(\bb_0), \balpha_2^{\star \rm T}\}\trans \|_2   o_p(1)  \\ 
&& + n^{-1/2}  \sumi \S\eff^{*\star}\{y_i, w_i, \delta_i,
\z_i;\bb_0, \balpha_1^*(\bb_0), \balpha_2^\star\} \\ 
&=&  \{\bA_1 + o_p(1)\} n^{1/2} \{\wh \balpha_1(\wh \bb) - \balpha_1^*(\bb_0)\} +
\{\bA_2 + o_p(1)\} n^{1/2} (\wh \balpha_2 - \balpha_2^\star)\\
&&+ \left(E\left[ \frac{\partial \S\eff^{*\star}\{Y, W, \Delta,
  \Z;\bb_0, \balpha_1^*(\bb_0), \balpha_2^\star\}}{\partial \bb\trans}\right]+o_p(1)\right) n^{1/2}(\wh
\bb - \bb_0) + o_p(1) \\ 
&& + n^{-1/2} \sumi \S\eff^{*\star}\{y_i, w_i, \delta_i,
\z_i;\bb_0, \balpha_1^*(\bb_0), \balpha_2^\star\} \\ 
&=& \{\bA_1 + o_p(1)\} n^{-1/2} \sumi \bphi_1\{y_i,
w_i, \delta_i, \z_i; \balpha_1^*(\bb_0), \bb_0\}\\
&&+ \{\bA_1 + o_p(1)\}
\left(E\left[\frac{d \bphi_1\{Y,W,\Delta,\Z; \balpha_1^*(\bb_0), \bb_0\}}{d \bb \trans}\right] + \frac{\partial \balpha_1^*(\bb_0)}{\partial \bb \trans} + o_p(1)\right) n^{1/2}(\wh \bb - \bb_0) \\
&&+ \{\bA_2 + o_p(1)\} n^{-1/2} \sumi \bphi_2(y_i, w_i, \delta_i, \z_i; \balpha_2^\star)\\
&&+ \left(E\left[ \frac{\partial \S\eff^{*\star}\{Y, W, \Delta,
  \Z;\bb_0, \balpha_1^*(\bb_0), \balpha_2^\star\}}{\partial \bb\trans}\right]+o_p(1)\right)  n^{1/2}(\wh
\bb - \bb_0) + o_p(1)\\
&& + n^{-1/2} \sumi \S\eff^{*\star}\{y_i, w_i, \delta_i,
\z_i;\bb_0, \balpha_1^*(\bb_0), \balpha_2^\star\} \\ 
&=&  n^{-1/2} \sumi \left[\S\eff^{*\star}\{y_i, w_i, \delta_i,
\z_i;\bb_0, \balpha_1^*(\bb_0), \balpha_2^\star\} + \bA_1 \bphi_1\{y_i,
w_i, \delta_i, \z_i; \balpha_1^*(\bb_0), \bb_0\} + \bA_2  \bphi_2(y_i, w_i, \delta_i, \z_i; \balpha_2^\star) \right]\\ 
&&+ \left\{ \bC+ o_p(1) \right\}n^{1/2}(\wh \bb - \bb_0) + o_p(1),
\ese
where $\bC = \bB + \bA_1 E\left[d \bphi_1\{Y,W,\Delta,\Z; \balpha_1^*(\bb_0), \bb_0\}/d \bb \trans\right]$.
Since $\bC$ is invertible, the continuous mapping theorem gives us $\{\bC + o_p(1)\}^{-1} = \bC^{-1} + o_p(1)$. Therefore,
\bse
n^{1/2} (\wh \bb - \bb_0) &=& - \bC^{-1} n^{-1/2} \sumi \left[\S\eff^{*\star}\{y_i, w_i, \delta_i, \z_i;\bb_0, \balpha_1^*(\bb_0),
  \balpha_2^\star\}\right. \\ 
&& \left.+ \bA_1 \bphi_1\{y_i, w_i, \delta_i, \z_i;\balpha_1^*(\bb_0), \bb_0\} + \bA_2
  \bphi_2(y_i, w_i, \delta_i, \z_i; \balpha_2^\star) \right\} + o_p(1). 
\ese

Recall that $E\{\S\eff^{*\star}(Y,W,\Delta, \Z ; \bb_0, \balpha_1, \balpha_2)\}=\0$ when either $\balpha_1 = \balpha_{10}$ or $\balpha_2 = \balpha_{20}$. First, assume that $\balpha_1^*(\bb_0) = \balpha_{10}$. Since $E\{\S\eff^{\star} (Y,W,\Delta, \Z ; \bb_0, \balpha_{10}, \balpha_2)\} = \0$ for any $\balpha_2$, we have $\bA_2 = E\{\partial \S\eff^{*\star} (Y,W,\Delta, \Z ; \bb_0, \balpha_{10}, \balpha_2^\star)/\partial \balpha_2 \trans\} =\0$.
By the central limit theorem, where $\stackrel{d}{\to}$ denotes convergence in distribution,
\bse
n^{1/2} (\wh \bb - \bb_0) &=& -\bC^{-1} n^{-1/2} \sumi  \left\{\S\eff^{\star}(y_i, w_i, \delta_i, \z_i;\bb_0, \balpha_{10}, \balpha_2^\star)+ \bA_1 \bphi_1(y_i, w_i, \delta_i, \z_i;\balpha_{10}, \bb_0)\right\} + o_p(1)\\
&\stackrel{d}{\to}& \Normal\{0, \bC^{-1}\bSigma_1 (\bC^{-1})\trans\}.
\ese

Likewise, if $\balpha_2^\star = \balpha_{20}$, then we have $\bA_1 = \0$ and $\bC = \bB$. Thus,
\bse
n^{1/2} (\wh \bb - \bb_0) &=& -\bB^{-1} n^{-1/2} \sumi  \left[\S\eff^{*}\{y_i, w_i, \delta_i, \z_i;\bb_0, \balpha_1^*(\bb_0), \balpha_{20}\}+ \bA_2 \bphi_2(y_i, w_i, \delta_i, \z_i; \balpha_{20})\right] + o_p(1)\\
&\stackrel{d}{\to}& N\{0, \bB^{-1}\bSigma_2 (\bB^{-1})\trans\}.
\ese

Finally, assume that both $\balpha_1^*(\bb_0) = \balpha_{10}$ and $\balpha_2^\star = \balpha_{20}$. Then $\bA_1= \0$ and $\bA_2= \0$. Note that since $E\{\S\eff(Y,W,\Delta, \Z ; \bb_0)\}=\0$, we have $\var\{\S\eff (Y,W,\Delta, \Z ; \bb_0)\} = E\{\S\eff^{\otimes 2}  (Y,W,\Delta, \Z ; \bb_0)\}$. Since $E_{\bb}\{\S\eff (Y,W,\Delta, \Z ; \bb)\}= \0$ for all $\bb$, differentiating with respect to $\bb$ yields
\bse
\bB &=& E\{\partial \S\eff (Y,W,\Delta, \Z ; \bb_0)/\partial \bb \trans\} \\
&=& -E\{\S\eff (Y,W,\Delta, \Z ; \bb_0)\S_\bb\trans (Y,W,\Delta, \Z ; \bb_0)\}\\
&=& -E\{\S\eff^{\otimes 2}  (Y,W,\Delta, \Z ; \bb_0)\},
\ese
which gives
\bse
n^{1/2} (\wh\bb -  \bb_0) &=& \bB^{-1} n^{-1/2} \sumi \S\eff(y_i, w_i, \delta_i, \z_i; \bb_0) + o_p(1) \\
&\stackrel{d}{\to}& \Normal(0, [E\{\S\eff^{\otimes 2} (Y,W,\Delta, \Z;\bb_0)\}]^{-1}).
\ese
\qed

\subsection{Technical Framework for Nonparametric Estimation}\label{sec:nonpar-details}

\subsubsection{Construction of Variability Functions}\label{sec:nonpar-variability-functions}

To analyze how nonparametric estimation errors affect SPYCE's asymptotic properties, we must identify the specific nonparametric components that are estimated and track how estimation errors propagate through SPYCE's efficient score function. While we previously defined the expectation operators $\wh E_1$ and $\wh E_2$ in a compact form in \eqref{eq:n15} and \eqref{eq:n16}, these operators are constructed from nonparametric estimators without requiring specification of the underlying nuisance models $\eta_1 = f_{X|\Z}$ and $\eta_2 = f_{C|Y,\Z}$. To apply standard asymptotic theory and derive the variability functions in Theorem \ref{th:nonpar}, we need to make the dependence on estimable components explicit.

Specifically, the expectation operators depend on two types of functional components that we can estimate nonparametrically: the conditional survival functions $S_{C|Y,\Z}$ and $S_{X|Y,\Z}$ (estimated using conditional Kaplan-Meier estimators), and empirical distributions of the observed data (incorporated through kernel-weighted averages). By expressing the expectation operators in terms of these estimable components, we can apply established asymptotic theory for conditional survival function estimators and kernel smoothing to derive the asymptotic properties of SPYCE.
We therefore rewrite the expectation operators to make their functional dependence explicit. The true expectation operator $E_{10}$ can be expressed as:
\be\label{eq:e1}
&&E_{10}\{g(y,X,c,\z)\mid y, c, \z;\bb\}\\
&=&E_1\{g(y,X,c,\z)\mid y, c, \z; \bb, S_{C|Y,\Z}, f_{\Delta, W,Y|\Z}\}\n\\
&=& \frac{1}{f_{Y|\Z}(y,\z;\bb)}\iiint \frac{\delta g(y,w,c,\z)}{S_{C|Y,\Z} (w,y',\z)}f_{\Delta,W,Y|\Z}(\delta,w,y',\z) f_{Y|X,\Z}(y,w,\z;\bb)d\delta dwdy',\n
\ee
and similarly for $E_{20}$:
\be\label{eq:e2}
&&E_{20}\{g(Y,x,C,\z)\mid x, \z; \bb\}\\
&=& E_2\{g(Y,x, C,\z) \mid x,\z; \bb, S_{X|Y,\Z}, f_{\Delta, W|Y,\Z}\} \n\\
&=& \int \frac{\iint (1-\delta) g(y,x,w,\z)f_{\Delta, W|Y,\Z}(\delta,w,y,\z) /S_{X|Y,\Z} (w,y,\z)d\delta dw}{\iint (1-\delta) f_{\Delta, W|Y,\Z}(\delta,w,y,\z) /S_{X|Y,\Z} (w,y,\z)d\delta dw}\n\\
&&\times f_{Y|X,\Z}(y,x,\z;\bb) dy.\n
\ee
Here, $f_{\Delta, W,Y|\Z}$ and $f_{\Delta, W|Y,\Z}$ represent empirical measures of the observed data $(\Delta, W, Y, \Z)$, which we will incorporate through kernel-weighted averages with Dirac delta functions defined below.

\eqref{eq:e1} and \eqref{eq:e2} decompose the expectation operators into their constituent nonparametric components, which allows asymptotic analysis of how estimation errors propagate through SPYCE. By expressing the operators in terms of the conditional survival functions $S_{C|Y,\Z}$ and $S_{X|Y,\Z}$ and empirical distributions, we can apply established asymptotic theory for conditional Kaplan-Meier estimators and kernel smoothing to derive the variability functions in Theorem \ref{th:nonpar}.

The construction proceeds by substituting the conditional survival functions with their conditional Kaplan-Meier estimators $\wh S_{C|Y,\Z}$ and $\wh S_{X|Y,\Z}$ given in \eqref{eq:hatS}, and incorporating the kernel-weighted averages: 
\bse
  \wh f_{\Delta, W,Y|\Z} (\delta,w,y,\z) &\equiv& \sum_{j=1}^n\frac{d_\0(\bo-\bo_j)K_{h_2}^{(m_2)}(\z-\z_j)}{ \sum_{k=1}^n K_{h_2}^{(m_2)}(\z-\z_k)}\\
  \wh f_{\Delta, W|Y,\Z} (\delta,w,y,\z) &\equiv& n^{-1} \sum_{j=1}^n \frac{d_\0(\bo-\bo_j) K_{h_3}^{(m_3)}(y-y_j,\z-\z_j)}{ f_{Y,\Z}(y,\z)},
  \ese
where $d_\0(\cdot)$ is the Dirac delta function. The resulting plug-in estimators are:
\bse
\wh E_1\{g(y,X,c,\z)\mid y, c, \z;\bb\} &=& E_1\{g(y,X,c,\z)\mid y, c, \z; \bb,\wh S_{C|Y,\Z}, \wh f_{\Delta, W,Y|\Z}\}\\
\wh E_2\{g(Y,x, C,\z) \mid x,\z; \bb\} &=& E_2\{g(Y,x, C,\z) \mid x,\z; \bb,\wh S_{X|Y,\Z}, \wh f_{\Delta, W|Y,\Z}\}.
\ese
The plug-in framework offers flexibility: alternative nonparametric estimators for the conditional survival functions, such as Nelson-Aalen estimators, can be substituted to create new expectation operator estimators with similar theoretical properties.

To derive the variability functions in Theorem \ref{th:nonpar}, we need to characterize how estimation errors in the conditional survival function estimators $\wh S_{C|Y,\Z}$ and $\wh S_{X|Y,\Z}$ propagate through SPYCE's efficient score function. This characterization relies on the asymptotic linearity of these conditional survival function estimators, which decomposes their estimation errors into manageable components that can be tracked through the asymptotic analysis.

Under Condition \ref{con:nker} (detailed in Section \ref{sec:nonpar-regularity-conditions}), by the multivariate extension of \cite{gm1994asymptotic}, Theorem 2.3, we have the asymptotic linearity of $\wh S_{C|Y,\Z}$ and $\wh S_{X|Y,\Z}$, given as
\be
&&\wh S_{C|Y,\Z}(t,  y, \z) - S_{C|Y,\Z}(t,  y, \z) \label{eq:n9}\\
&&= \frac{\sum_{j=1}^n K_{h_1}^{(m_1)}(y - y_j,\z - \z_j) \xi_C(w_j, \delta_j, t,
  y, \z)}{\sum_{k=1}^n K_{h_1}^{(m_1)}(y - y_k,\z - \z_k)} +
O_p\left\{\left(\frac{\log n}{nh_1^d}\right)^{3/4} + h_1^{m_1}\right\}, \n\\
&&\wh S_{X|Y,\Z}(t,  y, \z) - S_{X|Y,\Z}(t,  y, \z) \label{eq:n10}\\
&&= \frac{\sum_{j=1}^n K_{h_1}^{(m_1)}(y - y_j,\z - \z_j) \xi_X(w_j, \delta_j, t,
  y, \z)}{\sum_{k=1}^n K_{h_1}^{(m_1)}(y - y_k,\z - \z_k)} +
O_p\left\{\left(\frac{\log n}{nh_1^d}\right)^{3/4} +
  h_1^{m_1}\right\},\n
\ee
where $d$ is the dimension of the continuous part of $(y,\z\trans)\trans$, and the influence functions $\xi_C$ and $\xi_X$ are defined as
\bse
\xi_C(w_j, \delta_j, t, y, \z) &\equiv& S_{C|Y,\Z}(t,y,\z) \left\{-\int_{0}^{\min(w_j,t)} \frac{S_{X|Y,\Z}(s,y,\z) f_{C|Y,\Z} (s,y,\z)}{S_{W|Y,\Z}(s,y,\z)^2} ds + \frac{I(w_j \le t, \delta_j = 0)}{S_{W|Y,\Z}(w_j,y,\z)}\right\},\\
\xi_X(w_j, \delta_j, t, y, \z) &\equiv& S_{X|Y,\Z}(t,y,\z) \left\{-\int_{0}^{\min(w_j,t)} \frac{S_{C|Y,\Z}(s,y,\z) f_{X|Y,\Z} (s,y,\z)}{S_{W|Y,\Z}(s,y,\z)^2} ds + \frac{I(w_j \le t, \delta_j = 1)}{S_{W|Y,\Z}(w_j,y,\z)}\right\}.
\ese 
The convergence rates of the remainder terms depend on the bandwidth conditions: under Condition \ref{con:nkerbw}, the error terms satisfy $O_p\{(\log n)^{3/4}(nh_1^d)^{-3/4} + h_1^{m_1}\} =o_p(n^{-1/4})$, while under the stronger Condition \ref{con:nkerbw'}, they improve to $o_p(n^{-1/2})$. 

The influence functions $\xi_C$ and $\xi_X$ serve as the building blocks that connect estimation errors in the conditional survival functions to the variability functions in the asymptotic variance of $\wh\bb$ for Cases 1 and 2 respectively. These functions satisfy several useful properties that facilitate the asymptotic analysis, as detailed in the following remark.

\begin{Rem}[Properties of Influence Functions]
The influence functions $\xi_C$ and $\xi_X$ satisfy several key properties that support the asymptotic theory. First, they have zero conditional expectation:
\be
E\{\xi_C(W_j, \Delta_j, t, y, \z)\mid Y_j = y,\Z_j = \z\} &=& 0,\label{eq:n39}\\
E\{\xi_X(W_j, \Delta_j, t, y, \z)\mid Y_j = y,\Z_j = \z\} &=& 0. \label{eq:n40}
\ee
These zero-mean properties in \eqref{eq:n39} and \eqref{eq:n40} guarantee that the conditional survival function estimators $\wh S_{C|Y,\Z}$ and $\wh S_{X|Y,\Z}$ are consistent. The influence functions average to zero, so the estimation errors cancel out asymptotically.
The influence functions also satisfy computational identities that simplify the asymptotic variance calculations:
\be
E\left\{\frac{\xi_C(w_j, \delta_j, X, y, \z)}{ S_{C|Y,\Z}(X,y,\z)}\mid y,\z\right\}
&=& 1-\frac{\delta_j}{S_{C|Y,\Z}(w_j,y,\z)} , \label{eq:n41}\\
E\left\{\frac{\xi_X(w_j, \delta_j, C, y, \z)}{ S_{X|Y,\Z}(C,y,\z)}\mid y,\z\right\} &=& 1-\frac{1-\delta_j}{S_{X|Y,\Z}(w_j,y,\z)} .\label{eq:n42}
\ee  
These identities in  \eqref{eq:n41} and \eqref{eq:n42} provide computational shortcuts used extensively in deriving the explicit forms of the variability functions for $\wh\bb$.
\end{Rem}

The asymptotic linearity results in \eqref{eq:n9} and \eqref{eq:n10}, combined with the functional decomposition in  \eqref{eq:e1} and \eqref{eq:e2}, now allow us to construct the variability functions that appear in Table \ref{table:ta1} and Theorem \ref{th:nonpar}. These variability functions---denoted $\h_{1{\rm s}}^{\star}, \h_{1{\rm k}}^{\star}, \h_{2{\rm s}}^*, \h_{2{\rm k}}^*$---quantify precisely how estimation errors from each nonparametric component propagate through SPYCE's efficient score function to affect the asymptotic variance of $\wh\bb$. As outlined in Table \ref{table:ta1}, $\h_{1{\rm s}}^{\star}$ and $\h_{1{\rm k}}^{\star}$ capture the additional variability from conditional survival function estimation and kernel estimation in Case 1, while $\h_{2{\rm s}}^*$ and $\h_{2{\rm k}}^*$ capture the corresponding variability in Case 2.

The construction relies on the influence functions $\xi_C$ and $\xi_X$ derived above, which serve as the building blocks that connect estimation errors in the conditional survival functions to changes in SPYCE's estimating equations. To define the variability functions referred to in Theorem \ref{th:nonpar}, let $\bo_j\equiv (y_j,w_j,\delta_j,\z_j)$. We construct these functions systematically by source of estimation error. 

For Case 1, the functions $\bh_{1{\rm s}1}^\star$ and $\bh_{1{\rm k}1}^\star$ capture how errors in conditional survival function estimation ($\bh_{1{\rm s}1}^\star$, where "s" denotes survival) and kernel estimation ($\bh_{1{\rm k}1}^\star$, where "k" denotes kernel) propagate through the expectation operator $\wh E_1$:
\bse
\bh_{1{\rm s}1}^\star(y_j,w_j, \delta_j, \z_j)&\equiv&- E\left[ I(X>C) \{\S\eff^\star(Y,
    X, 1, \z_j;\bb_0) -\S\eff^\star(Y, C, 0, \z_j;\bb_0)\} \right.\n\\
&& \left.\times\frac{\xi_C (w_j, \delta_j, X,y_j,\z_j) f_{Y|X,\Z}(y_j,
    X, \z_j)}{S_{C|Y,\Z} (X,y_j,\z_j)f_{Y|\Z}(y_j, \z_j)} \mid \Z = \z_j,\bo_j\right],\\
\bh_{1{\rm k}1}^\star(y_j,w_j, \delta_j, \z_j)&\equiv&\frac{\delta_j }{S_{C|Y,\Z} (w_j,y_j,\z_j)}E\left[I(w_j>C)\{\S\eff^\star(Y,
    w_j, 1, \z_j;\bb_0) -\S\eff^\star(Y, C, 0, \z_j;\bb_0)\}\right.\n\\
    &&\left.\mid X = w_j, \Z = \z_j,\bo_j \right].
\ese
These functions capture the direct impact on $\wh E_1$. However, estimation errors also propagate indirectly through the function $\wh{\mathbf{a}}^{\star}$ that depends on $\wh E_1$. The functions $\bh_{1{\rm s}2}^\star$ and $\bh_{1{\rm k}2}^\star$ capture this indirect propagation:
\bse
\bh_{1{\rm s}2}^\star(y_j,w_j, \delta_j, \z_j)
&\equiv& E_2^\star\left[ I(X>C)\{\S\eff^\star(Y,X,1,\z_j;\bb_0) -\S\eff^\star(Y,C,0,\z_j;\bb_0) \}
\right.\n\\
&&\times \left.\frac{ \xi_C (w_j, \delta_j, X,y_j,\z_j)f_{Y|X,\Z}(y_j,X,\z_j)}{S_{C|Y,\Z} (X,y_j,\z_j)f_{Y|\Z}(y_j,\z_j)} \mid \Z = \z_j,\bo_j;\bb_0\right], \\
\bh_{1{\rm k}2}^\star(y_j,w_j, \delta_j, \z_j)&\equiv& -\frac{\delta_j }{S_{C|Y,\Z} (w_j,y_j,\z_j)}E_2^\star\left[I(w_j>C) \{\S\eff^\star(Y,w_j,1,\z_j;\bb_0)- \S\eff^\star(Y,C,0,\z_j;\bb_0)\}\right.\n\\
&&\left.\mid X=w_j, \Z = \z_j,\bo_j;\bb_0\right].
\ese 
The total variability functions for Case 1 combine both direct and indirect effects:
\bse
\bh_{1{\rm s}}^\star(y_j,w_j, \delta_j, \z_j)
&\equiv& \bh_{1{\rm s}1}^\star(y_j,w_j, \delta_j, \z_j) + \bh_{1{\rm s}2}^\star(y_j,w_j, \delta_j, \z_j),\\
\bh_{1{\rm k}}^\star(y_j,w_j, \delta_j, \z_j)&\equiv& \bh_{1{\rm k}1}^\star(y_j,w_j, \delta_j, \z_j) + \bh_{1{\rm k}2}^\star(y_j,w_j, \delta_j, \z_j).
\ese
For Case 2, where $\wh E_2$ is estimated nonparametrically while $E_1^*$ is a fixed working model, the variability functions $\bh_{2{\rm s}}^*$ and $\bh_{2{\rm k}}^*$ capture how estimation errors in $\wh E_2$ propagate indirectly through $\wh{\mathbf{a}}^*$:
\bse
&&\bh_{2{\rm s}}^*(y_j,w_j, \delta_j, \z_j)\n\\
&\equiv& E\left( \frac{ \xi_X (w_j, \delta_j, C,y_j,\z_j)}{S_{X|Y,\Z}(C,y_j,\z_j)} [I(X\le C)\S\eff^* (y_j,X,1,\z_j;\bb_0) + I(X>C)\S\eff^* (y_j,C,0,\z_j;\bb_0)\right.\n\\
&& - E\{I(X\le C)\S\eff^* (y_j,X,1,\z_j;\bb_0) + I(X>C)\S\eff^* (y_j,C,0,\z_j;\bb_0)\mid X, Y = y_j,\Z = \z_j\}]\n\\
&&\left.\mid Y = y_j,\Z = \z_j,\bo_j\right),
\ese
and
\bse
&&\bh_{2{\rm k}}^*(y_j,w_j, \delta_j, \z_j)\n\\
&\equiv& -\frac{1-\delta_j}{S_{X|Y,\Z} (w_j,y_j,\z_j)} E\left[ I(X \le w_j)\S\eff^*(y_j,X,1,\z_j;\bb_0) + I(X > w_j) \bS\eff^*(y_j,w_j,0,\z_j;\bb_0)\right.\n\\
&& - E\{I(X\le C)\S\eff^* (y_j,X,1,\z_j;\bb_0) + I(X>C)\S\eff^* (y_j,C,0,\z_j;\bb_0)\mid X, Y = y_j,\Z = \z_j\}\n\\
&& \left.\mid Y=y_j,\Z = \z_j,\bo_j\right].
\ese
The variability functions above are constructed using Gateaux derivative techniques to characterize how perturbations in the nonparametric components affect the efficient score function. Using Gateaux rather than Fréchet differentiability provides the necessary theoretical flexibility for infinite-dimensional nuisance model estimation while maintaining computational tractability.

\subsubsection{Regularity Conditions}\label{sec:nonpar-regularity-conditions}
The theoretical results for SPYCE with nonparametric estimation require several regularity conditions that ensure the nonparametric estimators behave well asymptotically. These conditions are standard in the kernel smoothing and semiparametric literature, but we briefly explain their roles in establishing SPYCE's properties.

Conditions \ref{con:nker}--\ref{con:nkerbw'} govern the kernel functions and bandwidth selection used in our conditional survival function estimators and expectation operators. Condition \ref{con:nker} specifies that the kernel functions have the smoothness and moment properties needed for consistent estimation, while Conditions \ref{con:nkerbw} and \ref{con:nkerbw'} control how the bandwidth shrinks with sample size to balance bias and variance. These bandwidth conditions determine the convergence rates we can achieve: the standard condition \ref{con:nkerbw} yields $o_p(n^{-1/4})$ rates, while the stronger condition \ref{con:nkerbw'} achieves $o_p(n^{-1/2})$ rates needed for asymptotic normality.

Conditions \ref{con:npbb}--\ref{con:nderivbddaway} impose smoothness and boundedness requirements on the underlying densities and conditional survival functions. These conditions ensure that our target parameter lies in a well-defined space (Condition \ref{con:npbb}), that key densities are bounded away from zero to avoid division-by-zero issues (Condition \ref{con:nbddaway}), and that the functions have sufficient smoothness for our conditional Kaplan-Meier estimators to achieve their required convergence rates (Condition \ref{con:nderivbddaway}).

Finally, Conditions \ref{con:nderivub} and \ref{con:nBinv2} require that SPYCE's efficient score function and its derivatives are sufficiently well-behaved to apply standard asymptotic theory. These are the nonparametric analogs of conditions needed in the parametric case and guarantee that our estimator achieves $n^{1/2}$-consistency and asymptotic normality.
\begin{enumerate}[label=(N\arabic*),ref=(N\arabic*),start=1]
    \item\label{con:nker}
    The $a$-variate kernel function $K^{(m)}(\cdot)$ is a product kernel function $K^{(m)}(\bt) = \prod_{i=1}^a k(t_i)$ with a univariate kernel function $k(\cdot)$. Here, $k(\cdot)$ is a smooth function
    on a compact support in $\mathbb{R}$. Then $\int k(u)du = 1$, $\int |k(u)|du <\infty$,
    $\int u^r k(u) du = 0$ for $r=1,...,m-1$, and $\int |u|^m k(u) du<
    \infty$ with $\int u^m k(u) du \ne 0$. Here, the  kernel order 
    $m$ satisfies $m>a/2$.  
    \item\label{con:nkerbw}
     Let $d$ and $d'$ be the dimensions of the continuous part of
     $(y,\z\trans)\trans$ and $\z$, respectively.
    The bandwidths $h_1$, $h_2$, and $h_3$ satisfy $n h_1^{4m_1} \to 0$, $nh_1^{2d}/(\log n)^2
    \to \infty$, $n h_2^{4m_2} \to 0$, $nh_2^{2d'}/(\log
    n)^2 \to \infty$, $n h_3^{4m_3} \to 0$, and $nh_3^{2d}/(\log n)^2 \to \infty$.
      \end{enumerate}
    \begin{enumerate}[label=(N\arabic*$'$),ref=(N\arabic*$'$), start=2]  
     \item \label{con:nkerbw'}
    The bandwidths $h_1$, $h_2$, and $h_3$ satisfy $n h_1^{2m_1} \to 0$, $nh_1^{3d}/(\log n)^3 \to \infty$, $n h_2^{2m_2} \to 0$, $nh_2^{2d'}/(\log n)^2 \to \infty$, $n h_3^{2m_3} \to 0$, and $nh_3^{2d}/(\log n)^2 \to \infty$.
\end{enumerate}
\begin{enumerate}[label=(N\arabic*),ref=(N\arabic*),start=3]
  \item \label{con:npbb}
    The true parameter $\bb_0$ is contained in a compact set $\Omega$.
 \item \label{con:nbddaway}
    $f_{Y,\Z}(y,\z; \bb)$ has compact support. Also, $f_{Y,\bZ}(y,\z; \bb)$ is uniformly bounded away from zero and
    uniformly bounded with respect to $\bb$, that is,
    \bse
    \inf_{\bb \in \Omega} \inf_{y,\z} f_{Y,\bZ}(y,\z; \bb) > 0,~ \sup_{\bb \in \Omega} \sup_{y,\z} f_{Y,\bZ}(y,\z; \bb) < \infty.
    \ese
    Also, $f_{Y|\bZ}(y,\z; \bb)$ is uniformly bounded away from zero with respect to $\bb$, that is,
     \bse
    \inf_{\bb \in \Omega} \inf_{y,\z} f_{Y|\bZ}(y,\z; \bb) > 0.
    \ese
    \item \label{con:nderivbddaway}
        For any $\bb \in \Omega$, $f_{Y,\Z}(y,\z; \bb)$,
        $S_{W|Y,\Z}(t,y,\z; \bb)$, $\int_t^\infty S_{C\mid
          Y,\Z}(s,y,\z)f_{X\mid Y,\Z}(s,y,\z;\bb)ds$, and $\int_t^\infty S_{X\mid
          Y,\Z}(s,y,\z;\bb)f_{C\mid Y,\Z}(s,y,\z)ds$
        have bounded continuous $m$-th derivatives with respect to
         the continuous part of  $(y,\z\trans)\trans$.
\end{enumerate}

    \begin{enumerate}[label=(N\arabic*),ref=(N\arabic*),start=6]
    \item \label{con:nderivub}
$\S\eff (Y,W,\Delta, \Z ;
\bb, E_1, \ba)$ is differentiable with respect to $\bb$, and
\bse
E\{\sup_{\bb \in \Omega}  \|\S\eff (Y,W,\Delta, \Z ;
\bb, E_{10}, \ba)\|_2\} < \infty,&&\\
E\{\sup_{\bb \in \Omega}  \|\partial \S\eff (Y,W,\Delta, \Z ;
\bb, E_{10}, \ba)/\partial \bb\trans\|_2\} < \infty.
\ese

\item\label{con:nBinv2}
 $\S\eff^\star(Y, W, \Delta,
  \Z; \bb_0, E_{10}, \ba_0^\star)$ and $\S\eff^*(Y, W, \Delta,
  \Z; \bb_0, E_1^*, \ba_0^*)$ are differentiable with respect to
  $\bb$. Moreover, the following matrices $\bB^\star$ and $\bB^*$, defined as
\bse
 \bB^\star &\equiv&  E\left\{\frac{d \S\eff^\star(Y, W, \Delta,
  \Z; \bb_0, E_{10},  \ba_0^\star)}{d \bb \trans}\right\}\\
  &=&E\left\{\frac{\partial \S\eff^\star(Y, W, \Delta,
  \Z; \bb_0, E_{10}, \ba_0^\star)}{\partial \bb \trans} +\frac{\partial \S\eff^\star(Y, W, \Delta,
  \Z; \bb_0, E_{10}, \ba_0^\star)}{\partial \ba} \frac{\partial \ba_0^\star(\bb_0)}{\partial \bb \trans}\right\},\\
\bB^* &\equiv& E\left\{\frac{d \S\eff^*(Y, W, \Delta,
  \Z; \bb_0, E_1^*, \ba_0^*)}{d \bb \trans}\right\}\\
  &=& E\left\{\frac{\partial \S\eff^*(Y, W, \Delta,
  \Z; \bb_0, E_1^*, \ba_0^*)}{\partial \bb \trans} + \frac{\partial \S\eff^*(Y, W, \Delta,
  \Z; \bb_0, E_1^*, \ba_0^*)}{\partial \ba} \frac{\partial \ba_0^*(\bb_0)}{\partial \bb \trans}\right\},
\ese
 are invertible.
 \end{enumerate}
Note that the stronger bandwidth condition \ref{con:nkerbw'} automatically satisfies condition \ref{con:nkerbw}. The bandwidth conditions control the convergence rates we can achieve for our nonparametric estimators. Under condition \ref{con:nkerbw}, the conditional survival function estimators $\wh S_{C|Y,\Z}$ and $\wh S_{X|Y,\Z}$ converge at rate $o_p(n^{-1/4})$ in uniform norm, which in turn ensures that our expectation operator estimators $\wh E_1$ and $\wh E_2$ converge to their true counterparts $E_{10}$ and $E_{20}$ at the same rate. This convergence rate is sufficient for Case 3 (where both expectation operators are estimated nonparametrically) to achieve semiparametric efficiency through standard Taylor expansion arguments.

\subsection{Proof of Theorem \ref{th:nonpar}}\label{sec:th2pf}
\subsubsection{Lemmas for Uniform Convergence Rate}\label{sec:l2-4pf}

\begin{Lem} \label{lem:l2}
      Under Conditions \ref{con:nker}--\ref{con:nderivbddaway},
     $\sup_{t,y,\z} |\wh S_{C|Y,\Z}(t,y,\z) - S_{C|Y,\Z}(t, y,\z)| =  o_p(n^{-1/4})$ and
     $\sup_{t,y,\z} |\wh S_{X|Y,\Z}(t,y,\z) - S_{X|Y,\Z}(t, y,\z)|
     = o_p(n^{-1/4})$.
\end{Lem}

Proof. 
Under Conditions \ref{con:nker}--\ref{con:nderivbddaway}, Corollary 2.2 of \cite{dabrowska1989uniform} implies that
\bse
\sup_{t,y,\z} |\wh S_{C|Y,\Z}(t,y,\z) - S_{C|Y,\Z}(t, y,\z)| = O_p\{(-d \log  h_1)^{1/2}n^{-1/2}h_1^{-d/2} + h_1^{m_1}\},\\
\sup_{t,y,\z} |\wh S_{X|Y,\Z}(t,y,\z) - S_{X|Y,\Z}(t, y,\z)| = O_p\{(-d \log  h_1)^{1/2}n^{-1/2}h_1^{-d/2}+ h_1^{m_1}\}.
\ese
Since $(-d \log h_1)^{1/2}n^{-1/2}h_1^{-d/2} = o_p(n^{-1/4})$ under the bandwidth assumption $nh_1^{2d}/(\log n)^2 \to \infty$ from Condition \ref{con:nkerbw}, the lemma is proven.
\qed

In Lemma \ref{lem:l3}, we establish the uniform convergence of $\wh E_1(\cdot\mid y,c,\z)$ to $E_{10} (\cdot\mid y,c,\z;\bb)$ and $\wh E_2(\cdot\mid y,x,\z)$ to $E_{20}(\cdot\mid y,x,\z)$.
\begin{Lem} \label{lem:l3}
Suppose that Conditions \ref{con:nker}--\ref{con:nderivbddaway}
hold. If a real function $g(y,x,c,\z;
\bb)$ is continuous with respect to $\bb$
and satisfies $\|g\|_\infty 
\equiv \sup_{\bb}\sup_{y,x,c,\z}|g(y,x,c,\z; \bb)| <\infty$, then
\bse
\sup_{\bb\in \Omega}\sup_{y,c,\z} |(\wh E_1 - E_{10}) \{g(y,X,c,\z ;\bb)| y,c,\z;\bb\}| &=& o_p(n^{-1/4}),\\
\sup_{\bb\in \Omega}\sup_{x,\z} |(\wh E_2 - E_{20}) \{g(Y,x,C,\z ; \bb)| x,\z;\bb\}| &=& o_p(n^{-1/4}).
\ese
\end{Lem}
Proof.
For a general function $f(\bt)$, let $\| f \|_\infty \equiv \sup_\bt |f(\bt)|$. For the first part, note that
\bse
&&\sup_{\bb,y,c,\z} \left|(\wh E_1- E_{10} )\{g(y,X,c,\z;\bb)|y, c,\z;\bb\}\right|\\
&=&\sup_{\bb,y,c,\z}  \left|\sumi\frac{ \delta_i g(y, w_i, c, \z_i;\bb) f_{Y|X,\Z}(y,w_i,\z_i;\bb) K_{h_2}^{(m_2)}(\z-\z_i)}{\wh S_{C|Y,\Z}(w_i,
  y_i, \z_i)f_{Y|\Z}(y,\z;\bb)\sum_{k=1}^n K_{h_2}^{(m_2)}(\z-\z_k)}\right.\\
  &&\left.-  E_{10}\left\{\frac{ \Delta g(y, W, c, \z; \bb) f_{Y|X,\Z}(y,W,\z;\bb) }{S_{C|Y,\Z}(W, Y, \z)f_{Y|\Z}(y,\z;\bb)}|\z\right\}\right|\\
  &\le&\sup_{\bb,y,c,\z}  \sumi\left|\frac{ \delta_i g(y, w_i, c, \z_i;\bb) f_{Y|X,\Z}(y,w_i,\z_i;\bb)}{\wh S_{C|Y,\Z}(w_i,
  y_i, \z_i)f_{Y|\Z}(y,\z;\bb)} - \frac{ \delta_i g(y, w_i, c, \z_i;\bb) f_{Y|X,\Z}(y,w_i,\z_i;\bb)}{S_{C|Y,\Z}(w_i,
  y_i, \z_i)f_{Y|\Z}(y,\z;\bb)}\right| \\
  &&\times\frac{|K_{h_2}^{(m_2)}(\z-\z_i)|}{|\sum_{k=1}^n K_{h_2}^{(m_2)}(\z-\z_k)|}\\
  &&+\sup_{\bb,y,c,\z} \left|\sumi\frac{ \delta_i g(y, w_i, c, \z_i;\bb) f_{Y|X,\Z}(y,w_i,\z_i;\bb)K_{h_2}^{(m_2)}(\z-\z_i)}{S_{C|Y,\Z}(w_i,
  y_i, \z_i)f_{Y|\Z}(y,\z;\bb)\sum_{k=1}^n K_{h_2}^{(m_2)}(\z-\z_k)}\right.\\
  &&\left.-  E_{10}\left\{\frac{ \Delta g(y, W, c, \z; \bb) f_{Y|X,\Z}(y,W,\z;\bb) }{S_{C|Y,\Z}(W, Y, \z)f_{Y|\Z}(y,\z;\bb)}|\z\right\}\right|\\
 &=&\sup_{\bb,y,c,\z} \sumi\left|\frac{ \delta_i g(y, w_i, c, \z_i;\bb) f_{Y|X,\Z}(y,w_i,\z_i;\bb)}{f_{Y|\Z}(y,\z;\bb)}\right| \left|\frac{S_{C|Y,\Z}(w_i,
  y_i, \z_i) - \wh S_{C|Y,\Z}(w_i,
  y_i, \z_i)}{\wh S_{C|Y,\Z}(w_i,
  y_i, \z_i)S_{C|Y,\Z}(w_i,
  y_i, \z_i)}\right|\\
  &&\times\frac{|K_{h_2}^{(m_2)}(\z-\z_i)|}{|\sum_{k=1}^n K_{h_2}^{(m_2)}(\z-\z_k)|} + o_p(n^{-1/4})\\
  &\le& O_p(1) \|g\|_\infty  \left\|S_{C|Y,\Z}-\wh
    S_{C|Y,\Z}\right\|_\infty   {\frac{\sup_{\z} n^{-1}\sumi
      |K_{h_2}^{(m_2)}(\z-\z_i)|}{ \inf_{\bb,y,\z}| n^{-1}\sum_{k=1}^n K_{h_2}^{(m_2)}(\z-\z_k)|f_{Y|\Z}(y,\z;\bb)}} +  o_p(n^{-1/4})\\
   &\le&   o_p(n^{-1/4})  {\frac{\sup_{\z}f_{\Z}(\z) \int |K^{(m_2)} (\bv)|d\bv + o_p(1)}{ \inf_{\bb,y,\z} |f_{Y,\Z}(y,\z;\bb)| + o_p(1)}} +o_p(n^{-1/4})\\
  &=&  o_p(n^{-1/4}).
\ese 
In the above derivation, the second equality follows from the uniform convergence rate $O_p\{h_2^{m_2} +(\log n)^{1/2}n^{-1/2}h_2^{-d'/2} \}$ of the kernel estimator for $$E_{10} \{\Delta g(y, W, c, \z; \bb)f_{Y|X,\Z}(y,W,\z;\bb)/{S_{C|Y,\Z}(W, Y, \z)}|\z;\bb\}$$ and Condition \ref{con:nkerbw}. The second inequality is established by Condition \ref{con:nbddaway} and Lemma \ref{lem:l2}. The third inequality follows because the denominator satisfies
\be \label{eq:n5}
&&\inf_{\bb,y,\z} n^{-1}\sum_{k=1}^n K_{h_2}^{(m_2)}(\z-\z_k)f_{Y|\Z}(y,\z;\bb)\\
&=& \inf_{\bb,y,\z} [f_{\bZ}(\z) + O_p\{(\log n )^{1/2}n^{-1/2}h_2^{d'/2}+h_2^{m_2}\}]f_{Y|\Z}(y,\z;\bb)\n\\
&=& \inf_{\bb,y,\z} f_{\bZ}(\z; \bb)f_{Y|\Z}(y,\z;\bb) + o_p(1)\n
\ee
under Conditions \ref{con:nker},
\ref{con:nkerbw}, and \ref{con:nbddaway}, and the numerator satisfies
    \be \label{eq:n6}
    \sup_{\z} n^{-1} \sum_{k=1}^n |K_{h_2}^{(m_2)}(\z-\z_k)|
&=& \sup_{\z}\int |K^{(m_2)} (\bv)| d\bv f_{\bZ}(\z) + O_p\{(\log n )^{1/2}n^{-1/2}h_2^{d'/2}+h_2^2\}\n\\
&=& \sup_{\z}f_{\Z}(\z) \int |K^{(m_2)} (\bv)|d\bv +o_p(1).
\ee
In \eqref{eq:n6}, the first equality follows from the uniform convergence rate of density estimation with kernel function $|K(\cdot)|/\int |K (u, \bv)| du d\bv$, and the second equality follows from Conditions \ref{con:nker} and \ref{con:nkerbw}.

The proof for the convergence of $\wh E_2$ uses the same properties of $S_{X|Y,\Z}$ as those established for $S_{C|Y,\Z}$. First, the uniform convergence rate of kernel estimation shows that
\bse
&&\sup_{\bb,y,x,\z} \left|n^{-1}\sumi \frac{ (1-\delta_i) g(y_i, x, w_i, \z_i; \bb) K_{h_3}^{(m_3)}(y-y_i,\z-\z_i)}{ S_{X|Y,\Z}(w_i,
  y_i, \z_i)f_{Y,\Z}(y,\z)}  - E_{20}\left\{\frac{ (1-\Delta) g(y, x, W, \z; \bb)}{S_{X|Y,\Z}(W, y, \z)}|y,\z\right\}\right|\\
  &=& O_p\{h_3^{m_3} +(\log n)^{1/2}n^{-1/2}h_3^{-d/2}\},
\ese
which is $o_p(n^{-1/4})$ under Condition \ref{con:nkerbw}.
Moreover, we have 
\bse
&&\sup_{\bb,y,x,\z} n^{-1} \sumi \left| \frac{ (1-\delta_i) g(y_i, x, w_i, \z_i; \bb) }{\wh S_{X|Y,\Z}(w_i,
  y_i, \z_i)} - \frac{ (1-\delta_i) g(y_i, x, w_i, \z_i; \bb) }{S_{X|Y,\Z}(w_i,
  y_i, \z_i)}\right| \frac{|K_{h_3}^{(m_3)}(y-y_i,\z-\z_i)|}{ f_{Y,\Z}(y,\z)}\\
  &\le& O_p(1) \|g\|_\infty  \left\|S_{X|Y,\Z}-\wh
    S_{X|Y,\Z}\right\|_\infty   {\frac{\sup_{y, \z} n^{-1}\sumi
      |K_{h_3}^{(m_3)}(y-y_i,\z-\z_i)|}{ \inf_{y, \z} f_{Y,\Z}(y,\z)}}\\
   &\le&   o_p(n^{-1/4})  {\frac{\sup_{y, \z}f_{Y,\Z}(y,\z) \int |K^{(m_3)} (u, \bv)|du d\bv}{ \inf_{y, \z} f_{Y,\Z}(y,\z) + o_p(1)}}\\
  &=&  o_p(n^{-1/4}),
\ese
where the second inequality follows for similar reasons as in \eqref{eq:n5} and \eqref{eq:n6} under Condition \ref{con:nkerbw}. These results give us
\bse
&&\sup_{\bb,y,x,\z} \left|n^{-1}\sumi \frac{ (1-\delta_i) g(y_i, x, w_i, \z_i; \bb) K_{h_3}^{(m_3)}(y-y_i,\z-\z_i)}{\wh S_{X|Y,\Z}(w_i,
  y_i, \z_i)f_{Y,\Z}(y,\z)}  - E_{20}\left\{\frac{ (1-\Delta) g(y, x, W, \z; \bb)}{S_{X|Y,\Z}(W, y, \z)}|y,\z\right\}\right|\\
  &=& o_p(n^{-1/4}).
\ese
Also, letting $g(y_i, x, w_i, \z_i; \bb) =1$, we have that 
\bse
&&\sup_{\bb,y,x,\z} \left|n^{-1}\sumi \frac{ (1-\delta_i) K_{h_3}^{(m_3)}(y-y_i,\z-\z_i)}{\wh S_{X|Y,\Z}(w_i,
  y_i, \z_i)f_{Y,\Z}(y,\z)}  - E_{20}\left\{\frac{ (1-\Delta)}{S_{X|Y,\Z}(W, y, \z)}|y,\z\right\}\right|\\
  &=& o_p(n^{-1/4}).
\ese

Then for any $g(y,x,c,\z;\bb)$,
  \bse
&&\sup_{\bb,x,\z} \left|(\wh E_2-E_{20} )\{g(Y,x,C,\z;\bb)|x, \z;\bb\}\right|\\
&=&\sup_{\bb,x,\z} \left|  \int \left[\frac{ \sumi (1-\delta_i) g(y_i, x, w_i, \z_i; \bb)K_{h_3}^{(m_3)}(y-y_i,\z-\z_i) /\wh S_{X|Y,\Z}(w_i,
  y_i, \z_i) }{\sumi (1-\delta_i) K_{h_3}^{(m_3)}(y-y_i,\z-\z_i) /\wh S_{X|Y,\Z}(w_i,
  y_i, \z_i)}\right.\right.\\
  &&\left.\left.- \frac{ E_{20}\{(1-\Delta) g(y, x, W, \z;
          \bb)/S_{X|Y,\Z}(W, y, \z)\mid y,\z\}}{ E_{20}\{(1-\Delta)/S_{X|Y,\Z}(W, y, \z)\mid y,\z\}}\right]f_{Y|X,\Z}(y,x,\z;\bb) dy\right|\\
  &\le&    \int\sup_{\bb,y,x,\z}  \left|\frac{ E_{20}\{(1-\Delta) g(y, x, W, \z;
          \bb)/S_{X|Y,\Z}(W, y, \z)\mid y,\z\}+o_p(n^{-1/4})}{ E_{20}\{(1-\Delta)/S_{X|Y,\Z}(W, y, \z)\mid y,\z\}+o_p(n^{-1/4})}\right.\\
  &&\left.- \frac{ E_{20}\{(1-\Delta) g(y, x, W, \z;
          \bb)/S_{X|Y,\Z}(W, y, \z)\mid y,\z\}}{ E_{20}\{(1-\Delta)/S_{X|Y,\Z}(W, y, \z)\mid y,\z\}}\right|f_{Y|X,\Z}(y,x,\z;\bb) dy\\
  &=& o_p(n^{-1/4}).
\ese
 This concludes the proof of the second part of Lemma \ref{lem:l3}.
\qed

Let $\|\bu\|_\infty
\equiv\sup_{j=1,...,p} \sup_{\bb,x,\z} |u_j(x,\z; \bb)|$, where
 $\bu(x,\z; \bb) = \{u_1(x,\z;
\bb),...,u_p(x,\z; \bb)\}\trans:\mathbb{R}^{\dim(\z)+1} \times \mathbb{R}^p \to
\mathbb{R}^p$.
Further, let $\G \equiv
\{\bu(x,\z; \bb): \|\bu\|_\infty < \infty\}$.
Define a linear operator $\L: \G \to \G$ such that
\be 
&&\L(\bu; E_1, E_2)(x,\z; \bb) \label{eq:n7}\\
&\equiv& E_2\{I(x\le C)|x,\z;\bb\} \bu (x,\bz;\bb) \n\\
&&+
E_2\left[\left.I(x>C)\frac{E_1\{I(X>C)\bu(X,\bz;\bb)
      |C,Y,\z;\bb\}}{E_1\{I(X>C)|C,Y,\z;\bb\}}\right|x,\bz;\bb\right].\n
\ee
Also, define $\bc(x,\bz;\bb, E_1, E_2)$ as
\bse
&&\bc(x,\bz;\bb, E_1, E_2)\\
&\equiv& E_2\{I(x\le C)\S_\bb^F(Y,x,\bz;\bb)|x,\z;\bb\}\n\\
&&+ E_2\left[\left.I(x>C)\frac{E_1\{I(X>C)\S_\bb^F(Y,X,\bz;\bb)
|C,Y,\z;\bb\}}{E_1\{I(X>C)|C,Y,\z;\bb\}}\right|x,\bz;\bb\right].\n
\ese
Then $\bc = \L(\ba;E_1,E_2)$, or equivalently, $\ba = \L^{-1}(\bc;E_1,E_2)$. 

\begin{Lem} \label{lem:l4}
Consider any function $g(y,x,c,\z ;\bb)$ that is continuous with respect to $\bb$ and satisfies $\|g\|_\infty < \infty$. Then
\bse
\sup_{\bb \in \Omega}\sup_{y,c,\z} |(\wh E_1 - E_{10}) \{g(y,X,c,\z ;\bb)| y,c,\z;\bb\}| &=&  o_p(n^{-1/4}),\\
\sup_{\bb \in \Omega}\sup_{x,\z} |(\wh E_2 - E_{20}) \{g(Y,x,C,\z ; \bb)| x,\z;\bb\}| &=& o_p(n^{-1/4}).
\ese
Then,  under Conditions \ref{con:nker}--\ref{con:nderivbddaway},
\begin{enumerate}[label=(\roman*),ref=(\roman*)]
\item \label{ncase1} for $\wh \ba^\star (x,\bz; \bb) \equiv \ba(x,\z;\bb, \wh E_1, E_2^\star)$ and $\ba_0^\star (x,\bz; \bb) \equiv \ba(x,\z;\bb, E_{10}, E_2^\star)$,
    \bse
\sup_{\bb \in \Omega}\sup_{x,\z}  \|(\wh \ba^\star - \ba_0^\star) (x,\bz; \bb)\|_\infty &=& o_p(n^{-1/4}).
\ese
\item \label{ncase2} for $\wh \ba^* (x,\bz; \bb) \equiv \ba(x,\z;\bb, E_1^*, \wh E_2)$ and $\ba_0^* (x,\bz; \bb) \equiv \ba(x,\z;\bb, E_1^*, E_{20})$,
    \bse
\sup_{\bb \in \Omega}\sup_{x,\z}  \|(\wh \ba^* - \ba_0^*) (x,\bz; \bb)\|_\infty &=& o_p(n^{-1/4}).
\ese
    
    \item \label{ncase3} for $\wh \ba (x,\bz; \bb) \equiv \ba(x,\z;\bb, \wh E_1, \wh E_2)$ and $\ba_0 (x,\bz; \bb) \equiv \ba(x,\z;\bb, E_{10}, E_{20}),$
\bse
\sup_{\bb \in \Omega}\sup_{x,\z} \|(\wh \ba - \ba_0) (x,\bz; \bb)\|_\infty &=& o_p(n^{-1/4}).
\ese
\end{enumerate}
\end{Lem}

Proof.
\textbf{Claim 1: For any $E_1$ and $E_2$, $\L^{-1}$, the inverse of
  $\L$, exists and is a bounded linear operator.} To establish invertibility, suppose that $\L(\bu; E_1, E_2) = 0$ for some $\bu \in \G$. Then both $\ba = \ba(x,\z; \bb, E_1, E_2)$ and $\ba' = \ba + \bu$ satisfy \eqref{eq:n1}. However, the efficient score function $\S\eff$ is uniquely determined as a function of $(y,w,\delta,\z;\bb)$ for each fixed $(E_1,E_2)$. Substituting $\delta = 1$ in \eqref{eq:n0} yields
\bse
\S_\bb(y,w,1,\z;\bb, E_1) - \ba(w, \bz; \bb)
&=& \S\eff (y,w,1, \z ; \bb, E_1, \ba)\\
&=& \S\eff (y,w,1, \z ; \bb, E_1, \ba')\\
&=& \S_\bb(y,w,1,\z;\bb, E_1) - \ba'(w, \bz; \bb).
\ese
Thus $\ba = \ba'$, which implies $\bu = \0$. Since $\L$ is a linear operator, the fact that $\L(\bu) = 0$ implies $\bu = \0$ proves that $\L$ is invertible for any $E_1$ and $E_2$.

For boundedness, we have $\|\L(\bu; E_1, E_2)\|_\infty \leq 2\|\bu\|_\infty$ for any $\bu \in \G$. By the bounded inverse theorem, $\L^{-1}: \L(\G) \to \G$ is a bounded linear operator.

\textbf{ Claim 2: For invertible linear operators $\wh \M$ and
  $\M$, if $\|\wh \M - \M\| = o_p(n^{-1/4})$ and $\|\wh \bc -
  \bc\|_\infty = o_p(n^{-1/4})$, then $\|\wh \M^{-1}(\wh \bc) -
  \M^{-1}(\bc)\|_\infty = o_p(n^{-1/4})$.}  This result directly
  follows from
\bse
\|\wh \M^{-1}(\wh \bc) - \M^{-1}(\bc)\|_\infty
&\le& \|\wh \M^{-1}(\wh \bc) - \M^{-1}(\wh \bc)\|_\infty + \|\M^{-1}(\wh \bc) - \M^{-1}(\bc)\|_\infty\\
&\le& \|\wh \M^{-1}(\wh \bc)\|_\infty \|\wh \M - \M\|\|\M^{-1}\| + \|\M^{-1}\|\|\wh \bc - \bc\|_\infty  \\
&=&o_p(n^{-1/4}).
\ese

We now prove the results for cases \ref{ncase1}--\ref{ncase3}.

First, consider case \ref{ncase1}. We only need to prove
$\|\L(\bu; \wh E_1, E_2^\star) - \L(\bu; E_{10}, E_2^\star)\|_\infty =
o_p(n^{-1/4})$  for $\bu \in \G$, and $\|\bc(\cdot, \wh E_1,
E_2^\star) - \bc(\cdot, E_{10}, E_2^\star)\| = o_p(n^{-1/4})$. 
For any $\bu \in \G$, 
\bse
&&\|\L(\bu; \wh E_1, E_2^\star) - \L(\bu; E_{10}, E_2^\star)\|_\infty\\
&=& 
\sup_{\bb, x,\bz}\left\|E_2^\star\left(\left.I(x>C)\left[\frac{\wh
          E_1\{I(X>C)\bu(X,\bz;\bb) 
      |C,Y,\z;\bb\}}{\wh E_1\{I(X>C)|C,Y,\z;\bb\}}\right.\right.\right.\right.\\
      &&\left.\left.\left.\left. - \frac{ E_{10}\{I(X>C)\bu(X,\bz;\bb)
      |C,Y,\z;\bb\}}{ E_{10}\{I(X>C)|C,Y,\z;\bb\}}\right]
\right|x,\bz;\bb\right)  \right\|_\infty\\ 
&\le& 
\sup_{\bb, c, y,\bz} \|(\wh E_1 - E_{10}) \{I(X>c)\bu(X,\bz;\bb)
      |y,\z;\bb\} \|_\infty O_p(1)\\
  &&+ \|\bu\|_\infty  \sup_{\bb, c, y,\bz} \|(\wh E_1 - E_{10}) \{I(X>c)
      |y,\z;\bb\} \|_\infty O_p(1)\\
      &=&o_p(n^{-1/4}).
\ese 
Similarly,
\bse
&&\|\bc(\cdot; \wh E_1, E_2^\star) - \bc(\cdot; E_{10}, E_2^\star)\|_\infty\\
&=& \sup_{\bb, x,\bz} \left\|E_2^\star \left(\left.I(x>C)\left[\frac{\wh E_1\{I(X>C)\S_\bb^F(Y,X,\bz;\bb)
|C,Y,\z;\bb\}}{\wh E_1\{I(X>C)|C,Y,\z;\bb\}} \right.\right.\right.\right.\\
&&\left.\left.\left.\left.- \frac{E_{10}\{I(X>C)\S_\bb^F(Y,X,\bz;\bb)
|C,Y,\z;\bb\}}{E_{10}\{I(X>C)|C,Y,\z;\bb\}}\right]\right|x,\bz;\bb\right)\right\|_\infty\\
&\le& \sup_{\bb, c,y,\bz} \left\|\frac{\wh E_1\{I(X>c)\S_\bb^F(y,X,\bz;\bb)
|y,\z;\bb\}}{\wh E_1\{I(X>c)|y,\z;\bb\}} - \frac{E_{10}\{I(X>c)\S_\bb^F(y,X,\bz;\bb)
|y,\z;\bb\}}{E_{10}\{I(X>c)|y,\z;\bb\}}\right\|_\infty\\
&\le& \sup_{\bb, c,y,\bz} \left\|\frac{\wh E_1\{I(X>c)\S_\bb^F(y,X,\bz;\bb)
|y,\z;\bb\}}{\wh E_1\{I(X>c)|y,\z;\bb\}} - \frac{E_{10}\{I(X>c)\S_\bb^F(y,X,\bz;\bb)
|y,\z;\bb\}}{\wh E_1\{I(X>c)|y,\z;\bb\}}\right\|_\infty\\
&&+\sup_{\bb, c,y,\bz} \left\|\frac{ E_{10}\{I(X>c)\S_\bb^F(y,X,\bz;\bb)
|y,\z;\bb\}}{\wh E_1\{I(X>c)|y,\z;\bb\}} - \frac{E_{10}\{I(X>c)\S_\bb^F(y,X,\bz;\bb)
|y,\z;\bb\}}{E_{10}\{I(X>c)|y,\z;\bb\}}\right\|_\infty\\
&\le& \sup_{\bb, c, y,\bz} \|(\wh E_1 - E_{10}) \{I(X>c)\S_\bb^F(y,X,\bz;\bb)
      |y,\z;\bb\} \|_\infty O_p(1)\\
  &&+ \|\S_\bb^F\|_\infty  \sup_{\bb, c, y,\bz} \|(\wh E_1 - E_{10}) \{I(X>c)
      |y,\z;\bb\} \|_\infty O_p(1)\\
      &=&o_p(n^{-1/4}).
\ese

For case \ref{ncase2}, we only need to prove $\|\L(\bu; E_1^*,
\wh E_2) - \L(\bu; E_1^*, E_{20})\|_\infty = o_p(n^{-1/4})$ for $\bu
\in \G$, and $\|\bc(\cdot; E_1^*, \wh E_2) - \bc(\cdot; E_1^*,
E_{20})\|_\infty = o_p(n^{-1/4})$. Indeed, for $\bu \in \G$,
\bse
&&\|\L(\bu; E_1^*, \wh E_2) - \L(\bu; E_1^*, E_{20})\|_\infty\\
&=& \sup_{\bb,x,\z}\left\|(\wh E_2 - E_{20})\{I(x\le C)|x,\z;\bb\} \bu (x,\bz;\bb) \right.\\
&&\left.+ (\wh E_2 - E_{20})\left[\left.I(x>C)\frac{E_1^*\{I(X>C)\bu(X,\bz;\bb)
      |C,Y,\z;\bb\}}{E_1^*\{I(X>C)|C,Y,\z;\bb\}}\right|x,\bz;\bb\right]\right\|_\infty\\
&\le& \sup_{\bb,x,\z}\left|(\wh E_2 - E_{20})\{I(x\le C)|x,\z;\bb\}\right| \|\bu \|_\infty \\
&&+ \sup_{\bb,x,\z}\left\|(\wh E_2 - E_{20})\left[\left.I(x>C)\frac{E_1^*\{I(X>C)\bu(X,\bz;\bb)
      |C,Y,\z;\bb\}}{E_1^*\{I(X>C)|C,Y,\z;\bb\}}\right|x,\bz;\bb\right]\right\|_\infty\\
&=&o_p(n^{-1/4}),
\ese
and
\bse
&&\|\bc(\cdot; E_1^*, \wh E_2) - \bc(\cdot; E_1^*, E_{20})\|_\infty\\
&=& \sup_{\bb,x,\z}\left\|(\wh E_2 - E_{20})\{I(x\le C)\S_\bb^F (Y,x,\bz;\bb)|x,\z;\bb\}  \right.\\
&&\left.+ (\wh E_2 - E_{20})\left[\left.I(x>C)\frac{E_1^*\{I(X>C)\S_\bb^F(Y, X,\bz;\bb)
      |C,Y,\z;\bb\}}{E_1^*\{I(X>C)|C,Y,\z;\bb\}}\right|x,\bz;\bb\right]\right\|_\infty\\
&=& o_p(n^{-1/4}).
\ese

Lastly, for case \ref{ncase3}, combining the results for cases \ref{ncase1} and \ref{ncase2}, we have that
\bse
&&\|\L(\bu; \wh E_1, \wh E_2) - \L(\bu; E_{10}, E_{20})\|_\infty\\
&\le& \|\L(\bu; \wh E_1, \wh E_2) - \L(\bu; \wh E_1, E_{20})\|_\infty + \|\L(\bu; \wh E_1, E_{20}) - \L(\bu; E_{10}, E_{20})\|_\infty\\
&=& o_p(n^{-1/4})
\ese
for any $\bu \in \G$, and
\bse
&&\|\bc(\cdot; \wh E_1, \wh E_2) - \bc(\cdot; E_{10}, E_{20})\|_\infty\\
&\le& \sup_{\bb, x,\z}\|\bc(x,\z; \bb, \wh E_1, \wh E_2) - \bc(x,\z; \bb, ; \wh E_1, E_{20})\|_\infty + \sup_{\bb, x,\z}\|\bc(x,\z; \bb, \wh E_1, E_{20}) - \bc(x,\z; \bb, ; E_{10}, E_{20})\|_\infty\\
&=& o_p(n^{-1/4}).
\ese
Together with \textbf{Claim 2}, we obtain the desired result for cases \ref{ncase1}--\ref{ncase3}.
\qed

\subsubsection{Variance Inflation under Model Misspecification}\label{sec:l5-7pf}
The expressions in \eqref{eq:e1} and \eqref{eq:e2} allow us to compute the Gateaux derivatives of $E_1\{g(y,X,c,\z)|y, c, \z; \bb, S_{C|Y,\Z}, f_{\Delta, W,Y|\Z}\}$ with respect to the functional parameters $S_{C|Y,\Z}$ and $f_{\Delta, W,Y|\Z}$:
\be\label{eq:d3}
&& \frac{\partial E_1\{g(y,X,c,\z)|y, c, \z; \bb, S_{C|Y,\Z}, f_{\Delta, W,Y|\Z}\}}{\partial S_{C|Y,\Z}}(u)\\
&=& -\frac{1}{f_{Y|\Z}(y,\z;\bb)}\iiint \frac{\delta g(y,w,c,\z)u(w,y',\z)}{\{S_{C|Y,\Z} (w,y',\z)\}^2}f_{\Delta,W,Y|\Z}(\delta,w,y',\z) \n\\
&&\times f_{Y|X,\Z}(y,w,\z;\bb)d\delta dwdy'\n\\
&=& - E_1\left\{g(y,X,c,\z)\frac{u(X,Y',\z)}{S_{C|Y,\Z} (X,Y',\z)}|y, c, \z;\bb, S_{C|Y,\Z}, f_{\Delta, W,Y|\Z}\right\}\n
\ee
for a bounded function $u (t,y,\z): \mathbb{R}^{\dim(\z)+2} \to
 \mathbb{R}$, and
\be \label{eq:d4}
&& \frac{\partial E_1\{g(y,X,c,\z)|y, c, \z; \bb,S_{C|Y,\Z}, f_{\Delta, W,Y|\Z}\}}{\partial f_{\Delta, W,Y|\Z}}(u)\\
&=& \frac{1}{f_{Y|\Z}(y,\z;\bb)}\iiint \frac{\delta g(y,w,c,\z)}{S_{C|Y,\Z} (w,y',\z)}u(\delta,w,y',\z) f_{Y|X,\Z}(y,w,\z;\bb)d\delta dwdy'\n\\
&=& E_1\{g(y,X,c,\z)|y, c, \z;\bb, S_{C|Y,\Z}, u\}.\n
\ee
Above, we treat $u$ as a probability density function conditional on $\z$ and incorporate the computation in \eqref{eq:e1}. Similarly, the Gateaux derivatives of $E_2\{g(Y,x,C,\z)|x, \z; \bb, S_{X|Y,\Z}, f_{\Delta, W|Y,\Z}\}$ with respect to $S_{X|Y,\Z}$ and $f_{\Delta, W|Y,\Z}$ are computed as
\be
 &&\frac{\partial E_2\{g(Y,x,C,\z)|x, \z; \bb, S_{X|Y,\Z}, f_{\Delta, W|Y,\Z}\}}{\partial S_{X|Y,\Z}}(u)\label{eq:d5}\\
 &=& - \int \frac{\iint (1-\delta) g(y,x,w,\z)u(w,y,\z)f_{\Delta, W|Y,\Z}(\delta,w,y,\z) /S_{X|Y,\Z} (w,y,\z)^2 d\delta dw}{\iint (1-\delta) f_{\Delta, W|Y,\Z}(\delta,w,y,\z) /S_{X|Y,\Z} (w,y,\z)d\delta dw}\n\\
 &&\times f_{Y|X,\Z}(y,x,\z;\bb) dy\n\\
 && + \int \left[\frac{\iint (1-\delta) g(y,x,w,\z) f_{\Delta, W|Y,\Z}(\delta,w,y,\z) /S_{X|Y,\Z} (w,y,\z) d\delta dw}{\{\iint (1-\delta) f_{\Delta, W|Y,\Z}(\delta,w,y,\z) /S_{X|Y,\Z} (w,y,\z)d\delta dw\}^2}\right.\n\\
 &&\left.\times \iint (1-\delta) u(w,y,\z)f_{\Delta, W|Y,\Z}(\delta,w,y,\z) /S_{X|Y,\Z} (w,y,\z)^2 d\delta dw\right]\n\\
 &&\times f_{Y|X,\Z}(y,x,\z;\bb) dy\n\\
 &=& -E_2\left\{g(Y,x,C,\z)\frac{u(C,Y,\z)}{S_{X|Y,\Z} (C,Y,\z)}|x, \z; \bb, S_{X|Y,\Z}, f_{\Delta, W|Y,\Z}\right\}\n\\
 &&+ E\left[E_2 \{g(Y,x,C,\z)\mid Y,\z; S_{X|Y,\Z}, f_{\Delta, W|Y,\Z}\} \right.\n\\
 &&\left. E_2 \left\{\frac{u(C,Y,\z)}{S_{X|Y,\Z} (C,Y,\z)}\mid Y,\z;S_{X|Y,\Z}, f_{\Delta, W|Y,\Z}\right\} \mid x, \z; \bb\right] \n\\
 &=&-E\left([g(Y,x,C,\z) - E_2 \{g(Y,x,C,\z)\mid Y,\z; S_{X|Y,\Z}, f_{\Delta, W|Y,\Z}\}]\right.\n\\
 &&\left.\times\frac{u(C,Y,\z)}{S_{X|Y,\Z} (C,Y,\z)}\mid x, \z; \bb\right)  \n
\ee
for a bounded function $u (t,y,\z): \mathbb{R}^{\dim(\z)+2} \to
 \mathbb{R}$, and 
\be\label{eq:d6}
&&\frac{\partial E_2\{g(Y,x,C,\z)|x, \z; \bb, S_{X|Y,\Z}, f_{\Delta, W|Y,\Z}\}}{\partial f_{\Delta, W|Y,\Z}}(u)\\
&=& \int \frac{\iint (1-\delta) g(y,x,w,\z)u(\delta,w,y,\z) /S_{X|Y,\Z} (w,y,\z) d\delta dw}{\iint (1-\delta) f_{\Delta, W|Y,\Z}(\delta,w,y,\z) /S_{X|Y,\Z} (w,y,\z)d\delta dw}f_{Y|X,\Z}(y,x,\z;\bb) dy\n\\
 && - \int \left[\frac{\iint (1-\delta) g(y,x,w,\z) f_{\Delta, W|Y,\Z}(\delta,w,y,\z) /S_{X|Y,\Z} (w,y,\z) d\delta dw}{\{\iint (1-\delta) f_{\Delta, W|Y,\Z}(\delta,w,y,\z) /S_{X|Y,\Z} (w,y,\z)d\delta dw\}^2}\right.\n\\
 &&\left.\times \iint (1-\delta) u(\delta,w,y,\z) /S_{X|Y,\Z} (w,y,\z) d\delta dw\right]f_{Y|X,\Z}(y,x,\z;\bb) dy\n\\
 &=&E[E_2 \{g(Y,x,C,\z)\mid Y,\z; S_{X|Y,\Z}, u\} \n\\
 &&- E_2 \{g(Y,x,C,\z)\mid Y,\z;S_{X|Y,\Z}, f_{\Delta, W|Y,\Z} \}  E_2 \{1 \mid Y,\z; S_{X|Y,\Z}, u\}\mid x, \z; \bb], \n
 \ee
 where 
 \bse
 E_2 \{h(y,C,\z)\mid y,\z; S_{X|Y,\Z}, u\} \equiv \iint \frac{(1-\delta) h(y,w,\z)}{S_{X|Y,\Z} (w,y,\z)}u(\delta,w,y,\z)  d\delta dw
 \ese
in the last expression of \eqref{eq:d6}. Here, we again treat $u$ as a probability density function conditional on $y,\z$ and incorporate the computation in \eqref{eq:e2}. These relationships will be useful in proving the following lemmas.

Finally, recall that from $\S_\bb(y, w, \delta, \z; \bb)$ and \eqref{eq:n0}, the efficient score function $\S\eff$ can be written as
\bse
\S\eff (y,w,\delta, \z ; \bb, E_1, \ba) &=& \delta\{\S_\bb^F(y,w,\z;\bb, E_1) - \ba(w, \bz; \bb)\}\\
&&+ (1-\delta) \frac{ E_1[I(X>
        w)\{\S_\bb^F(y,X,\z;\bb, E_1) - \ba(X,\z;\bb)\} |y,\z;\bb]}{ E_1\{I(X> w)|y,\z;\bb\}}.
\ese 
Then $\S\eff$ can be formulated into 
\be
&&\S\eff\{y, w, \delta,
  \z; \bb, E_1(\cdot; S_{C|Y,\Z}, f_{\Delta, W,Y|\Z}), \ba\}\label{eq:t2}\\
&=& \delta\{\S_\bb^F(y,w,\z;\bb, E_1) - \ba(w, \bz; \bb)\}+ (1-\delta)\n\\
&&\times \frac{ E_1[I(X>
w)\{\S_\bb^F(y,X,\z;\bb, E_1) - \ba(X,\z;\bb)\} |y,\z;\bb, S_{C|Y,\Z}, f_{\Delta, W,Y|\Z}]}{ E_1\{I(X> w)|y,\z;\bb, S_{C|Y,\Z}, f_{\Delta, W,Y|\Z}\}}\n\\
&=& \delta\{\S_\bb^F(y,w,\z;\bb, E_1) - \ba(w, \bz; \bb)\} \n\\
&&+ (1-\delta) \iiint \delta' I(x>w)\{\S_\bb^F(y,x,\z;\bb) - \ba(x,\z;\bb)\} \n\\
&&\times\frac{f_{\Delta, W,Y|\Z}(\delta',x,y',\z)}{ S_{C|Y,\Z} (x,y',\z)} f_{Y|X,\Z}(y,x,\z;\bb)d\delta' dxdy'\n\\
&&\times \left\{\iiint \delta' I(x>w)\frac{f_{\Delta, W,Y|\Z}(\delta',x,y',\z)}{S_{C|Y,\Z} (x,y',\z) }f_{Y|X,\Z}(y,x,\z;\bb)d\delta' dxdy'\right\}^{-1}. \n
\ee 
This equivalent expression of
  $\S\eff\{y, w, \delta,
  \z; \bb_0, E_1(\cdot; S_{C|Y,\Z}, f_{\Delta, W,Y|\Z}), \ba\}$ allows
  us to treat $S_{C|Y,\Z}$ and  $f_{\Delta, W,Y|\Z}$ as two free
  functional parameters.

\begin{Lem} \label{lem:l8} 
Under Conditions \ref{con:nker}, \ref{con:nkerbw'}, \ref{con:nbddaway}, and 
\ref{con:nderivbddaway}, 
    \bse
&&n^{-1/2} \sumi \frac{\partial \S\eff^\star\{y_i, w_i, \delta_i,
  \z_i; \bb_0, E_1(\cdot; S_{C|Y,\Z}, f_{\Delta, W, Y|\Z}), \ba_0^\star\}}{\partial S_{C|Y,\Z}}
(\wh S_{C|Y,\Z} - S_{C|Y,\Z})\\
&&= n^{-1/2}\sum_{j=1}^n \bh_{1{\rm s}1}^\star(y_j,w_j, \delta_j, \z_j) +o_p(1),\\
&&n^{-1/2} \sumi \frac{\partial \S\eff^\star\{y_i, w_i, \delta_i,
  \z_i; \bb_0, E_1(\cdot; S_{C|Y,\Z}, f_{\Delta, W,Y|\Z}), \ba_0^\star\}}{\partial f_{\Delta, W,Y|\Z}}
(\wh f_{\Delta, W,Y|\Z} - f_{\Delta, W,Y|\Z})\\
&&= n^{-1/2}  \sum_{j=1}^n \bh_{1{\rm k}1}^\star(y_j,w_j, \delta_j, \z_j) + o_p(1),\\
&&n^{-1/2} \sumi \frac{\partial \S\eff^\star(y_i, w_i, \delta_i,
  \z_i; \bb_0, E_{10}, \ba_0^\star)}{\partial \ba}
(\wh \ba^\star - \ba_0^\star)\\
&&= n^{-1/2} \sum_{j=1}^n \{\bh_{1{\rm s}2}^\star(y_j,w_j, \delta_j, \z_j) + \bh_{1{\rm k}2}^\star(y_j,w_j, \delta_j, \z_j)\} +o_p(1),\\
&&n^{-1/2} \sumi \frac{\partial \S\eff^*(y_i, w_i, \delta_i,
  \z_i; \bb_0, E_1^*, \ba_0^*)}{\partial \ba}
(\wh \ba^* - \ba_0^*)\\
&&= n^{-1/2}\sum_{j=1}^n \{\bh_{2{\rm s}}^*(y_j,w_j, \delta_j, \z_j) + \bh_{2{\rm k}}^*(y_j,w_j, \delta_j, \z_j)\} + o_p(1).
\ese
\end{Lem}
 Proof. We prove Lemma \ref{lem:l8} by decomposing it into Lemmas \ref{lem:l5}--\ref{lem:l7}. For clarity, we use the subscript `${\rm s}$' to denote terms arising from the estimation of conditional survival functions $\wh S_{C|Y,\Z}$ or $\wh S_{X|Y,\Z}$, and the subscript `${\rm k}$' to denote terms arising from the kernel estimation of $\wh f_{\Delta, W,Y|\Z}$ or $\wh f_{\Delta, W|Y,\Z}$. Each lemma analyzes a pair of estimators---either $(\wh S_{C|Y,\Z}, \wh f_{\Delta, W,Y|\Z})$ or $(\wh S_{X|Y,\Z}, \wh f_{\Delta, W|Y,\Z})$---and therefore addresses both `${\rm s}$' and `${\rm k}$' components.
 \begin{Lem} \label{lem:l5}
    Let $\o_j\equiv (\delta_j,w_j,y_j,\z_j)$.
Under Conditions \ref{con:nker}, \ref{con:nkerbw'}, \ref{con:nbddaway},  and
\ref{con:nderivbddaway}, 
    \bse
&&n^{-1/2} \sumi \frac{\partial \S\eff^\star\{y_i, w_i, \delta_i,
  \z_i; \bb_0, E_1(\cdot; S_{C|Y,\Z}, f_{\Delta, W, Y|\Z}), \ba_0^\star\}}{\partial S_{C|Y,\Z}}
(\wh S_{C|Y,\Z} - S_{C|Y,\Z})\\
&=& n^{-1/2}\sum_{j=1}^n \bh_{1{\rm s}1}^\star(y_j,w_j, \delta_j, \z_j) +o_p(1),
\ese
and 
\bse
&&n^{-1/2} \sumi \frac{\partial \S\eff^\star\{y_i, w_i, \delta_i,
  \z_i; \bb_0, E_1(\cdot; S_{C|Y,\Z}, f_{\Delta, W,Y|\Z}), \ba_0^\star\}}{\partial f_{\Delta, W,Y|\Z}}
(\wh f_{\Delta, W,Y|\Z} - f_{\Delta, W,Y|\Z})\\
&=& n^{-1/2}  \sum_{j=1}^n \bh_{1{\rm k}1}^\star(y_j,w_j, \delta_j, \z_j) + o_p(1),
\ese
where $\bh_{1{\rm s}1}^\star(y_j,w_j, \delta_j, \z_j)$ and $\bh_{1{\rm k}1}^\star(y_j,w_j, \delta_j, \z_j)$ are defined as
\be
\bh_{1{\rm s}1}^\star(y_j,w_j, \delta_j, \z_j)&\equiv&- E\left[ I(X>C) \{\S\eff^\star(Y,
    X, 1, \z_j;\bb_0) -\S\eff^\star(Y, C, 0, \z_j;\bb_0)\} \right.\n\\
&& \left.\times\frac{\xi_C (w_j, \delta_j, X,y_j,\z_j) f_{Y|X,\Z}(y_j,
    X, \z_j)}{S_{C|Y,\Z} (X,y_j,\z_j)f_{Y|\Z}(y_j, \z_j)} \mid \Z = \z_j,\bo_j\right], \label{eq:h1v1}\\
\bh_{1{\rm k}1}^\star(y_j,w_j, \delta_j, \z_j)&\equiv&\frac{\delta_j }{S_{C|Y,\Z} (w_j,y_j,\z_j)}E\left[I(w_j>C)\right.\n\\
&&\times \{\S\eff^\star(Y,
    w_j, 1, \z_j;\bb_0) -\S\eff^\star(Y, C, 0, \z_j;\bb_0)\}\n\\
    &&\left.\mid X = w_j, \Z = \z_j,\bo_j \right]. \label{eq:h1q1}
\ee
\end{Lem}
Proof. 
 Define {\rm(A)} and {\rm(B)} as
 \bse
{\rm(A)}&\equiv&n^{-1/2} \sum_{i=1}^n \frac{\partial \S\eff^\star\{y_i, w_i, \delta_i,
  \z_i; \bb_0, E_1(\cdot; S_{C|Y,\Z}, f_{\Delta, W, Y|\Z}), \ba_0^\star\}}{\partial S_{C|Y,\Z}}
(\wh S_{C|Y,\Z} - S_{C|Y,\Z}),\\
{\rm(B)} &\equiv& n^{-1/2} \sumi \frac{\partial \S\eff^\star\{y_i, w_i, \delta_i,
  \z_i; \bb_0, E_1(\cdot; S_{C|Y,\Z}, f_{\Delta, W,Y|\Z}), \ba_0^\star\}}{\partial f_{\Delta, W,Y|\Z}}
(\wh f_{\Delta, W,Y|\Z} - f_{\Delta, W,Y|\Z}).
\ese
First of all, from \eqref{eq:t2}, the Gateaux derivative of $\S\eff$
with respect to $S_{C|Y,\Z}$ is
\bse
&&\frac{\partial \S\eff \{y,w,\delta, \z ; \bb, E_1(\cdot; S_{C|Y,\Z}, f_{\Delta, W,Y|\Z}), \ba\}}{\partial S_{C|Y,\Z}} (u)\\
&=& (1-\delta) \frac{\partial}{\partial S_{C|Y,\Z}}\left[\left\{\iiint \delta' I(x>w)\frac{f_{\Delta, W,Y|\Z}(\delta',x,y',\z)}{S_{C|Y,\Z} (x,y',\z) }f_{Y|X,\Z}(y,x,\z;\bb)d\delta' dxdy'\right\}^{-1}\right.\\
&&\times \left.\iiint \delta' I(x>w)\{\S_\bb^F(y,x,\z;\bb) - \ba(x,\z;\bb)\} \frac{f_{\Delta, W,Y|\Z}(\delta',x,y',\z)}{ S_{C|Y,\Z} (x,y',\z)} f_{Y|X,\Z}(y,x,\z;\bb)d\delta' dxdy'\right](u)\\
&=& -(1-\delta) \left\{\iiint \delta' I(x>w)\frac{f_{\Delta, W,Y|\Z}(\delta',x,y',\z)}{S_{C|Y,\Z} (x,y',\z) }f_{Y|X,\Z}(y,x,\z;\bb)d\delta' dxdy'\right\}^{-1} \\
&&\times\iiint \delta' I(x>w)\{\S_\bb^F(y,x,\z;\bb) - \ba(x,\z;\bb)\} \frac{u(x,y',\z)f_{\Delta, W,Y|\Z}(\delta',x,y',\z)}{ S_{C|Y,\Z} (x,y',\z)^2} f_{Y|X,\Z}(y,x,\z;\bb)d\delta' dxdy'\\
&&+(1-\delta) \left\{\iiint \delta' I(x>w)\frac{f_{\Delta, W,Y|\Z}(\delta',x,y',\z)}{S_{C|Y,\Z} (x,y',\z) }f_{Y|X,\Z}(y,x,\z;\bb)d\delta' dxdy'\right\}^{-2}\\
&&\times \iiint \delta' I(x>w)\frac{u(x,y',\z)f_{\Delta, W,Y|\Z}(\delta',x,y',\z)}{S_{C|Y,\Z} (x,y',\z)^2 }f_{Y|X,\Z}(y,x,\z;\bb)d\delta' dxdy'\\
&&\times \iiint \delta' I(x>w)\{\S_\bb^F(y,x,\z;\bb) - \ba(x,\z;\bb)\} \frac{f_{\Delta, W,Y|\Z}(\delta',x,y',\z)}{ S_{C|Y,\Z} (x,y',\z)} f_{Y|X,\Z}(y,x,\z;\bb)d\delta' dxdy' 
\ese
for a bounded function $u (t,y,\z): \mathbb{R}^{\dim(\z)+2} \to
 \mathbb{R}$. 
Letting $\ba=\ba_0^\star$ and $u=\wh  S_{C|Y,\Z}  -  S_{C|Y,\Z}$,  from \eqref{eq:e1}, we
have 
\bse
{\rm(A)} &=& - n^{-1/2} \sum_{i=1}^n (1-\delta_i)E\left\{ I(X_i>w_i)\mid y_i,\z_i\right\}^{-1} E\left[ I(X_i>w_i)\{\S_\bb^F(y_i,X_i,\z_i;\bb_0) - \ba_0^\star(X_i,\z_i;\bb_0)\} \right. \n\\
&& \left. \times E\left\{\frac{(\wh  S_{C|Y,\Z}  -  S_{C|Y,\Z})(X_i,Y_i',\z_i)}{ S_{C|Y,\Z} (X_i,Y_i',\z_i)}\mid X_i,\z_i\right\}\mid y_i,\z_i\right]  \n\\
&&+ n^{-1/2} \sum_{i=1}^n (1-\delta_i) \frac{E\left[I(X_i>w_i)\{\S_\bb^F(y_i,X_i,\z_i;\bb_0) - \ba_0^\star(X_i,\z_i;\bb_0)\} \mid y_i,\z_i\right]}{E\left\{ I(X_i>w_i)\mid y_i,\z_i\right\}^2}\n\\
&&\times  E\left\{ I(X_i>w_i) E\left\{\frac{(\wh  S_{C|Y,\Z}  -  S_{C|Y,\Z})(X_i,Y_i',\z_i)}{ S_{C|Y,\Z} (X_i,Y_i',\z_i)}\mid X_i,\z_i\right\}\mid y_i,\z_i\right\}.
\ese 
From  \eqref{eq:n9} and Taylor's theorem, we have
\be\label{eq:n11}
&&E\left\{\frac{(\wh  S_{C|Y,\Z}  -  S_{C|Y,\Z})(X_i,Y_i',\z_i)}{ S_{C|Y,\Z} (X_i,Y_i',\z_i)}\mid X_i,\z_i\right\}\\
&=& E\left\{\frac{\sum_{j=1}^n \xi_C (w_j, \delta_j, X_i,Y_i',\z_i) K_{h_1}^{(m_1)}(Y_i' - y_j, \z_i - \z_j)}{S_{C|Y,\Z} (X_i,Y_i',\z_i)\sum_{k=1}^n K_{h_1}^{(m_1)}(Y_i' - y_k, \z_i - \z_k)}\mid X_i, \z_i\right\}\n\\
&&+O_p\{(\log n)^{3/4} n^{-3/4} h_1^{-3d/4} + h_1^{m_1}\}\n\\
&=& E\left[\frac{E\left\{\xi_C (W_j, \Delta_j, X_i,Y_i',\z_i)\mid X_i, Y_j = Y_i',\Z_j = \z_i\right\} }{S_{C|Y,\Z} (X_i,Y_i',\z_i)}\mid X_i, \z_i\right]\n\\
&&+n^{-1}E\left\{\frac{\sum_{j=1}^n \xi_C (w_j, \delta_j, X_i,Y_i',\z_i) K_{h_1}^{(m_1)}(Y_i' - y_j, \z_i - \z_j)}{S_{C|Y,\Z} (X_i,Y_i',\z_i)f_{Y,\Z}(Y_i', \z_i)}\mid X_i, \z_i\right\}\n\\
&&-n^{-1}E\left\{\frac{E\left\{\xi_C (W_j, \Delta_j, X_i,Y_i',\z_i)\mid X_i, Y_j = Y_i',\Z_j = \z_i\right\}}{S_{C|Y,\Z} (X_i,Y_i',\z_i)f_{Y,\Z}(Y_i', \z_i)}\right.\n\\
&&\left.\sum_{k=1}^n K_{h_1}^{(m_1)}(Y_i' - y_k, \z_i - \z_k)\mid X_i, \z_i\right\}\n\\
&&+O_p\{(\log n)^{3/4} n^{-3/4} h_1^{-3d/4} + h_1^{m_1}\} + O_p\{(\log n) n^{-1} h_1^{-d} + h_1^{2m_1}\}\n
\ee
uniformly for $X_i,\z_i$. Under Condition \ref{con:nkerbw'}, we have $O_p\{(\log n)^{3/4} n^{-3/4} h_1^{-3d/4} + h_1^{m_1}\} + O_p\{(\log n) n^{-1} h_1^{-d} + h_1^{2m_1}\} = o_p(n^{-1/2})$. Using \eqref{eq:n39}, expression \eqref{eq:n11} becomes
\be\label{eq:n12}
&&E\left\{\frac{(\wh  S_{C|Y,\Z}  -  S_{C|Y,\Z})(X_i,Y_i',\z_i)}{ S_{C|Y,\Z} (X_i,Y_i',\z_i)}\mid X_i,\z_i\right\}\\
&=& n^{-1}E\left\{\frac{\sum_{j=1}^n \xi_C (w_j, \delta_j, X_i,Y_i',\z_i) K_{h_1}^{(m_1)}(Y_i' - y_j, \z_i - \z_j)}{S_{C|Y,\Z} (X_i,Y_i',\z_i)f_{Y,\Z}(Y_i', \z_i)}\mid X_i, \z_i\right\}\n\\
&&+o_p(n^{-1/2})\n\\
&=& n^{-1}\sum_{j=1}^n \frac{\xi_C (w_j, \delta_j, X_i,y_j,\z_i) f_{Y|X,\Z}(y_j, X_i, \z_i)K_{h_1}^{(m_1)}(\z_i - \z_j)}{S_{C|Y,\Z} (X_i,y_j,\z_i)f_{Y,\Z}(y_j, \z_i)}+o_p(n^{-1/2}),\n
\ee
where the last line holds since $O_p(h_1^{m_1}) = o_p(n^{-1/2})$ under
Condition \ref{con:nkerbw'}. 
Thus, we can rewrite (A) as
\be \label{eq:n21}
&&{\rm(A)}\\
&=& - n^{-3/2} \sum_{i=1}^n \sum_{j=1}^n (1-\delta_i) E\left\{ I(X_i>w_i)\mid y_i,\z_i\right\}^{-1}\n\\
&&\times E\left[ I(X_i>w_i)\{\S_\bb^F(y_i,X_i,\z_i;\bb_0) - \ba_0^\star(X_i,\z_i;\bb_0)\} \right. \n\\
&& \left. \times\frac{\xi_C (w_j, \delta_j, X_i,y_j,\z_i) f_{Y|X,\Z}(y_j, X_i, \z_i)K_{h_1}^{(m_1)}(\z_i - \z_j)}{S_{C|Y,\Z} (X_i,y_j,\z_i)f_{Y,\Z}(y_j, \z_i)}\mid y_i,\z_i\right] \n\\
&&+ n^{-3/2} \sum_{i=1}^n \sum_{j=1}^n (1-\delta_i)\n\\
&&\times\frac{E\left[I(X_i>w_i)\{\S_\bb^F(y_i,X_i,\z_i;\bb_0) - \ba_0^\star(X_i,\z_i;\bb_0)\} \mid y_i,\z_i\right]}{E\left\{ I(X_i>w_i)\mid y_i,\z_i\right\}^2}\n\\
&&\times  E\left\{ I(X_i>w_i)\frac{\xi_C (w_j, \delta_j, X_i,y_j,\z_i) f_{Y|X,\Z}(y_j, X_i, \z_i)K_{h_1}^{(m_1)}(\z_i - \z_j)}{S_{C|Y,\Z} (X_i,y_j,\z_i)f_{Y,\Z}(y_j, \z_i)}\mid y_i,\z_i\right\}\n\\
&&+o_p(1)\n\\
&=&  n^{-3/2} \sum_{i=1}^n \sum_{j=1}^n \frac{\bv(\bo_i,\bo_j) K_{h_1}^{(m_1)}(\z_i - \z_j)}{f_{\Z}(\z_i)}+o_p(1),\n
\ee
where, for $\bo_i = (y_i,w_i,\delta_i,\z_i)$ and $\bo_j =
(y_j,w_j,\delta_j,\z_j)$, $\bv(\bo_i,\bo_j)$ is defined as 
\be
\bv(\bo_i,\bo_j) &\equiv& - (1-\delta_i) E\left\{ I(X_i>w_i)\mid y_i,\z_i\right\}^{-1}\n\\
&&\times E\left[ I(X_i>w_i)\{\S_\bb^F(y_i,X_i,\z_i;\bb_0) - \ba_0^\star(X_i,\z_i;\bb_0)\} \right. \n\\
&& \left. \times\frac{\xi_C (w_j, \delta_j, X_i,y_j,\z_i) f_{Y|X,\Z}(y_j, X_i, \z_i)}{S_{C|Y,\Z} (X_i,y_j,\z_i)f_{Y|\Z}(y_j, \z_i)}\mid y_i,\z_i\right] \n\\
&&+ (1-\delta_i) \frac{E\left[I(X_i>w_i)\{\S_\bb^F(y_i,X_i,\z_i;\bb_0) - \ba_0^\star(X_i,\z_i;\bb_0)\} \mid y_i,\z_i\right]}{E\left\{ I(X_i>w_i)\mid y_i,\z_i\right\}^2}\n\\
&&\times  E\left\{ I(X_i>w_i)\frac{\xi_C (w_j, \delta_j, X_i,y_j,\z_i) f_{Y|X,\Z}(y_j, X_i, \z_i)}{S_{C|Y,\Z} (X_i,y_j,\z_i)f_{Y|\Z}(y_j, \z_i)}\mid y_i,\z_i\right\}.\label{eq:n17}
\ee 
Since $E\{\bv(\bo_i,\bO_j) \mid w_i, \delta_i, y_i, \Z_j = \z_i\}=\0$ by \eqref{eq:n39}, we can apply Hoeffding's U-statistics theory to show that
\be
{\rm(A)} &=& n^{-1/2}  \sumi E\left\{\frac{\bv(\bo_i,\bO_j) K_{h_1}^{(m_1)}(\z_i - \Z_j)}{f_{\Z}(\z_i)}\mid \bo_i\right\}\n\\
&&+  n^{-1/2}  \sum_{j=1}^n E\left\{\frac{\bv(\bO_i,\bo_j) K_{h_1}^{(m_1)}(\Z_i - \z_j)}{f_{\Z}(\Z_i)} \mid \bo_j \right\}\n\\
&& - n^{1/2}   E\left\{\frac{\bv(\bO_i,\bO_j) K_{h_1}^{(m_1)}(\Z_i - \Z_j)}{f_{\Z}(\Z_i)}\right\} + O_p\{n^{-1/2} h_1^{-d'/2}\} + o_p(1)
\n\\
&=&  n^{-1/2}  \sum_{j=1}^n E\left\{\bv(\bO_i,\bo_j) \mid \o_j, \Z_i = \z_j \right\}+ o_p(1), \label{eq:n19}
\ee
where the last equality holds since $O_p\{n^{-1/2} h_1^{-d'/2} +
n^{1/2}h_1^{m_1}\}$ is $o_p(1)$ under \ref{con:nkerbw'}.  Then,
  plugging in the expression of 
$\bv(\bo_i,\bo_j)$ from \eqref{eq:n17} into the above, we obtain
\bse
{\rm(A)} &=& n^{-1/2}  \sum_{j=1}^n E\left(-(1-\Delta_i) E\left\{ I(X_i>W_i)\mid W_i,Y_i,\Z_i\right\}^{-1}\right.\\
&&\left.\times E\left[ I(X_i>W_i)\{\S_\bb^F(Y_i,X_i,\Z_i;\bb_0) - \ba_0^\star(X_i,\Z_i;\bb_0)\} \right. \right.\n\\
&& \left. \times\frac{\xi_C (w_j, \delta_j, X_i,y_j,\Z_i) f_{Y|X,\Z}(y_j, X_i, \Z_i)}{S_{C|Y,\Z} (X_i,y_j,\Z_i)f_{Y|\Z}(y_j, \Z_i)}\mid W_i,Y_i,\Z_i\right]\n\\
&&+ (1-\Delta_i) \frac{E\left[I(X_i>W_i)\{\S_\bb^F(Y_i,X_i,\Z_i;\bb_0) - \ba_0^\star(X_i,\Z_i;\bb_0)\} \mid W_i,Y_i,\Z_i\right]}{E\left\{ I(X_i>W_i)\mid W_i,Y_i,\Z_i\right\}^2}\n\\
&&\left.\times  E\left\{ I(X_i>W_i)\frac{\xi_C (w_j, \delta_j, X_i,y_j,\Z_i) f_{Y|X,\Z}(y_j, X_i, \Z_i)}{S_{C|Y,\Z} (X_i,y_j,\Z_i)f_{Y|\Z}(y_j, \Z_i)}\mid W_i,Y_i,\Z_i\right\}\mid \Z_i = \z_j,\bo_j\right)+o_p(1)\\
&=& n^{-1/2}  \sum_{j=1}^n E\left( -E\left[ I(X_i>C_i)\{\S_\bb^F(Y_i,X_i,\z_j;\bb_0) - \ba_0^\star(X_i,\z_j;\bb_0)\} \right. \right.\n\\
&& \left. \times\frac{\xi_C (w_j, \delta_j, X_i,y_j,\z_j) f_{Y|X,\Z}(y_j, X_i, \z_j)}{S_{C|Y,\Z} (X_i,y_j,\z_j)f_{Y|\Z}(y_j, \z_j)}\mid C_i,Y_i,\Z_i = \z_j\right]\n\\
&&+ \frac{E\left[I(X_i>C_i)\{\S_\bb^F(Y_i,X_i,\z_j;\bb_0) - \ba_0^\star(X_i,\z_j;\bb_0)\} \mid C_i,Y_i,\Z_i=\z_j\right]}{E\left\{ I(X_i>C_i)\mid C_i,Y_i,\Z_i=\z_j\right\}}\n\\
&&\left.\times  E\left\{ I(X_i>C_i)\frac{\xi_C (w_j, \delta_j,
      X_i,y_j,\z_j) f_{Y|X,\Z}(y_j, X_i, \z_j)}{S_{C|Y,\Z}
      (X_i,y_j,\z_j)f_{Y|\Z}(y_j, \z_j)}\mid
    C_i,Y_i,\Z_i=\z_j\right\}\mid \Z_i = \z_j,\bo_j\right)+o_p(1)\\
&=& - n^{-1/2}  \sum_{j=1}^n E\left(E\left[ I(X_i>C_i)\{\S\eff^\star(Y_i,
    X_i, 1, \z_j;\bb_0) -\S\eff^\star(Y_i, C_i, 0, \z_j;\bb_0)\}\right.\right.\\
    &&\left.\left.\frac{\xi_C (w_j, \delta_j, X_i,y_j,\z_j)
          f_{Y|X,\Z}(y_j, X_i, \z_j)}{S_{C|Y,\Z}
          (X_i,y_j,\z_j)f_{Y|\Z}(y_j, \z_j)}\mid C_i,Y_i,\Z_i =
        \z_j\right]\mid \Z_i = \z_j,\bo_j \right) + o_p(1),
\ese
which gives the fist part of Lemma \ref{lem:l5}. 

 For the second part, from \eqref{eq:t2}, we compute the Gateaux derivative of $\S\eff$ with respect to $f_{\Delta, W,Y|\Z}$ as
\bse
&&\frac{\partial \S\eff \{y,w,\delta, \z ; \bb, E_1(\cdot; S_{C|Y,\Z}, f_{\Delta, W,Y|\Z}), \ba\}}{\partial f_{\Delta, W,Y|\Z}} (u)\\
&=& (1-\delta) \frac{\partial}{\partial f_{\Delta, W,Y|\Z}}\left[\left\{\iiint \delta' I(x>w)\frac{f_{\Delta, W,Y|\Z}(\delta',x,y',\z)}{S_{C|Y,\Z} (x,y',\z) }f_{Y|X,\Z}(y,x,\z;\bb)d\delta' dxdy'\right\}^{-1}\right.\\
&&\times \left.\iiint \delta' I(x>w)\{\S_\bb^F(y,x,\z;\bb) - \ba(x,\z;\bb)\} \frac{f_{\Delta, W,Y|\Z}(\delta',x,y',\z)}{ S_{C|Y,\Z} (x,y',\z)} f_{Y|X,\Z}(y,x,\z;\bb)d\delta' dxdy'\right](u)\\
&=& (1-\delta) \left\{\iiint \delta' I(x>w)\frac{f_{\Delta, W,Y|\Z}(\delta',x,y',\z)}{S_{C|Y,\Z} (x,y',\z) }f_{Y|X,\Z}(y,x,\z;\bb)d\delta' dxdy'\right\}^{-1} \\
&&\times\iiint \delta' I(x>w)\{\S_\bb^F(y,x,\z;\bb) - \ba(x,\z;\bb)\} \frac{u(\delta',x,y',\z)}{ S_{C|Y,\Z} (x,y',\z)} f_{Y|X,\Z}(y,x,\z;\bb)d\delta' dxdy'\\
&&-(1-\delta) \left\{\iiint \delta' I(x>w)\frac{f_{\Delta, W,Y|\Z}(\delta',x,y',\z)}{S_{C|Y,\Z} (x,y',\z) }f_{Y|X,\Z}(y,x,\z;\bb)d\delta' dxdy'\right\}^{-2}\\
&&\times \iiint \delta' I(x>w)\frac{u(\delta',x,y',\z)}{S_{C|Y,\Z} (x,y',\z) }f_{Y|X,\Z}(y,x,\z;\bb)d\delta' dxdy'\\
&&\times \iiint \delta' I(x>w)\{\S_\bb^F(y,x,\z;\bb) - \ba(x,\z;\bb)\} \frac{f_{\Delta, W,Y|\Z}(\delta',x,y',\z)}{ S_{C|Y,\Z} (x,y',\z)} f_{Y|X,\Z}(y,x,\z;\bb)d\delta' dxdy'
\ese
for a bounded function $u (\delta, x,y,\z): \mathbb{R}^{\dim(\z)+3} \to
 \mathbb{R}$. Setting $\bb=\bb_0$, $\ba=\ba_0^\star$, and $u=\wh f_{\Delta, W,Y|\Z} - f_{\Delta, W,Y|\Z}$, and using \eqref{eq:e1} and the definition of $\wh f_{\Delta, W,Y|\Z}$, we obtain
\bse
&&\frac{\partial \S\eff \{y,w,\delta, \z ; \bb_0, E_1(\cdot; S_{C|Y,\Z}, f_{\Delta, W,Y|\Z}), \ba_0^\star\}}{\partial f_{\Delta, W,Y|\Z}} (\wh f_{\Delta, W,Y|\Z} - f_{\Delta, W,Y|\Z})\\
&=&\frac{\partial \S\eff \{y,w,\delta, \z ; \bb_0, E_1(\cdot; S_{C|Y,\Z}, f_{\Delta, W,Y|\Z}), \ba_0^\star\}}{\partial f_{\Delta, W,Y|\Z}} (\wh f_{\Delta, W,Y|\Z})\\
&=& (1-\delta) [f_{Y|\Z}(y,\z) E\{ I(X>w)\mid y,\z\}]^{-1} \\
&&\times \sum_{j=1}^n \frac{\delta_j I(w_j>w)\{\S_\bb^F(y,w_j,\z_j;\bb_0) - \ba_0^\star(w_j,\z_j;\bb_0)\}f_{Y|X,\Z}(y,w_j,\z_j) K_{h_2}^{(m_2)}(\z-\z_j)}{ S_{C|Y,\Z} (w_j,y_j,\z_j)\sum_{k=1}^n K_{h_2}^{(m_2)}(\z-\z_k)} \\
&&-(1-\delta) [f_{Y|\Z}(y,\z) E\{ I(X>w)\mid y,\z\}]^{-2}\\
&&\times \sum_{j=1}^n \frac{\delta_j I(w_j>w)f_{Y|X,\Z}(y,w_j,\z_j)K_{h_2}^{(m_2)}(\z-\z_j)}{ S_{C|Y,\Z} (w_j,y_j,\z_j)\sum_{k=1}^n K_{h_2}^{(m_2)}(\z-\z_k)} \\
&&\times f_{Y|\Z}(y,\z) E[ I(X>w)\{\S_\bb^F(y,X,\z;\bb_0) - \ba_0^\star(X,\z;\bb_0)\} \mid y,\z].
\ese
Then we can expand (B) as
\be
{\rm(B)} &=& n^{-1/2} \sum_{i=1}^n (1-\delta_i) \sum_{j=1}^n \frac{\delta_j I(w_j>w_i)\{\S_\bb^F(y_i,w_j,\z_j;\bb_0) - \ba_0^\star(w_j,\z_j;\bb_0)\}}{f_{Y|\Z}(y_i,\z_i)E\{ I(X_i>w_i)\mid y_i,\z_i\} S_{C|Y,\Z} (w_j,y_j,\z_j)}\n\\
&& \times\frac{f_{Y|X,\Z}(y_i,w_j,\z_j)K_{h_2}^{(m_2)}(\z_i-\z_j)}{\sum_{k=1}^n K_{h_2}^{(m_2)}(\z_i-\z_k)}-- n^{-1/2} \sum_{i=1}^n (1-\delta_i) \n\\
&&\times \frac{E\left[I(X_i>w_i)\{\S_\bb^F(y_i,X_i,\z_i;\bb_0) - \ba_0^\star(X_i,\z_i;\bb_0)\} \mid y_i,\z_i\right]}{f_{Y|\Z}(y_i,\z_i)E\left\{ I(X_i>w_i)\mid y_i,\z_i\right\}^2}\n\\
&&\times  \sum_{j=1}^n \frac{\delta_j I(w_j>w_i)f_{Y|X,\Z}(y_i,w_j,\z_j)K_{h_2}^{(m_2)}(\z_i-\z_j)}{ S_{C|Y,\Z} (w_j,y_j,\z_j)\sum_{k=1}^n K_{h_2}^{(m_2)}(\z_i-\z_k)} \n\\
&=& n^{-1/2} \sumi\frac{ \sum_{j=1}^n \bq(\bo_i,\bo_j) K_{h_2}^{(m_2)}(\z_i-\z_j)}{ \sum_{k=1}^n K_{h_2}^{(m_2)}(\z_i-\z_k)},\label{eq:n22}
\ee
where $\bq(\bo_i,\bo_j)$ denotes
\be
\bq(\bo_i,\bo_j) &=& (1-\delta_i) \frac{\delta_j I(w_j>w_i)\{\S_\bb^F(y_i,w_j,\z_j;\bb_0) - \ba_0^\star(w_j,\z_j;\bb_0)\}f_{Y|X,\Z}(y_i,w_j,\z_j)}{f_{Y|\Z}(y_i,\z_i)E\{ I(X_i>w_i)\mid y_i,\z_i\} S_{C|Y,\Z} (w_j,y_j,\z_j)}\n\\
&&- (1-\delta_i) \frac{E\left[I(X_i>w_i)\{\S_\bb^F(y_i,X_i,\z_i;\bb_0) - \ba_0^\star(X_i,\z_i;\bb_0)\} \mid y_i,\z_i\right]}{f_{Y|\Z}(y_i,\z_i) E\left\{ I(X_i>w_i)\mid y_i,\z_i\right\}^2}\n\\
&&\times \frac{\delta_j I(w_j>w_i)f_{Y|X,\Z}(y_i,w_j,\z_j)}{ S_{C|Y,\Z} (w_j,y_j,\z_j)},\label{eq:n20}
\ee
for $\bo_i = (y_i,w_i,\delta_i,\z_i)$ and $\bo_j = (y_j,w_j,\delta_j,\z_j)$.
Meanwhile, $E\{\bq(\bo_i,\bO_j) \mid w_i, \delta_i, y_i, \Z_j = \z_i\} = \0$ since 
\bse
&&E\{\bq(\bo_i,\bO_j) \mid w_i, \delta_i, y_i, \Z_j = \z_i\}\\
&=& E\left( (1-\delta_i) \frac{\Delta_j I(W_j>w_i)\{\S_\bb^F(y_i,W_j,\Z_j;\bb_0) - \ba_0^\star(W_j,\Z_j;\bb_0)\}f_{Y|X,\Z}(y_i,W_j,\Z_j)}{f_{Y|\Z}(y_i,\z_i)E\{ I(X_i>w_i)\mid y_i,\z_i\} S_{C|Y,\Z} (W_j,Y_j,\Z_j)}\right.\n\\
&&- (1-\delta_i) \frac{E\left[I(X_i>w_i)\{\S_\bb^F(y_i,X_i,\z_i;\bb_0) - \ba_0^\star(X_i,\z_i;\bb_0)\} \mid y_i,\z_i\right]}{f_{Y|\Z}(y_i,\z_i)E\left\{ I(X_i>w_i)\mid y_i,\z_i\right\}^2}\n\\
&&\left.\times \frac{\Delta_j I(W_j>w_i)f_{Y|X,\Z}(y_i,W_j,\Z_j)}{ S_{C|Y,\Z} (W_j,Y_j,\Z_j)}\mid w_i, \delta_i, y_i, \Z_j = \z_i \right)\\
&=& \frac{1-\delta_i}{E\{ I(X_i>w_i)\mid y_i,\z_i\}}\left(E\left[ I(X_j>w_i)\{\S_\bb^F(y_i,X_j,\Z_j;\bb_0) - \ba_0^\star(X_j,\Z_j;\bb_0)\}\mid Y_j = y_i, \Z_j = \z_i \right]\right.\n\\
&&- \frac{E\left[I(X_i>w_i)\{\S_\bb^F(y_i,X_i,\z_i;\bb_0) - \ba_0^\star(X_i,\z_i;\bb_0)\} \mid y_i,\z_i\right]}{E\left\{ I(X_i>w_i)\mid y_i,\z_i\right\}}\n\\
&&\left.\times E\left\{I(X_j>w_i)\mid Y_j = y_i, \Z_j = \z_i \right\}\right)\\
&=&\0.
\ese
Then by Taylor's theorem, we have
\be
{\rm (B)} &=& n^{-1/2} \sumi E \left\{\bq(\bo_i,\bO_j)\mid \o_i, \Z_j = \z_i\right\}\n\\
&&+n^{-3/2} \sumi \sum_{j=1}^n \frac{ \bq(\bo_i,\bo_j) K_{h_2}^{(m_2)}(\z_i-\z_j)}{ f_{\Z}(\z_i)}\n\\
&&-n^{-3/2} \sumi \sum_{k=1}^n \frac{E \left\{\bq(\bo_i,\bO_j)\mid \bo_i, \Z_j = \z_i\right\}}{ f_{\Z}(\z_i)} K_{h_2}^{(m_2)}(\z_i-\z_k)\n\\
&&+O_p\{(\log n) n^{-1/2} h_2^{-d'} + n^{1/2}h_2^{2m_2}\}\n\\
&=& n^{-3/2} \sumi \sum_{j=1}^n \frac{ \bq(\bo_i,\bo_j) K_{h_2}^{(m_2)}(\z_i-\z_j)}{ f_{\Z}(\z_i)}+ o_p(1), \label{eq:n13}
\ee
where the last equality follows since $O_p\{(\log n) n^{-1/2} h_2^{-d'} + n^{1/2}h_2^{2m_2}\}$ is $o_p(1)$ under Condition \ref{con:nkerbw}. Therefore, by the U-statistic argument as in \eqref{eq:n19}, we obtain
\bse
{\rm (B)} &=& n^{-1/2} \sumi E\left\{\frac{ \bq(\bo_i,\bO_j) K_{h_2}^{(m_2)}(\z_i-\Z_j)}{ f_{\Z}(\z_i)}\mid \bo_i\right\}\\
&&+n^{-1/2} \sum_{j=1}^n E\left\{\frac{ \bq(\bO_i,\bo_j) K_{h_2}^{(m_2)}(\Z_i-\z_j)}{ f_{\Z}(\Z_i)}\mid \bo_j\right\}\\
&&-n^{1/2} E\left\{\frac{ \bq(\bO_i,\bO_j) K_{h_2}^{(m_2)}(\Z_i-\Z_j)}{ f_{\Z}(\Z_i)}\right\}+ O_p\{n^{-1/2} h_2^{-d'/2}\} +o_p(1)\\
&=&  n^{-1/2}  \sum_{j=1}^n E\left\{\bq(\bO_i,\bo_j) \mid \Z_i = \z_j,\bo_j \right\}+ o_p(1),
\ese
where the last equality holds since $O_p\{n^{-1/2} h_2^{-d'/2} + n^{1/2}h_2^{m_2}\}$ is $o_p(1)$
under \ref{con:nkerbw'}. 
From \eqref{eq:n20}, we finally have
\bse
{\rm(B)} &=& n^{-1/2}  \sum_{j=1}^ n E\left((1-\Delta_i) \frac{\delta_j I(w_j>W_i)\{\S_\bb^F(Y_i,w_j,\z_j;\bb_0) - \ba_0^\star(w_j,\z_j;\bb_0)\}f_{Y|X,\Z}(Y_i,w_j,\z_j)}{f_{Y|\Z}(Y_i,\Z_i)E\{ I(X_i>W_i)\mid W_i,Y_i,\Z_i\} S_{C|Y,\Z} (w_j,y_j,\z_j)}\right.\n\\
&&- (1-\Delta_i) \frac{E\left[I(X_i>W_i)\{\S_\bb^F(Y_i,X_i,\Z_i;\bb_0) - \ba_0^\star(X_i,\Z_i;\bb_0)\} \mid W_i,Y_i,\Z_i\right]}{f_{Y|\Z}(Y_i,\Z_i)E\left\{ I(X_i>W_i)\mid W_i,Y_i,\Z_i\right\}^2}\n\\
&&\left.\times \frac{\delta_j I(w_j>W_i)f_{Y|X,\Z}(Y_i,w_j,\z_j)}{ S_{C|Y,\Z} (w_j,y_j,\z_j)}\mid \Z_i = \z_j,\bo_j \right)+o_p(1)\\
 &=& n^{-1/2}  \sum_{j=1}^ n E\left(\frac{\delta_j I(w_j>C_i)\{\S_\bb^F(Y_i,w_j,\z_j;\bb_0) - \ba_0^\star(w_j,\z_j;\bb_0)\}f_{Y|X,\Z}(Y_i,w_j,\z_j)}{f_{Y|\Z}(Y_i,\z_j)S_{C|Y,\Z} (w_j,y_j,\z_j)}\right.\n\\
&&- \frac{E\left[I(X_i>C_i)\{\S_\bb^F(Y_i,X_i,\Z_i;\bb_0) - \ba_0^\star(X_i,\Z_i;\bb_0)\} \mid C_i,Y_i,\Z_i\right]}{E\left\{ I(X_i>C_i)\mid C_i,Y_i,\Z_i\right\}}\n\\
&&\left.\times \frac{\delta_j I(w_j>C_i)f_{Y|X,\Z}(Y_i,w_j,\z_j)}{f_{Y|\Z}(Y_i,\z_j) S_{C|Y,\Z} (w_j,y_j,\z_j)}\mid \Z_i = \z_j,\bo_j \right)+o_p(1)\\
&=& n^{-1/2}  \sum_{j=1}^n E\left[\frac{\delta_j I(w_j>C_i)f_{Y|X,\Z}(Y_i,w_j,\z_j)}{f_{Y|\Z}(Y_i,\z_j)S_{C|Y,\Z} (w_j,y_j,\z_j)}\{\S\eff^\star(Y_i,
    w_j, 1, \z_j;\bb_0) -\S\eff^\star(Y_i, C_i, 0, \z_j;\bb_0)\}\right.\\
    &&\left.\mid \Z_i = \z_j,\bo_j  \right] + o_p(1),
\ese
and this completes the proof of the second part of Lemma \ref{lem:l5}
since the following equation 
\bse
E\left\{g(C,Y,x,\z) \frac{f_{Y|X,\Z}(Y,x,\z)}{f_{Y|\Z}(Y,\z)}\mid \z\right\} = E\{g(C,Y,x,\z) \mid x,\z\}
\ese
holds under $X\indep C|Y,\Z$.
\qed

From the definition of $\L$ in \eqref{eq:n7}, we can derive that
\bse 
&&E_1\{\L(\bu; E_1, E_2)(X,\z; \bb) \mid\z;\bb\}\\
&=& E_1\left(E_2\{I(X\le C)|X,\z;\bb\} \bu (X,\bz;\bb)\right.\\
&&\left.+E_2\left[\left.I(X>C)\frac{E_1\{I(X>C)\bu(X,\bz;\bb)
      |C,Y,\z;\bb\}}{E_1\{I(X>C)|C,Y,\z;\bb\}}\right|X,\bz;\bb\right]\mid \z;\bb\right)\\
&=& E_1\left\{I(X\le C)\bu (X,\bz;\bb)+I(X>C)\bu(X,\bz;\bb)\mid \z;\bb\right\}\\
&=& E_1\left\{\bu (X,\bz;\bb)\mid \z;\bb\right\}, 
\ese
which leads to 
\be 
E_1\{\L^{-1}(\bu; E_1, E_2)(X,\z; \bb) \mid\z;\bb\} &=& E_1\left\{\bu (X,\bz;\bb)\mid \z;\bb\right\}. \label{eq:n37}
\ee

\begin{Lem} \label{lem:l6} 
    Let $\o_j\equiv (\delta_j,w_j,y_j,\z_j)$. Under Conditions \ref{con:nker}, \ref{con:nkerbw'}, \ref{con:nbddaway}, and 
\ref{con:nderivbddaway}, 
 \bse
&&n^{-1/2} \sumi \frac{\partial \S\eff^\star(y_i, w_i, \delta_i,
  \z_i; \bb_0, E_{10}, \ba_0^\star)}{\partial \ba}
(\wh \ba^\star - \ba_0^\star)\\
&=& n^{-1/2} \sum_{j=1}^n \{\bh_{1{\rm s}2}^\star(y_j,w_j, \delta_j, \z_j) + \bh_{1{\rm k}2}^\star(y_j,w_j, \delta_j, \z_j)\} +o_p(1),
\ese
where $\bh_{1{\rm s}2}^\star(y_j,w_j, \delta_j, \z_j)$ and $\bh_{1{\rm k}2}^\star(y_j,w_j, \delta_j, \z_j)$ are defined as 
\be
\bh_{1{\rm s}2}^\star(y_j,w_j, \delta_j, \z_j)
&\equiv& E_2^\star\left[ I(X>C)\{\S\eff^\star(Y,X,1,\z_j;\bb_0) -\S\eff^\star(Y,C,0,\z_j;\bb_0) \}
\right.\n\\
&&\times \left.\frac{ \xi_C (w_j, \delta_j, X,y_j,\z_j)f_{Y|X,\Z}(y_j,X,\z_j)}{S_{C|Y,\Z} (X,y_j,\z_j)f_{Y|\Z}(y_j,\z_j)} \mid \Z = \z_j,\bo_j;\bb_0\right], \label{eq:h1v2}\\
\bh_{1{\rm k}2}^\star(y_j,w_j, \delta_j, \z_j)&\equiv& -\frac{\delta_j }{S_{C|Y,\Z} (w_j,y_j,\z_j)}E_2^\star\left[I(w_j>C) \{\S\eff^\star(Y,w_j,1,\z_j;\bb_0)- \S\eff^\star(Y,C,0,\z_j;\bb_0)\}\right.\n\\
&&\left.\mid X=w_j, \Z = \z_j,\bo_j;\bb_0\right\}.\label{eq:h1q2}
\ee
\end{Lem}
Proof.
The Gateaux derivative of $\S\eff$ with respect to $\ba$ is
 \be\label{eq:n25}
 &&\frac{\partial \S\eff (y,w,\delta, \z ; \bb, E_1, \ba)}{\partial \ba} (\bu)\\
 &=& -\delta \bu(w,\z; \bb) - (1-\delta) \frac{ E_1\{I(X>w)\bu(X,\z;\bb) |y,\z;\bb\}}{E_1\{I(X> w)|y,\z;\bb\}}\n 
\ee
for a bounded function $\bu(x,\z; \bb): \mathbb{R}^{\dim(\z)+1} \times \mathbb{R}^p \to
\mathbb{R}^p$.
Let $\L_0^\star(\cdot) = \L(\cdot; E_{10}, E_2^\star)$ and $\wh
\L^\star(\cdot) = \L(\cdot; \wh E_1, E_2^\star)$. Then 
\bse
&& (\wh \ba^\star - \ba_0^\star)(x,\bz;\bb_0)\\
&=& \{\wh \L^{\star -1}(\wh\bc^\star) - \L_0^{\star-1}(\bc_0^\star)\} (x,\bz;\bb_0)\\
&=& - \L_0^{\star -1}(\wh \L^\star - \L_0^\star) \L_0^{\star -1}(\bc_0^\star) (x,\bz;\bb_0) + \L_0^{\star -1}(\wh\bc^\star - \bc_0^\star) (x,\bz;\bb_0) + o_p(n^{-1/2})\\
&=& - \L_0^{\star -1}\left\{E_2^\star\left(\left.I(x>C)\left[\frac{\wh E_1\{I(X>C)\ba_0^\star(X,\bz;\bb_0)
|C,Y,\z;\bb_0\}}{\wh E_1\{I(X>C)|C,Y,\z;\bb_0\}} \right.\right.\right.\right.\\
&&\left.\left.\left.\left.- \frac{ E_{10}\{I(X>C)\ba_0^\star(X,\bz;\bb_0)
|C,Y,\z;\bb_0\}}{E_{10}\{I(X>C)|C,Y,\z;\bb_0\}}\right]\right|x,\bz;\bb_0\right)\right\} (x,\bz;\bb_0) \\
&&+ \L_0^{\star -1}\left\{E_2^\star\left(\left.I(x>C)\left[\frac{\wh E_1\{I(X>C)\S_\bb^F(Y,X,\bz;\bb_0)
|C,Y,\z;\bb_0\}}{\wh E_1\{I(X>C)|C,Y,\z;\bb_0\}} \right.\right.\right.\right.\\
&&\left.\left.\left.\left.- \frac{ E_{10}\{I(X>C)\S_\bb^F(Y,X,\bz;\bb_0)
|C,Y,\z;\bb_0\}}{E_{10}\{I(X>C)|C,Y,\z;\bb_0\}}\right]\right|x,\bz;\bb_0\right)\right\} (x,\bz;\bb_0) + o_p(n^{-1/2})\\
&=& \L_0^{\star -1}\left[E_2^\star\left\{\left.I(x>C)\left(\frac{\wh E_1[I(X>C)\{\S_\bb^F(Y,X,\bz;\bb_0) - \ba_0^\star(X,\bz;\bb_0)\}
|C,Y,\z;\bb_0]}{\wh E_1\{I(X>C)|C,Y,\z;\bb_0\}} \right.\right.\right.\right.\\
&&\left.\left.\left.\left.- \frac{ E_{10}[I(X>C)\{\S_\bb^F(Y,X,\bz;\bb_0) - \ba_0^\star(X,\bz;\bb_0)\}
|C,Y,\z;\bb_0]}{E_{10}\{I(X>C)|C,Y,\z;\bb_0\}}\right)\right|x,\bz;\bb_0\right\}\right] (x,\bz;\bb_0) + o_p(n^{-1/2}),
\ese
uniformly for $x$ and $\z$. The second equality follows from Taylor's theorem and Lemma \ref{lem:l3}. Using \eqref{eq:e1} and noting that both $\|\wh S_{C|Y,\Z} - S_{C|Y,\Z}\|_\infty^2$ and $\|\wh f_{\Delta, W,Y|\Z} - f_{\Delta, W,Y|\Z}\|_\infty^2$ are $o_p(n^{-1/2})$ by Lemma \ref{lem:l2} and Condition \ref{con:nkerbw}, we obtain
\be
(\wh \ba^\star - \ba_0^\star)(x,\bz;\bb_0)
 &=& \L_0^{\star -1}(E_2^\star[I(x>C)\{\br_{1{\rm s}}(C,Y,\z;\bb_0) + \br_{1{\rm k}}(C,Y,\z;\bb_0)\label{eq:n26}\\
 &&+o_p(n^{-1/2})\} \mid x,\z;\bb_0])(x,\z;\bb_0)\n 
\ee
by Taylor's theorem, where
\bse
\br_{1{\rm s}}(c,y,\z;\bb) &\equiv& \frac{\partial}{\partial S_{C|Y,\Z}}(E_1[I(X>c)\{\S_\bb^F(y,X,\bz;\bb) - \ba_0^\star(X,\bz;\bb)\}
|c,y,\z;\bb, S_{C|Y,\Z}, f_{\Delta, W,Y|\Z}]\\
&&\times [E_1\{I(X>c)|y,\z;\bb,  S_{C|Y,\Z}, f_{\Delta, W,Y|\Z}\}]^{-1}) (\wh S_{C|Y,\Z} - S_{C|Y,\Z}), \\
\br_{1{\rm k}}(c,y,\z;\bb) &\equiv& \frac{\partial}{\partial f_{\Delta, W,Y|\Z}}(E_1[I(X>c)\{\S_\bb^F(y,X,\bz;\bb) - \ba_0^\star(X,\bz;\bb)\}
|y,\z;\bb, S_{C|Y,\Z}, f_{\Delta, W,Y|\Z}]\\
&&\times [E_1\{I(X>c)|y,\z;\bb,  S_{C|Y,\Z}, f_{\Delta, W,Y|\Z}\}]^{-1}) (\wh f_{\Delta, W,Y|\Z} - f_{\Delta, W,Y|\Z}).
\ese
Applying the calculation in \eqref{eq:d3}, we can show 
\be
\br_{1{\rm s}}(c,y,\z;\bb_0) &=& - E\left[I(X>c)\{\S_\bb^F(y,X,\bz;\bb_0) - \ba_0^\star(X,\bz;\bb_0)\}\frac{ (\wh S_{C|Y,\Z} - S_{C|Y,\Z})(X,Y',\z)}{S_{C|Y,\Z}(X,Y',\z)}
|c,y,\z\right]\n\\
&&\times [E\{I(X>c)|y,\z\}]^{-1} \n\\
&&+E[I(X>c)\{\S_\bb^F(y,X,\bz;\bb_0) - \ba_0^\star(X,\bz;\bb_0)\}
|c,y,\z]\n\\
&&\times E\left\{I(X>c)\frac{ (\wh S_{C|Y,\Z} - S_{C|Y,\Z})(X,Y',\z)}{S_{C|Y,\Z}(X,Y',\z)}\mid y,\z\right\}[E\{I(X>c)|y,\z\}]^{-2}, \n
\ee 
and employing the expression of
\bse
E\left\{\frac{ (\wh S_{C|Y,\Z} - S_{C|Y,\Z})(X,Y',\z)}{S_{C|Y,\Z}(X,Y',\z)}\mid X,\z\right\}
\ese
from \eqref{eq:n12}, we obtain 
\be
\br_{1{\rm s}}(c,y,\z;\bb_0) &=& n^{-1}\sum_{j=1}^n\frac{\bv_1 (c,y,\z,\bo_j) K_{h_1}^{(m_1)}(\z - \z_j)}{f_{\Z}(\z)} + o_p(n^{-1/2}),\label{eq:n23}
\ee
where $\bv_1(c,y,\z,\bo_j)$ is defined as
\bse
\bv_1(c,y,\z,\bo_j)&\equiv& -  E\left[ I(X>c)\{\S_\bb^F(y,X,\z;\bb) - \ba_0^\star(X,\z;\bb_0)\} \frac{\xi_C (w_j, \delta_j, X,y_j,\z)f_{Y|X,\Z}(y_j,X,\z)}{S_{C|Y,\Z} (X,y_j,\z)f_{Y|\Z}(y_j,\z)} \mid y,\z\right] \n\\
&&\times [E\left\{ I(X>c)\mid y,\z\right\}]^{-1} \n\\
&&+ \frac{E\left[I(X>c)\{\S_\bb^F(y,X,\z;\bb_0) - \ba_0^\star(X,\z;\bb_0)\} \mid y,\z\right]}{E\left\{ I(X>c)\mid y,\z\right\}^2}\n\\
&&\times  E\left\{ I(X>c) \frac{\xi_C (w_j, \delta_j, X,y_j,\z)f_{Y|X,\Z}(y_j,X,\z)}{S_{C|Y,\Z} (X,y_j,\z)f_{Y|\Z}(y_j,\z)}\mid y,\z\right\}
\ese
for $\bo_j = (y_j,w_j,\delta_j,\z_j)$. 
In a similar way, using the definition of $\wh f_{\Delta, W,Y|\Z}$ and \eqref{eq:d4}, we can write
\bse
\br_{1{\rm k}}(c,y,\z;\bb_0) &=& E_1\left[I(X>c)\{\S_\bb^F(y,X,\bz;\bb_0) - \ba_0^\star(X,\bz;\bb_0)\}
|c,y,\z;\bb_0, S_{C|Y,\Z},\wh f_{\Delta, W,Y|\Z}-f_{\Delta, W,Y|\Z}\right]\n\\
&&\times [E\{I(X>c)|y,\z\}]^{-1} \n\\
&&-E_1\left\{I(X>c)\mid y,\z;\bb_0, S_{C|Y,\Z},\wh f_{\Delta, W,Y|\Z}-f_{\Delta, W,Y|\Z}\right\}
\n\\
&&\times E[I(X>c)\{\S_\bb^F(y,X,\bz;\bb_0) - \ba_0^\star(X,\bz;\bb_0)\}|c,y,\z][E\{I(X>c)|y,\z\}]^{-2} \n\\
&=& E_1\left[I(X>c)\{\S_\bb^F(y,X,\bz;\bb_0) - \ba_0^\star(X,\bz;\bb_0)\}
|c,y,\z;\bb_0, S_{C|Y,\Z},\wh f_{\Delta, W,Y|\Z}\right]\n\\
&&\times [E\{I(X>c)|y,\z\}]^{-1} \n\\
&&-E_1\left\{I(X>c)\mid y,\z;\bb_0, S_{C|Y,\Z},\wh f_{\Delta, W,Y|\Z}\right\}
\n\\
&&\times E[I(X>c)\{\S_\bb^F(y,X,\bz;\bb_0) - \ba_0^\star(X,\bz;\bb_0)\}|c,y,\z][E\{I(X>c)|y,\z\}]^{-2} \n\\
&=& \sum_{j=1}^n\frac{\bq_1 (c,y,\z,\bo_j) K_{h_2}^{(m_2)}(\z - \z_j)}{\sum_{k=1}^n K_{h_2}^{(m_2)}(\z - \z_k)} + o_p(n^{-1/2}),
\ese
where $\bq_1 (c,y,\z,\bo_j)$ denotes
\bse
\bq_1 (c,y,\z,\bo_j) &\equiv&  \frac{\delta_j
  I(w_j>c)\{\S_\bb^F(y,w_j,\z_j;\bb_0) - \ba_0^\star(w_j,
    \z_j;\bb_0)\}f_{Y|X,\Z}(y,w_j,\z_j)}{ S_{C|Y,\Z} (w_j,y_j,\z_j) f_{Y|\Z}(y,\z) } [E\{ I(X>c)\mid y,\z\}]^{-1}\n\\
&& -  \frac{\delta_j I(w_j>c)f_{Y|X,\Z}(y,w_j,\z_j) }{S_{C|Y,\Z} (w_j,y_j,\z_j) f_{Y|\Z}(y,\z)} \frac{E\left[I(X>c)\{\S_\bb^F(y,X,\z;\bb_0) - \ba_0^\star(X,\z;\bb_0)\} \mid y,\z\right]}{E\left\{ I(X>c)\mid y,\z\right\}^2}.
\ese
From \eqref{eq:n39}, we have
  \be
  E\{\bv_1(c,y,\z,\bO_j)\mid \Z_j=\z\} &=& \0.\label{eq:n18}
  \ee
Also, since
  \bse
  && E\left[\frac{\Delta_j  I(W_j>c)\{\S_\bb^F(y,W_j,\Z_j;\bb_0) - \ba_0^\star(W_j,
\Z_j;\bb_0)\} f_{Y|X,\Z}(y,W_j,\Z_j) }{S_{C|Y,\Z} (W_j,Y_j,\Z_j) f_{Y|\Z}(y,\z)} \mid \Z_j = \z\right]\\
&=& E[I(X>c)\{\S_\bb^F(y,X,\z;\bb_0) - \ba_0^\star(X,\z;\bb_0)\}\mid y,\z],
\ese
and
\bse
E\left\{\frac{\Delta_j  I(W_j>c) f_{Y|X,\Z}(y,W_j,\Z_j) }{S_{C|Y,\Z} (W_j,Y_j,\Z_j) f_{Y|\Z}(y,\z)} \mid \Z_j = \z\right\} &=& E\{I(X>c)\mid y,\z\},
  \ese
we have
\be
E\{\bq_1(c,y,\z,\bO_j)\mid \Z_j=\z\} &=& \0.\label{eq:n24}
\ee
Then, as in \eqref{eq:n13}, Taylor's theorem leads to 
\be
\br_{1{\rm k}}(c,y,\z;\bb_0) &=&  n^{-1}  \sum_{j=1}^n\frac{\bq_1 (c,y,\z,\bo_j) K_{h_2}^{(m_2)}(\z - \z_j)}{f_{\Z}(\z)} + o_p(n^{-1/2}),\label{eq:n27}
\ee
uniformly for $(c,y,\z)$ under Condition \ref{con:nkerbw}. 

For notational simplicity, denote 
\bse
{\rm (C)}&\equiv&n^{-1/2}\sumi \frac{\partial \S\eff^\star (y_i,w_i,\delta_i, \z_i ; \bb_0, E_{10}, \ba_0^\star)}{\partial \ba} (\wh \ba^\star - \ba_0^\star).
\ese
Combining \eqref{eq:n25}, \eqref{eq:n26}, \eqref{eq:n23}, and \eqref{eq:n27}, we obtain
\bse
&&{\rm (C)}\\
&=& n^{-1/2}\sumi\left[-\delta_i (\wh \ba^\star - \ba_0^\star)(w_i,\z_i; \bb_0) - (1-\delta_i) \frac{ E\{I(X_i>w_i)(\wh \ba^\star - \ba_0^\star)(X_i,\z_i;\bb_0) |y_i,\z_i\}}{E\{I(X_i> w_i)|y_i,\z_i\}}\right]\\
&=& -n^{-1/2}\sumi \left[\delta_i \L_0^{\star -1}(E_2^\star[I(
  w_i>C_i)\{\br_{1{\rm s}}(C_i,Y_i,\z_i;\bb_0) + \br_{1{\rm k}}(C_i,Y_i, \z_i;\bb_0)\} \mid w_i, \z_i;\bb_0])( w_i,\z_i;\bb_0)\right.\\
&&+\frac{1-\delta_i}{E\{I(X_i> w_i)|y_i,\z_i\}} E\left\{I(X_i>w_i)\right.\\
&&\left.\left.\times \L_0^{\star -1}(E_2^\star[I(X_i>C_i)\{\br_{1{\rm s}}(C_i,Y_i, \z_i;\bb_0) + \br_{1{\rm k}}(C_i,Y_i, \z_i;\bb_0)\} \mid X_i, \z_i;\bb_0])(X_i,\z_i;\bb_0)|y_i,\z_i
\right\}\right] + o_p(1)\\
&=&  -n^{-3/2}\sumi  \sum_{j=1}^n \left\{ \delta_i  \L_0^{\star -1}\left[E_2^\star\left\{
\frac{ I( w_i >C_i) \bv_1(C_i,Y_i,\z_i,\bo_j)K_{h_1}^{(m_1)}(\z_i - \z_j)}{f_{\Z}(\z_i)}\mid w_i, \z_i;\bb_0\right\}\right](w_i,\z_i;\bb_0)\right.\\
 &&+\frac{1-\delta_i}{E\{I(X_i> w_i)|y_i,\z_i\}} E\left(I(X_i>w_i)\right.\\
&&\left.\left.\times \L_0^{\star -1}\left[E_2^\star\left\{
\frac{ I(X_i>C_i) \bv_1(C_i,Y_i, \z_i,\bo_j)K_{h_1}^{(m_1)}(\z_i - \z_j)}{f_{\Z} (\z_i)}\mid  X_i, \z_i;\bb_0\right\}\right](X_i,\z_i;\bb_0)\mid y_i,\z_i\right)\right\}\\
&&-n^{-3/2}\sumi  \sum_{j=1}^n \left\{ \delta_i  \L_0^{\star -1}\left[E_2^\star\left\{
\frac{ I( w_i >C_i) \bq_1(C_i,Y_i,\z_i,\bo_j)K_{h_2}^{(m_2)}(\z_i - \z_j)}{f_{\Z}(\z_i)}\mid w_i, \z_i;\bb_0\right\}\right](w_i,\z_i;\bb_0)\right.\\
 &&+\frac{1-\delta_i}{E\{I(X_i> w_i)|y_i,\z_i\}} E\left(I(X_i>w_i)\right.\\
&&\left.\left.\times \L_0^{\star -1}\left[E_2^\star\left\{
\frac{ I(X_i>C_i) \bq_1(C_i,Y_i, \z_i,\bo_j)K_{h_2}^{(m_2)}(\z_i - \z_j)}{f_{\Z} (\z_i)}\mid  X_i, \z_i;\bb_0\right\}\right](X_i,\z_i;\bb_0)\mid y_i,\z_i\right)\right\}+ o_p(1).
\ese 
From \eqref{eq:n18} and \eqref{eq:n24}, we have 
\bse
&&E\left(\L_0^{\star -1} \left[E_2^\star\left\{\frac{ I(x>C)
        \bv_1(C,Y,\z,\bO_j)K_{h_1}^{(m_1)}(\z-\Z_j)}{f_{\Z} (\z)}\mid x,\z,\bO_j;\bb_0\right\}\right](x,\z; \bb_0, \bO_j)\mid x, \z\right)\\
&=&\L_0^{\star -1}\left(E_2^\star \left[\frac{ I(x>C)
        E\{\bv_1(C,Y,\z,\bO_j)K_{h_1}^{(m_1)}(\z-\Z_j)\mid C,Y,\z\}}{f_{\Z} (\z)} \mid x, \z;\bb_0\right]\right)(x,\z; \bb_0)\\
&=& O_p(h_1^{m_1})
\ese
and 
\bse
&&E\left(\L_0^{\star -1} \left[E_2^\star\left\{\frac{ I(x>C)
        \bq_1(C,Y,\z,\bO_j)K_{h_2}^{(m_2)}(\z-\Z_j)}{f_{\Z} (\z)}\mid x,\z,\bO_j;\bb_0\right\}\right](x,\z; \bb_0, \bO_j)\mid x, \z\right)\\
&=&\L_0^{\star -1}\left(E_2^\star \left[\frac{ I(x>C)
        E\{\bq_1(C,Y,\z,\bO_j)K_{h_2}^{(m_2)}(\z-\Z_j)\mid C,Y,\z\}}{f_{\Z} (\z)} \mid x, \z;\bb_0\right]\right)(x,\z; \bb_0)\\
&=& O_p(h_2^{m_2}),
\ese
uniformly for $(x,\z)$. Here, since $\L_0^{\star -1}$ is a linear
operator of functions with respect to $x,\z$, we can switch the order
of $\L_0^{\star -1}$ and expectation of $\bO_j$. Note that
$O_p(h_1^{m_1})$ and $O_p(h_2^{m_2})$ are $o_p(n^{-1/2})$ under
\ref{con:nkerbw'}.
Hence, applying the U-statistic argument used in \eqref{eq:n19}, we can simplify (C) to
\bse
&&{\rm (C)}\\
&=&  -n^{-1/2} \sum_{j=1}^n E\left\{ \Delta_i  \L_0^{\star -1}\left[E_2^\star\left\{
\frac{ I( W_i >C_i) \bv_1(C_i,Y_i,\Z_i,\bo_j)K_{h_1}^{(m_1)}(\Z_i - \z_j)}{f_{\Z}(\Z_i)}\mid W_i, \Z_i;\bb_0\right\}\right](W_i,\Z_i;\bb_0)\right.\\
 &&+\frac{1-\Delta_i}{E\{I(X_i> W_i)|W_i,Y_i,\Z_i\}} E\left(I(X_i>W_i)\right.\\
&&\left.\left.\times \L_0^{\star -1}\left[E_2^\star\left\{
\frac{ I(X_i>C_i) \bv_1(C_i,Y_i, \Z_i,\bo_j)K_{h_1}^{(m_1)}(\Z_i - \z_j)}{f_{\Z} (\Z_i)}\mid  X_i, \Z_i;\bb_0\right\}\right](X_i,\Z_i;\bb_0)\mid W_i,Y_i,\Z_i\right)\mid \bo_j\right\}\\
&&-n^{-1/2}  \sum_{j=1}^n E\left\{ \Delta_i  \L_0^{\star -1}\left[E_2^\star\left\{
\frac{ I( W_i >C_i) \bq_1(C_i,Y_i,\Z_i,\bo_j)K_{h_2}^{(m_2)}(\Z_i - \z_j)}{f_{\Z}(\Z_i)}\mid W_i, \Z_i;\bb_0\right\}\right](W_i,\Z_i;\bb_0)\right.\\
 &&+\frac{1-\Delta_i}{E\{I(X_i> W_i)|W_i,Y_i,\Z_i\}} E\left(I(X_i>W_i)\right.\\
&&\left.\left.\times \L_0^{\star -1}\left[E_2^\star\left\{
\frac{ I(X_i>C_i) \bq_1(C_i,Y_i, \Z_i,\bo_j)K_{h_2}^{(m_2)}(\Z_i - \z_j)}{f_{\Z} (\Z_i)}\mid  X_i, \Z_i;\bb_0\right\}\right](X_i,\Z_i;\bb_0)\mid W_i,Y_i,\Z_i\right)\mid \bo_j\right\}\\
&&+ o_p(1)\\
&=&  -n^{-1/2} \sum_{j=1}^n E\left( \L_0^{\star -1}\left[E_2^\star\left\{
\frac{ I( X_i >C_i) \bv_1(C_i,Y_i,\Z_i,\bo_j)K_{h_1}^{(m_1)}(\Z_i - \z_j)}{f_{\Z}(\Z_i)}\mid X_i, \Z_i;\bb_0\right\}\right](X_i,\Z_i;\bb_0)\mid \bo_j\right)\\
&&-n^{-1/2} \sum_{j=1}^n E\left( \L_0^{\star -1}\left[E_2^\star\left\{
\frac{ I( X_i >C_i) \bq_1(C_i,Y_i,\Z_i,\bo_j)K_{h_2}^{(m_2)}(\Z_i - \z_j)}{f_{\Z}(\Z_i)}\mid X_i, \Z_i;\bb_0\right\}\right](X_i,\Z_i;\bb_0)\mid \bo_j\right)\\
&=& -n^{-1/2} \sum_{j=1}^n E\left\{ \L_0^{\star -1}\left(E_2^\star\left[
 I( X_i >C_i) \{\bv_1(C_i,Y_i,\z_j,\bo_j) + \bq_1(C_i,Y_i,\z_j,\bo_j)\}\mid X_i, \Z_i = \z_j;\bb_0\right]\right)\right.\\
 &&\left.(X_i,\z_j;\bb_0)\mid \Z_i = \z_j,\bo_j\right\} + o_p(1),
 \ese
   where the last equality holds since $O_p(n^{1/2}h_1^{m_1})$ and $O_p(n^{1/2}h_2^{m_2})$ are $o_p(1)$ under Condition \ref{con:nkerbw}.
   Applying \eqref{eq:n37}, we can further simplify (C) to be
\be
{\rm (C)}
&=&-n^{-1/2} \sum_{j=1}^n E_2^\star\left[
 I( X_i >C_i) \{\bv_1(C_i,Y_i,\z_j,\bo_j) + \bq_1(C_i,Y_i,\z_j,\bo_j)\}\right.\n\\
&&\left.\mid \Z_i = \z_j,\bo_j;\bb_0\right]+ o_p(1)\n\\
&=&-n^{-1/2} \sum_{j=1}^n E_2^\star[
 S_{X|Y,\Z}(C_i,Y_i,\z_j) \{\bv_1(C_i,Y_i,\z_j,\bo_j) + \bq_1(C_i,Y_i,\z_j,\bo_j)\}\n\\
&& \mid \Z_i = \z_j,\bo_j;\bb_0]+ o_p(1).
 \label{eq:n38}
\ee
Noting that
\bse
\bv_1(c,y,\z,\bo_j)&=& -  E\left[ \frac{ I(X>c) \xi_C (w_j, \delta_j, X,y_j,\z)f_{Y|X,\Z}(y_j,X,\z)}{S_{C|Y,\Z} (X,y_j,\z)S_{X|Y,\Z} (c,y,\z)f_{Y|\Z}(y_j,\z)}\right.\\
&&\times \left.\{\S\eff^\star(y,X,1,\z;\bb_0) -\S\eff^\star(y,c,0,\z;\bb_0) \} \mid y,\z,\bo_j\right], \\
\bq_1(c,y,\z,\bo_j)&=&\frac{\delta_j I(w_j>c)f_{Y|X,\Z}(y,w_j,\z)}{S_{C|Y,\Z} (w_j,y_j,\z)S_{X|Y,\Z} (c,y,\z)f_{Y|\Z}(y,\z)} \{\S\eff^\star(y,w_j,1,\z;\bb_0)- \S\eff^\star(y,c,0,\z;\bb_0)\},
\ese
and inserting 
\bse
\bh_{1{\rm s}2}^\star(y_j,w_j, \delta_j, \z_j)&=&- E_2^\star\left\{
 S_{X|Y,\Z}(C_i,Y_i,\z_j) \bv_1(C_i,Y_i,\z_j,\bo_j) \mid \Z_i = \z_j,\bo_j;\bb_0\right\},\\
 \bh_{1{\rm k}2}^\star(y_j,w_j, \delta_j, \z_j)&=&- E_2^\star\left\{
  S_{X|Y,\Z}(C_i,Y_i,\z_j) \bq_1(C_i,Y_i,\z_j,\bo_j) \mid \Z_i = \z_j,\bo_j;\bb_0\right\}
\ese
into \eqref{eq:n38},
we obtain the desired result of Lemma \ref{lem:l6}.
\qed

\begin{Lem} \label{lem:l7}
Let $\o_j\equiv (\delta_j,w_j,y_j,\z_j)$. Under Conditions \ref{con:nker}, \ref{con:nkerbw'}, \ref{con:nbddaway}, and 
\ref{con:nderivbddaway}, 
    \bse
&&n^{-1/2} \sumi \frac{\partial \S\eff^*(y_i, w_i, \delta_i,
  \z_i; \bb_0, E_1^*, \ba_0^*)}{\partial \ba}
(\wh \ba^* - \ba_0^*)\\
&=& n^{-1/2}\sum_{j=1}^n \{\bh_{2{\rm s}}^*(y_j,w_j, \delta_j, \z_j) + \bh_{2{\rm k}}^*(y_j,w_j, \delta_j, \z_j)\} + o_p(1),
\ese
where $\bh_{2{\rm s}}^*(y_j,w_j, \delta_j, \z_j)$ and $\bh_{2{\rm k}}^*(y_j,w_j, \delta_j, \z_j)$ are defined as
\be\label{eq:h2v}
&&\bh_{2{\rm s}}^*(y_j,w_j, \delta_j, \z_j)\\
&\equiv& E\left( \frac{ \xi_X (w_j, \delta_j, C,y_j,\z_j)}{S_{X|Y,\Z}(C,y_j,\z_j)} [I(X\le C)\S\eff^* (y_j,X,1,\z_j;\bb_0)\right.\n\\
&&\left.+ I(X>C)\S\eff^* (y_j,C,0,\z_j;\bb_0)\right.\n\\
&& - E\{I(X\le C)\S\eff^* (y_j,X,1,\z_j;\bb_0)\n\\
&&+ I(X>C)\S\eff^* (y_j,C,0,\z_j;\bb_0)\mid X, Y = y_j,\Z = \z_j\}]\n\\
&&\left.\mid Y = y_j,\Z = \z_j,\bo_j\right),\n
\ee
and
\be\label{eq:h2q}
&&\bh_{2{\rm k}}^*(y_j,w_j, \delta_j, \z_j)\\
&\equiv& -\frac{1-\delta_j}{S_{X|Y,\Z} (w_j,y_j,\z_j)} E\left[ I(X \le w_j)\S\eff^*(y_j,X,1,\z_j;\bb_0) \right.\n\\
&&+ I(X > w_j) \bS\eff^*(y_j,w_j,0,\z_j;\bb_0)\n\\
&& - E\{I(X\le C)\S\eff^* (y_j,X,1,\z_j;\bb_0) \n\\
&&+ I(X>C)\S\eff^* (y_j,C,0,\z_j;\bb_0)\mid X, Y = y_j,\Z = \z_j\}\left.\mid Y=y_j,\Z = \z_j,\bo_j\right].\n
\ee
\end{Lem}
Proof. 
We prove Lemma \ref{lem:l7} using a method similar to that used in the proof of Lemma \ref{lem:l6}.
 Let
  $\L_0^*(\cdot) = \L(\cdot; E_1^*, E_{20})$ and $\wh
  \L^*(\cdot) = \L(\cdot; E_1^*, \wh E_2)$.
Then
\bse
&& (\wh \ba^* - \ba_0^*)(x,\bz;\bb_0)\\
&=& \{\wh \L^{* -1}(\wh\bc^*) - \L_0^{*-1}(\bc_0^*)\} (x,\bz;\bb_0)\\
&=& - \L_0^{* -1}(\wh \L^* - \L_0^*) \L_0^{* -1}(\bc_0^*) (x,\bz;\bb_0) + \L_0^{* -1}(\wh\bc^* - \bc_0^*) (x,\bz;\bb_0) + o_p(n^{-1/2})\\
&=& - \L_0^{* -1}\left\{(\wh E_2-E_{20})\{I(x\le C)|x,\z;\bb\} \a_0^* (x,\bz;\bb_0)\right.\\
&&\left.+(\wh E_2-E_{20})\left(\left.I(x>C)\left[\frac{E_1^*\{I(X>C)\ba_0^*(X,\bz;\bb_0)
|C,Y,\z;\bb_0\}}{E_1^*\{I(X>C)|C,Y,\z;\bb_0\}}\right]\right|x,\bz;\bb_0\right)\right\} (x,\bz;\bb_0) \\
&&+ \L_0^{* -1}\left\{(\wh E_2-E_{20})\{I(x\le C)\S_\bb^F(Y,x,\bz;\bb_0)|x,\z;\bb_0\}\right.\\
&&\left.+(\wh E_2-E_{20})\left(\left.I(x>C)\left[\frac{E_1^*\{I(X>C)\S_\bb^F(Y,X,\bz;\bb_0)
|C,Y,\z;\bb_0\}}{E_1^*\{I(X>C)|C,Y,\z;\bb_0\}}\right]\right|x,\bz;\bb_0\right)\right\} (x,\bz;\bb_0)\\
&&+ o_p(n^{-1/2})\\
&=& \L_0^{* -1}[(\wh E_2-E_{20})\{\bt(Y,x,C,\z) \mid x,\bz;\bb_0\}] (x,\bz;\bb_0)+ o_p(n^{-1/2}),
\ese
uniformly for $x$ and $\z$, where $\bt(y,x,c,\z)$ denotes
\bse
\bt(y,x,c,\z)&\equiv& I(x\le c)\{\S_\bb^F(y,x,\bz;\bb_0) - \ba_0^*(x,\bz;\bb_0)\}\\
&&+I(x>c)\frac{E_1^*[I(X>c)\{\S_\bb^F(y,X,\bz;\bb_0) - \ba_0^*(X,\bz;\bb_0)\}
|y,\z;\bb_0]}{E_1^*\{I(X>c)|y,\z;\bb_0\}}.
\ese
Here, the second equality holds by Taylor's theorem and Lemma
\ref{lem:l3}. 
Since both $\|\wh S_{X|Y,\Z} - S_{X|Y,\Z}\|_\infty^2$ and $\|\wh
f_{\Delta, W|Y,\Z} - f_{\Delta, W|Y,\Z}\|_\infty^2$ are
$o_p(n^{-1/2})$ by Lemma \ref{lem:l2} and Condition \ref{con:nkerbw},
we have 
\be \label{eq:n30}
&&(\wh \ba^* - \ba_0^*)(x,\bz;\bb_0)\\
 &&= \L_0^{* -1}\{\br_{2{\rm s}}(x,\z;\bb_0) + \br_{2{\rm k}}(x,\z;\bb_0)\}(x,\z;\bb_0)+o_p(n^{-1/2}),\n
\ee
where $\br_{2{\rm s}}(x,\z;\bb_0)$ and $\br_{2{\rm k}}(x,\z;\bb_0)$ are defined as
\bse
\br_{2{\rm s}}(x,\z;\bb_0) &\equiv& \frac{\partial}{\partial S_{X|Y,\Z}}E_2\{\bt(Y,x,C,\z)\mid x,\bz;\bb_0,  S_{X|Y,\Z}, f_{\Delta, W|Y,\Z}\} (\wh S_{X|Y,\Z} - S_{X|Y,\Z}), \\
\br_{2{\rm k}}(x,\z;\bb_0) &\equiv& \frac{\partial}{\partial f_{\Delta, W|Y,\Z}}E_2\{\bt(Y,x,C,\z)\mid x,\bz;\bb_0,  S_{X|Y,\Z}, f_{\Delta, W|Y,\Z}\} (\wh f_{\Delta, W|Y,\Z} - f_{\Delta, W|Y,\Z}).
\ese 
First, $\br_{2{\rm s}}(x,\z;\bb_0)$ can be expanded, using the expression in \eqref{eq:d5}, as 
\bse
&&\br_{2{\rm s}}(x,\z;\bb_0)\\
&=& -E\left([\bt(Y,x,C,\z) - E\{\bt(Y,x,C,\z)\mid Y,\z\}]\frac{\wh S_{X|Y,\Z}(C,Y,\z) - S_{X|Y,\Z}(C,Y,\z)}{S_{X|Y,\Z}(C,Y,\z)}\mid x,\bz\right)\\
&=& -E\left([\bt(Y,x,C,\z) - E\{\bt(Y,x,C,\z)\mid Y,\z\}]\frac{\sum_{j=1}^n \xi_X (w_j, \delta_j, C,Y,\z) K_{h_1}^{(m_1)}(Y - y_j, \z - \z_j)}{S_{X|Y,\Z}(C,Y,\z)\sum_{k=1}^n K_{h_1}^{(m_1)}(Y - y_k, \z - \z_k)}\mid x,\bz\right)\\
&&+O_p\{(\log n)^{3/4} n^{-3/4} h_1^{-3d/4} + h_1^{m_1}\},
\ese
where the third equality holds by \eqref{eq:n10}. Note that 
$O_p\{(\log n)^{3/4} n^{-3/4} h_1^{-3d/4} + h_1^{m_1}\}$ is $o_p(n^{-1/2})$
under \ref{con:nkerbw'}. 
Then
\bse
&&\br_{2{\rm s}}(x,\z;\bb_0)\\
&=&-E\left([\bt(Y,x,C,\z) - E\{\bt(Y,x,C,\z)\mid Y,\z\}]\frac{E \{\xi_X (W_j, \Delta_j, C,Y,\z)\mid Y_j = Y,\Z_j=\z\}}{S_{X|Y,\Z}(C,Y,\z)}\mid x,\bz\right)\\
&&-n^{-1} E\left([\bt(Y,x,C,\z) - E\{\bt(Y,x,C,\z)\mid Y,\z\}]\frac{\sum_{j=1}^n \xi_X (w_j, \delta_j, C,Y,\z) K_{h_1}^{(m_1)}(Y - y_j, \z - \z_j)}{S_{X|Y,\Z}(C,Y,\z)f_{Y,\Z}(Y,\z)}\mid x,\bz\right)\\
&&+n^{-1}E\left([\bt(Y,x,C,\z) - E\{\bt(Y,x,C,\z)\mid Y,\z\}]\right.\\
&&\left.\times \frac{E \{\xi_X (W_j, \Delta_j, C,Y,\z)\mid Y_j = Y,\Z_j=\z\}\sum_{k=1}^n K_{h_1}^{(m_1)}(Y - y_k, \z - \z_k)}{S_{X|Y,\Z}(C,Y,\z)f_{Y,\Z}(Y,\z)}\mid x,\bz\right)\\
&&+ O_p\{(\log n) n^{-1}h_1^{-d} + h_1^{2m_1}\} + o_p(n^{-1/2}),
\ese
uniformly for $(x,\z)$ by Taylor's theorem. Using \eqref{eq:n40} and noting that $O_p\{(\log n) n^{-1}h_1^{-d} + h_1^{2m_1}\}$ is $o_p(n^{-1/2})$ under Condition \ref{con:nkerbw}, we obtain
\be\label{eq:n28}
&& \br_{2{\rm s}}(x,\z;\bb_0) \\
&&= n^{-1}\sum_{j=1}^n E\left\{\frac{\bv_2(Y,x,C,\z,\bo_j) K_{h_1}^{(m_1)}(Y - y_j, \z - \z_j)}{f_{Y,\Z}(Y,\z)}\mid x,\bz\right\}+o_p(n^{-1/2}),\n
\ee
where $\bv_2(y,x,c,\z,\bo_j)$ is defined as
\be \label{eq:n35}
&&\bv_2(y,x,c,\z,\bo_j)\\
&&\equiv -\frac{[\bt(y,x,c,\z) - E\{\bt(y,x,C,\z)\mid y,\z\}] \xi_X (w_j, \delta_j, c,y,\z)}{S_{X|Y,\Z}(c,y,\z)}.\n
\ee

Next, using \eqref{eq:d6}, $\br_{2{\rm k}}(x,\z;\bb_0)$ can be written as 
\bse
&&\br_{2{\rm k}}(x,\z;\bb_0)\\
&=& E[E_2 \{\bt(Y,x,C,\z)\mid Y,\z; S_{X|Y,\Z}, (\wh f_{\Delta, W|Y,\Z} - f_{\Delta, W|Y,\Z})\} \n\\
 &&- E \{\bt(Y,x,C,\z)\mid Y,\z\}  E_2 \{1 \mid Y,\z; S_{X|Y,\Z}, (\wh f_{\Delta, W|Y,\Z} - f_{\Delta, W|Y,\Z})\}\mid x, \z; \bb_0] \\
&=&n^{-1} E\left[\sum_{j=1}^n\frac{ (1-\delta_j) \bt(y_j,x,w_j,\z_j)K_{h_3}^{(m_3)}(Y - y_j, \z - \z_j)}{S_{X|Y,\Z} (w_j,y_j,\z_j)f_{Y,\Z}(Y,\z)}\right.\\
&&\left. - E\{\bt(Y,x,C,\z)\mid Y,\z\}\sum_{j=1}^n\frac{ (1-\delta_j)K_{h_3}^{(m_3)}(Y - y_j, \z - \z_j)}{S_{X|Y,\Z} (w_j,y_j,\z_j)f_{Y,\Z}(Y,\z)}\mid x,\bz;\bb_0\right]\\
&=&n^{-1}\sum_{j=1}^n E\left(\frac{ (1-\delta_j) [\bt(y_j,x,w_j,\z_j)-E\{\bt(Y,x,C,\z)\mid Y,\z\}]K_{h_3}^{(m_3)}(Y - y_j, \z - \z_j)}{S_{X|Y,\Z} (w_j,y_j,\z_j)f_{Y,\Z}(Y,\z)}\mid x,\bz\right),
\ese
where the second equality holds since $E\{\bt(Y,x,C,\z)\mid
x,\z\} = \0$ by the definition of $\ba_0^*(x,\bz;\bb_0)$.
\be\label{eq:n29}
\br_{2{\rm k}}(x,\z;\bb_0)&=& n^{-1} \sum_{j=1}^n E\left\{\frac{\bq_2(Y,x,\z,\bo_j) K_{h_3}^{(m_3)}(Y - y_j, \z - \z_j)}{f_{Y,\Z}(Y, \z )}\mid x,\bz\right\} \\
&&+ o_p(n^{-1/2}),\n
\ee
uniformly for $(x,\z)$, where $\bq_2(y,x,\z,\bo_j)$ is defined as 
\be\label{eq:n36}
\bq_2(y,x,\z,\bo_j)&\equiv&\frac{(1-\delta_j) [\bt(y_j,x,w_j,\z_j)-E\{\bt(y,x,C,\z)\mid y,\z\}]}{S_{X|Y,\Z} (w_j,y_j,\z_j)}. 
\ee
Now, denote
\bse
{\rm (D)}&=&n^{-1/2}\sumi \frac{\partial \S\eff^* (y_i,w_i,\delta_i, \z_i ; \bb_0, E_1^*, \ba_0^*)}{\partial \ba} (\wh \ba^* - \ba_0^*).
\ese
Then (D) can be transformed into
\bse
{\rm (D)} &=& -n^{-1/2}\sumi \left(\delta_i \L_0^{* -1}\{\br_{2{\rm s}}(w_i,\z_i;\bb_0) + \br_{2{\rm k}}(w_i,\z_i;\bb_0)\}(w_i,\z_i;\bb_0)\right.\\
&&\left.+ (1-\delta_i) \frac{ E_1^*[I(X_i>w_i)\L_0^{* -1}\{\br_{2{\rm s}}(X_i,\z_i;\bb_0) + \br_{2{\rm k}}(X_i,\z_i;\bb_0)\}(X_i,\z_i;\bb_0) |y_i,\z_i;\bb_0]}{E_1^*\{I(X_i> w_i)|y_i,\z_i;\bb_0\}}\right) + o_p(1)\\
&=& - n^{-3/2} \sumi \sum_{j=1}^n \left\{\delta_i \L_0^{*-1}\left[ E\left\{\frac{ \bv_2(Y_i,w_i,C_i,\z_i,\bo_j)K_{h_1}^{(m_1)}(Y_i - y_j, \z_i - \z_j)}{f_{Y,\Z}(Y_i, \z_i)}\mid w_i,\bz_i\right\}\right](w_i,\z_i;\bb_0)\right.\\
&&+ \frac{1-\delta_i}{E_1^*\{I(X_i> w_i)|y_i,\z_i;\bb_0\}} E_1^*\left(I(X_i>w_i)\right.\\
&&\left.\left.\times \L_0^{* -1}\left[E\left\{\frac{ \bv_2(Y_i,X_i,C_i,\z_i,\bo_j)K_{h_1}^{(m_1)}(Y_i - y_j, \z_i - \z_j)}{f_{Y,\Z}(Y_i, \z_i )}\mid X_i,\bz_i\right\}\right](X_i,\z_i;\bb_0) |y_i,\z_i;\bb_0\right)\right\}\\
&&- n^{-3/2} \sumi \sum_{j=1}^n \left\{\delta_i \L_0^{*-1}\left[ E\left\{\frac{ \bq_2(Y_i,w_i,\bz_i,\bo_j)K_{h_3}^{(m_3)}(Y_i - y_j, \z_i - \z_j)}{f_{Y,\Z}(Y_i, \z_i)}\mid w_i,\bz_i\right\}\right](w_i,\z_i;\bb_0)\right.\\
&&+ \frac{1-\delta_i}{E_1^*\{I(X_i> w_i)|y_i,\z_i;\bb_0\}} E_1^*\left(I(X_i>w_i)\right.\\
&&\left.\left.\times \L_0^{* -1}\left[E\left\{\frac{\bq_2(Y_i,X_i,\bz_i,\bo_j)K_{h_3}^{(m_3)}(Y_i - y_j, \z_i - \z_j)}{f_{Y,\Z}(Y_i, \z_i )}\mid X_i,\bz_i\right\}\right](X_i,\z_i;\bb_0) |y_i,\z_i;\bb_0\right)\right\}\\
&&+ o_p(1),
\ese
where the first line follows by \eqref{eq:n25} and \eqref{eq:n30}, and
the second equality holds by \eqref{eq:n28} and \eqref{eq:n29}. Note that
\bse
&&E\left\{\frac{\bv_2(Y,x,C,\z,\bO_j)K_{h_1}^{(m_1)}(Y - Y_j, \z - \Z_j)}{f_{Y,\Z}(Y, \z)} \mid x,\z\right\}\\
&=& -E\left[\frac{\bt(Y,x,C,\z) - E\{\bt(Y,x,C,\z)\mid Y,\z\} }{S_{X|Y,\Z}(C,Y,\z)}E\{\xi_X (W_j, \Delta_j, C,Y,\z)\mid x,C, Y_j = Y, \Z_j = \z\}\mid x,\z\right]\\
&&+ O_p(h_1^{m_1})\\
&=& O_p(h_1^{m_1})
\ese
from  \eqref{eq:n40}
and that
\bse
&&E\left\{\frac{\bq_2(Y,x,z,\bO_j)K_{h_3}^{(m_3)}(Y - Y_j, \z - \Z_j)}{f_{Y,\Z}(Y, \z)} \mid x,\z\right\}\\
&=& E\left\{E\left(\frac{(1-\Delta_j) [\bt(Y_j,x,W_j,\Z_j)-E\{\bt(Y,x,C,\z)\mid Y,\z\}]}{S_{X|Y,\Z} (W_j,Y_j,\Z_j)}\mid x,C, Y_j = Y, \Z_j = \z\right)\mid x,\z\right\}+ O_p(h_3^{m_3})\\
&=& E\left[E\{ \bt(Y_j,x,C_j,\Z_j)\mid Y_j = Y, \Z_j = \z\} -E\{\bt(Y,x,C,\z)\mid Y,\z\} \mid x,\z\right] + O_p(h_3^{m_3})\\
&=& O_p(h_3^{m_3}).
\ese
Here, $O_p(h_1^{m_1})$ and $O_p(h_3^{m_3})$ are $o_p(n^{-1/2})$ under Condition \ref{con:nkerbw'}. Applying the U-statistic argument from \eqref{eq:n19} yields
\bse
{\rm (D)} &=& - n^{-1/2} \sum_{j=1}^n E\left\{\Delta_i \L_0^{*-1}\left[ E\left\{\frac{ \bv_2(Y_i,W_i,C_i,\Z_i,\bo_j)K_{h_1}^{(m_1)}(Y_i - y_j, \Z_i - \z_j)}{f_{Y,\Z}(Y_i, \Z_i)}\mid W_i,\bZ_i\right\}\right]\right.\\
&&(W_i,\Z_i;\bb_0)+ \frac{1-\Delta_i}{E_1^*\{I(X_i> W_i)|W_i,Y_i,\Z_i;\bb_0\}} E_1^*\left(I(X_i>W_i)\right.\\
&&\left.\left.\times \L_0^{* -1}\left[ E\left\{\frac{ \bv_2(Y_i,X_i,C_i,\Z_i,\bo_j)K_{h_1}^{(m_1)}(Y_i - y_j, \Z_i - \z_j)}{f_{Y,\Z}(Y_i, \Z_i)}\mid X_i,\bZ_i\right\}\right](X_i,\Z_i;\bb_0)\right.\right.\\
&&\left.\left.|W_i,Y_i,\Z_i;\bb_0\right)\mid \bo_j\right\}\\
&& - n^{-1/2} \sum_{j=1}^n E\left\{\Delta_i \L_0^{*-1}\left[ E\left\{\frac{ \bq_2(Y_i,W_i,\bZ_i,\bo_j)K_{h_3}^{(m_3)}(Y_i - y_j, \Z_i - \z_j)}{f_{Y,\Z}(Y_i, \Z_i)}\mid W_i,\bZ_i\right\}\right]\right.\\
&&(W_i,\Z_i;\bb_0)+ \frac{1-\Delta_i}{E_1^*\{I(X_i> W_i)|W_i,Y_i,\Z_i;\bb_0\}} E_1^*\left(I(X_i>W_i)\right.\\
&&\left.\left.\times \L_0^{* -1}\left[ E\left\{\frac{\bq_2(Y_i,X_i,\bZ_i,\bo_j)K_{h_3}^{(m_3)}(Y_i - y_j, \Z_i - \z_j)}{f_{Y,\Z}(Y_i, \Z_i)}\mid X_i,\bZ_i\right\}\right](X_i,\Z_i;\bb_0)\right.\right.\\
&&\left.\left.|W_i,Y_i,\Z_i;\bb_0\right)\mid \bo_j\right\}+ o_p(1),
\ese
and by the definition of $\L_0^*$, we can further simplify (D) to
\bse
{\rm (D)} &=&  - n^{-1/2} \sum_{j=1}^n E\left[E\left\{\frac{ \bv_2(Y_i,X_i,C_i,\Z_i,\bo_j)K_{h_1}^{(m_1)}(Y_i - y_j, \Z_i - \z_j)}{f_{Y,\Z}(Y_i, \Z_i)}\mid X_i,\bZ_i\right\}\mid \bo_j\right] \\
&&- n^{-1/2} \sum_{j=1}^n E\left[E\left\{\frac{\bq_2(Y_i,X_i,\bZ_i,\bo_j)K_{h_3}^{(m_3)}(Y_i - y_j, \Z_i - \z_j)}{f_{Y,\Z}(Y_i, \Z_i)}\mid X_i,\bZ_i\right\}\mid \bo_j\right] +o_p(1)\\
&=&  - n^{-1/2} \sum_{j=1}^n E\left\{  \bv_2(Y_i,X_i,C_i,\Z_i,\bo_j) +\bq_2(Y_i,X_i,\bZ_i,\bo_j) \mid Y_i = y_j,\Z_i = \z_j,\bo_j\right\} +o_p(1).
\ese
Noting that
\bse
\bt(y,x,c,\z)&=& I(x\le c) \S\eff^*(y,x,1,\z;\bb_0)+I(x>c)\S\eff^*(y,c,0,\z;\bb_0),
\ese
and combining these results with \eqref{eq:n35} and \eqref{eq:n36}, we obtain
\bse
\bh_{2{\rm s}}^*(y_j,w_j, \delta_j, \z_j)&=&  -E\left\{  \bv_2(Y_i,X_i,C_i,\Z_i,\bo_j) \mid Y_i = y_j,\Z_i = \z_j,\bo_j\right\},\\
\bh_{2{\rm k}}^*(y_j,w_j, \delta_j, \z_j) &=& -E\left\{ \bq_2(Y_i,X_i,\bZ_i,\bo_j)\mid Y_i = y_j,\Z_i = \z_j,\bo_j\right\}.
\ese
Substituting the relationships from the last expression of (D), we obtain the result of Lemma \ref{lem:l7}.
\qed

Let $\bh_{1{\rm s}}^\star(y_j,w_j, \delta_j, \z_j)$ and $\bh_{1{\rm k}}^\star(y_j,w_j, \delta_j, \z_j)$ be
\bse
\bh_{1{\rm s}}^\star(y_j,w_j, \delta_j, \z_j)
&\equiv& \bh_{1{\rm s}1}^\star(y_j,w_j, \delta_j, \z_j) + \bh_{1{\rm s}2}^\star(y_j,w_j, \delta_j, \z_j),\\
\bh_{1{\rm k}}^\star(y_j,w_j, \delta_j, \z_j)&\equiv& \bh_{1{\rm k}1}^\star(y_j,w_j, \delta_j, \z_j) + \bh_{1{\rm k}2}^\star(y_j,w_j, \delta_j, \z_j),
\ese
where $\bh_{1{\rm s}1}^\star(y_j,w_j, \delta_j, \z_j)$ and $\bh_{1{\rm k}1}^\star(y_j,w_j, \delta_j, \z_j)$ are defined in Lemma \ref{lem:l5} as \eqref{eq:h1v1} and \eqref{eq:h1q1}, respectively, and $\bh_{1{\rm s}2}^\star(y_j,w_j, \delta_j, \z_j)$ and $\bh_{1{\rm k}2}^\star(y_j,w_j, \delta_j, \z_j)$ are defined in Lemma \ref{lem:l6} as \eqref{eq:h1v2} and \eqref{eq:h1q2}, respectively. Furthermore, let $\bh_{2{\rm s}}^*(y_j,w_j, \delta_j, \z_j)$ and $\bh_{2{\rm k}}^*(y_j,w_j, \delta_j, \z_j)$ be as defined in \eqref{eq:h2v} and \eqref{eq:h2q}, respectively.

From Lemmas \ref{lem:l5} and \ref{lem:l6}, we see that
\bse
&&n^{-1/2} \sumi \frac{\partial \S\eff^\star\{y_i, w_i, \delta_i,
  \z_i; \bb_0, E_1(\cdot; S_{C|Y,\Z}, f_{\Delta, W,Y|\Z}), \ba_0^\star\}}{\partial S_{C|Y,\Z}}
(\wh S_{C|Y,\Z} - S_{C|Y,\Z})\\
&&+n^{-1/2} \sumi \frac{\partial \S\eff^\star\{y_i, w_i, \delta_i,
  \z_i; \bb_0, E_1(\cdot; S_{C|Y,\Z}, f_{\Delta, W,Y|\Z}), \ba_0^\star\}}{\partial f_{\Delta, W,Y|\Z}}(\wh f_{\Delta, W,Y|\Z} - f_{\Delta, W,Y|\Z})\\
&&+n^{-1/2} \sumi \frac{\partial \S\eff^\star(y_i, w_i, \delta_i,
  \z_i; \bb_0, E_{10}, \ba_0^\star)}{\partial \ba}
(\wh \ba^\star - \ba_0^\star)\\
&=& n^{-1/2} \sum_{j=1}^n \{\bh_{1{\rm s}}^\star(y_j,w_j, \delta_j, \z_j) + \bh_{1{\rm k}}^\star(y_j,w_j, \delta_j, \z_j)\} + o_p(1),
\ese
and, by Lemma \ref{lem:l7}, that
 \bse
&&n^{-1/2} \sumi \frac{\partial \S\eff^*(y_i, w_i, \delta_i,
  \z_i; \bb_0, E_1^*, \ba_0^*)}{\partial \ba}
(\wh \ba^* - \ba_0^*)\\
&=& n^{-1/2} \sum_{j=1}^n\{\bh_{2{\rm s}}^*(y_j,w_j, \delta_j, \z_j) + \bh_{2{\rm k}}^*(y_j,w_j, \delta_j, \z_j)\} + o_p(1). 
\ese

\subsubsection{Proof of Proposition \ref{pro:5}}\label{sec:pro5pf}

When $E_2^\star = E_{20}$, we have $\bh_{1{\rm s}1}(y_j,w_j, \delta_j, \z_j) = -\bh_{1{\rm s}2}(y_j,w_j, \delta_j, \z_j)$ and $\bh_{1{\rm k}1}(y_j,w_j, \delta_j, \z_j) = -\bh_{1{\rm k}2}(y_j,w_j, \delta_j, \z_j)$. As a result, $\h_{1{\rm s}} = \h_{1{\rm k}} = 0$. 
 
Now, we will show $\h_{2{\rm s}} = \h_{2{\rm k}} = 0$. Suppose $E_1^*= E_{10}$, and consider
$\bh_{2{\rm s}}(y_j,w_j, \delta_j, \z_j)$ and $\bh_{2{\rm k}}(y_j,w_j,
\delta_j, \z_j)$. We first utilize the fact that 
 \be\label{eq:n43}
&& E\{I(X\le c)\S\eff (y,X,1,\z;\bb_0) + I(X>c)\S\eff (y,c,0,\z;\bb_0)\mid c,y,\z\}\\
&=& E\left[  I(X\le c) \{ \S_\bb^F(y,X,\z;\bb_0) - \ba_0(X,\z;\bb_0)\}\mid y,\z\right]\n\\
&&+E\{I(X> c)\mid y,\z\}\frac{E[I(X> c)\{\S_\bb^F(y,X,\z;\bb_0) - \ba_0(X,\z;\bb_0)\}\mid y,\z]}{E\{I(X> c)\mid y,\z\}}\n\\
&=& E\{ \S_\bb^F(y,X,\z;\bb_0) - \ba_0(X,\z;\bb_0)\mid y,\z\}.\n
\ee 
Then $\bh_{2{\rm s}}(y_j,w_j, \delta_j, \z_j) = {\rm (a)} - {\rm (b)}$, where {\rm (a)} and {\rm (b)} are defined as
\bse
{\rm(a)} &\equiv& E\left[ \frac{ \xi_X (w_j, \delta_j, C,y_j,\z_j)}{S_{X|Y,\Z}(C,y_j,\z_j)} \{I(X\le C)\S\eff (y_j,X,1,\z_j;\bb_0) + I(X>C)\S\eff (y_j,C,0,\z_j;\bb_0)\}\right.\\
&&\left.\mid Y = y_j,\Z = \z_j,\bo_j\right],\\
{\rm (b)}&\equiv& E\left[ \frac{ \xi_X (w_j, \delta_j, C,y_j,\z_j)}{S_{X|Y,\Z}(C,y_j,\z_j)} E\{I(X\le C)\S\eff (y_j,X,1,\z_j;\bb_0)\right.\\
&& \left. + I(X>C)\S\eff (y_j,C,0,\z_j;\bb_0)\mid X, Y = y_j,\Z = \z_j\}\mid Y = y_j,\Z = \z_j,\bo_j\right].
\ese 
It is enough to show ${\rm(a)} = {\rm(b)}$. This result  holds since
\bse
&&{\rm(a)}\\
&=&  E\left[ \frac{ \xi_X (w_j, \delta_j, C,y_j,\z_j)}{S_{X|Y,\Z}(C,y_j,\z_j)} E \{ \S_\bb^F(y_j,X,\z_j;\bb_0) - \ba_0(X,\z_j;\bb_0)\mid Y=y_j,\Z=\z_j\}\mid Y = y_j,\Z = \z_j,\bo_j\right]\\
 &=& E\left\{ \frac{ \xi_X (w_j, \delta_j, C,y_j,\z_j)}{S_{X|Y,\Z}(C,y_j,\z_j)} \mid Y = y_j,\Z = \z_j,\bo_j\right\}E\{ \S_\bb^F(y_j,X,\z_j;\bb_0) - \ba_0(X,\z_j;\bb_0)\mid Y = y_j,\Z = \z_j\},
\ese
where the first equality follows by \eqref{eq:n43}, and 
\bse
&&{\rm (b)}\\
&=& E\left[E\left\{ \frac{ \xi_X (w_j, \delta_j, C,y_j,\z_j)}{S_{X|Y,\Z}(C,y_j,\z_j)}\mid X, Y = y_j,\Z = \z_j,\bo_j\right\}E\{I(X\le C)\S\eff (y_j,X,1,\z_j;\bb_0)\right.\\
&&\left.+ I(X>C)\S\eff (y_j,C,0,\z_j;\bb_0)\mid X, Y = y_j,\Z = \z_j\}\mid Y = y_j,\Z = \z_j,\bo_j\right]\\
&=& E\left[E\left\{ \frac{ \xi_X (w_j, \delta_j, C,y_j,\z_j)}{S_{X|Y,\Z}(C,y_j,\z_j)}\mid Y = y_j,\Z = \z_j,\bo_j\right\}E\{I(X\le C)\S\eff (y_j,X,1,\z_j;\bb_0)\right.\\
&&\left.+ I(X>C)\S\eff (y_j,C,0,\z_j;\bb_0)\mid X, Y = y_j,\Z = \z_j\}\mid Y = y_j,\Z = \z_j,\bo_j\right]\\
&=& E\left\{ \frac{ \xi_X (w_j, \delta_j, C,y_j,\z_j)}{S_{X|Y,\Z}(C,y_j,\z_j)}\mid Y = y_j,\Z = \z_j,\bo_j\right\}\\
&&\times E\{I(X\le C)\S\eff (y_j,X,1,\z_j;\bb_0)+ I(X>C)\S\eff (y_j,C,0,\z_j;\bb_0)\mid Y = y_j,\Z = \z_j\}\\
 &=& E\left\{ \frac{ \xi_X (w_j, \delta_j, C,y_j,\z_j)}{S_{X|Y,\Z}(C,y_j,\z_j)} \mid Y = y_j,\Z = \z_j,\bo_j\right\}E\{ \S_\bb^F(y_j,X,\z_j;\bb_0) - \ba_0(X,\z_j;\bb_0)\mid Y = y_j,\Z = \z_j\},
\ese
where the last equality holds by \eqref{eq:n43}. Thus,
  $\bh_{2{\rm s}}(y_j,w_j, \delta_j, \z_j)=\0$.  Lastly, $\bh_{2{\rm k}}(y_j,w_j, \delta_j, \z_j) = \0$ since
\bse
&&\bh_{2{\rm k}}(y_j,w_j, \delta_j, \z_j)\\
&=& -\frac{1-\delta_j}{S_{X|Y,\Z} (w_j,y_j,\z_j)} [E\{ I(X \le w_j)\S\eff(y_j,X,1,\z_j;\bb_0) \\
&&+ I(X > w_j) \bS\eff(y_j,w_j,0,\z_j;\bb_0)\mid Y=y_j,\Z = \z_j,\bo_j\}\n\\
&& - E\{I(X\le C)\S\eff (y_j,X,1,\z_j;\bb_0) + I(X>C)\S\eff (y_j,C,0,\z_j;\bb_0)\mid Y = y_j,\Z = \z_j\}]\\
&=& -\frac{1-\delta_j}{S_{X|Y,\Z} (w_j,y_j,\z_j)}[E\{ \S_\bb^F(y_j,X,\z_j;\bb_0) - \ba_0(X,\z_j;\bb_0)\mid Y = y_j,\Z = \z_j\}\\
&&-E\{ \S_\bb^F(y_j,X,\z_j;\bb_0) - \ba_0(X,\z_j;\bb_0)\mid Y = y_j,\Z = \z_j\}]\\
&=&\0,
\ese
where the second equality holds by \eqref{eq:n43}.

\qed

\subsubsection{Main Proof of Theorem \ref{th:nonpar}}\label{sec:mainth2pf}
For notational clarity, we denote $\wh\bb^\star$ and
  $\wh\bb^*$ as the estimator $\wh\bb$ for cases \ref{ncase1} and
  \ref{ncase2}, respectively, and keep the notation $\wh\bb$ only for
  case \ref{ncase3}.
It follows from the proof of
Theorem \ref{th:doublyrobust} that 
\bse
E\{\S\eff^*(Y,W,\Delta, \Z ; \bb_0, E_1^*, E_{20})\}&=&\0,\\
E\{\S\eff^\star(Y,W,\Delta, \Z ; \bb_0, E_{10}, E_2^\star)\}&=&\0.
\ese
Define functions
\bse 
Q_0^\star(\bb) &\equiv& -\|E[\S\eff^{\star} \{Y,W,\Delta, \Z ; \bb, E_{10}, \ba_0^\star(\bb)\}]\|_2^2,\\
\wh Q_n^\star (\bb) &\equiv& -\|n^{-1}\sumi
\S\eff^\star\{y_i, w_i, \delta_i, \z_i;\bb, \wh E_1, \wh\ba^\star(\bb)\}\|_2^2.
\ese 
By the inverse function theorem and Condition \ref{con:nBinv2}, the solution to $E[\S\eff^{\star} \{Y,W,\Delta, \Z ; \bb, E_{10}, \ba_0^\star(\bb)\}]=\0$ is unique in a neighborhood of $\bb = \bb_0$. Therefore, $Q_0^\star(\bb)$ is uniquely maximized at $\bb = \bb_0$ in this neighborhood. We restrict our analysis to a compact subset of this neighborhood.

Under Condition \ref{con:nderivub}, the mapping $\bb \mapsto E[\S\eff^{\star} \{Y,W,\Delta, \Z ; \bb, E_{10},\ba_0^\star(\bb)\}]$ is continuous. Using a similar argument as in the proof of Lemma \ref{lem:l1}, and by Lemmas \ref{lem:l3} and \ref{lem:l4}, we obtain uniform convergence in probability of $n^{-1}\sumi \S\eff^\star\{y_i, w_i, \delta_i, \z_i;\bb, \wh E_1, \wh\ba^\star(\bb)\}$ to $E[\S\eff^{\star} \{Y,W,\Delta, \Z ; \bb, E_{10}, \ba_0^\star(\bb)\}]$ for $\bb \in \Omega$. Since $-\|t\|_2^2$ is continuous as a function of $t$, the uniform convergence implies that $\wh Q_n^\star(\bb)$ converges uniformly in probability to $Q_0^\star(\bb)$.

Since $\wh\bb^\star$ maximizes $\wh Q_n^\star(\bb)$, Theorem 2.1 of \cite{newey1994large} establishes that $\wh\bb^\star$ is consistent for $\bb_0$. Similarly, we can show that both $\wh\bb^*$ and $\wh\bb$ are consistent for $\bb_0$ by applying the same theorem to the function pairs $(Q_0^*(\bb), \wh Q_n^*(\bb))$ and $(Q_0(\bb), \wh Q_n(\bb))$, where
\bse
Q_0^*(\bb) &\equiv& -\|E[\S\eff^{\star} \{Y,W,\Delta, \Z ; \bb, E_1^*, \ba_0^*(\bb)\}]\|_2^2,\\
\wh Q_n^* (\bb) &\equiv& -\|n^{-1}\sumi
\S\eff^*\{y_i, w_i, \delta_i, \z_i;\bb, E_1^*, \wh\ba^*(\bb)\}\|_2^2,\\
Q_0(\bb) &\equiv& -\|E[\S\eff^{\star} \{Y,W,\Delta, \Z ; \bb, E_{10},\ba_0(\bb)\}]\|_2^2,\\
\wh Q_n (\bb) &\equiv& -\|n^{-1}\sumi
\S\eff\{y_i, w_i, \delta_i, \z_i;\bb, \wh E_1, \wh\ba(\bb)\}\|_2^2.
\ese 

First, we consider case \ref{th2case1}. We have
\be\label{eq:n33}
\0 &=& n^{-1/2} \sumi \S\eff^\star(y_i, w_i, \delta_i,
\z_i; \wh\bb^\star, \wh E_1, \wh \ba^\star)  \n\\
&=& n^{-1/2} \sumi \frac{\partial \S\eff^\star\{y_i, w_i, \delta_i,
  \z_i; \bb_0, E_1(\cdot; S_{C|Y,\Z}, f_{\Delta, W,Y|\Z}), \ba_0^\star\}}{\partial S_{C|Y,\Z}}
(\wh S_{C|Y,\Z} - S_{C|Y,\Z})\n\\
&&+n^{-1/2} \sumi \frac{\partial \S\eff^\star\{y_i, w_i, \delta_i,
  \z_i; \bb_0, E_1(\cdot; S_{C|Y,\Z}, f_{\Delta, W,Y|\Z}),
  \ba_0^\star\}}{\partial f_{\Delta, W|Y,\Z}}(\wh f_{\Delta, W,Y|\Z} - f_{\Delta, W,Y|\Z})\n\\
&&+n^{-1/2} \sumi \frac{\partial \S\eff^\star(y_i, w_i, \delta_i,
  \z_i; \bb_0, E_{10}, \ba_0^\star)}{\partial \ba}
(\wh \ba^\star - \ba_0^\star)\n\\
&&+ n^{-1/2} \sumi \frac{d \S\eff^\star(y_i, w_i, \delta_i,
  \z_i; \bb_0, E_{10}, \ba_0^\star)}{d \bb \trans}
(\wh\bb^\star - \bb_0)  \n\\
&& + n^{-1/2}  \sumi \S\eff^\star(y_i, w_i, \delta_i,
\z_i; \bb_0, E_{10}, \ba_0^\star)+ o_p(n^{1/2}\|\wh\bb ^\star - \bb_0\|_2) + o_p(1)\n\\
&=&n^{-1/2} \sumi \{\S\eff^\star(y_i, w_i, \delta_i,
\z_i; \bb_0, E_{10}, \ba_0^\star) + \bh_{1{\rm s}}^\star(y_i, w_i, \delta_i, \z_i) + \bh_{1{\rm k}}^\star(y_i, w_i, \delta_i, \z_i)\}\n\\
&&\qquad +\{ \bB^\star  + o_p(1)\} n^{1/2}(\wh\bb^\star - \bb_0) +o_p(1), 
\ee
where the second equality follows from Taylor's theorem combined with Lemmas \ref{lem:l3} and \ref{lem:l4}, and the third equality follows from Lemmas \ref{lem:l5} and \ref{lem:l6} and the weak law of large numbers. Since $\bB^\star$ is invertible under Condition \ref{con:nBinv2}, the continuous mapping theorem gives $\{ \bB^\star  + o_p(1)\}^{-1} =  \bB^{\star-1} + o_p(1)$. By the central limit theorem, the asymptotic distribution of $\wh\bb^\star$ is
\bse
&&n^{1/2}(\wh\bb^\star - \bb_0)\\
&=& -\bB^{\star-1}n^{-1/2} \sumi \{\S\eff^\star(y_i, w_i, \delta_i,
\z_i; \bb_0, E_{10}, \ba_0^\star) + \bh_{1{\rm s}}^\star(y_i, w_i, \delta_i, \z_i)+\bh_{1{\rm k}}^\star(y_i, w_i, \delta_i, \z_i)\} +o_p(1)\\
&\stackrel{d}{\to}& \Normal\{\0,\bB^{\star-1}\bSigma_1 (\bB^{\star-1})\trans\}.
\ese

For case \ref{th2case2}, we have 
\be\label{eq:n34}
\0 &=& n^{-1/2} \sumi \S\eff^*(y_i, w_i, \delta_i,
\z_i; \wh\bb^*, E_1^*, \wh \ba^*)  \n\\ 
&=& n^{-1/2} \sumi \frac{\partial \S\eff^*(y_i, w_i, \delta_i,
  \z_i; \bb_0, E_1^*, \ba_0^*)}{\partial \ba} (\wh \ba^* - \ba_0^*)\n\\
  &&+ n^{-1/2} \sumi \frac{d \S\eff^*(y_i, w_i, \delta_i,
  \z_i; \bb_0, E_1^*, \ba_0^*)}{d \bb \trans}
(\wh\bb^* - \bb_0) \n\\ 
&& + n^{-1/2}  \sumi \S\eff^*(y_i, w_i, \delta_i,
\z_i; \bb_0, E_1^*, \ba_0^*) + o_p(n^{1/2}\|\wh\bb^* - \bb_0\|_2) + o_p(1) \n\\ 
&=& n^{-1/2} \sumi \{\S\eff^*(y_i, w_i, \delta_i,
\z_i; \bb_0, E_1^*, \ba_0^*)+\bh_{2{\rm s}}^*(y_i, w_i, \delta_i, \z_i) + \bh_{2{\rm k}}^*(y_i, w_i, \delta_i, \z_i) \}\n\\ 
&&\qquad +\{ \bB^*  + o_p(1)\}n^{1/2} 
(\wh\bb^* - \bb_0)+ o_p(1), 
\ee
where the second equality follows from Taylor's theorem combined with Lemmas \ref{lem:l3} and \ref{lem:l4}, and the third equality follows from Lemma \ref{lem:l7} and the weak law of large numbers. By the central limit theorem, $\wh\bb^*$ has the asymptotic distribution
\bse
&&n^{1/2}(\wh\bb^* - \bb_0)\\
&=& -\bB^{*-1}n^{-1/2} \sumi \{\S\eff^*(y_i, w_i, \delta_i,
\z_i; \bb_0, E_1^*, \ba_0^*)+\bh_{2{\rm s}}^*(y_i, w_i, \delta_i, \z_i)+\bh_{2{\rm k}}^*(y_i, w_i, \delta_i, \z_i)\} +o_p(1)\\
&\to& N\{\0,\bB^{*-1}\bSigma_2 (\bB^{*-1})\trans\}.
\ese 
Lastly, for case \ref{th2case3}, Taylor's theorem gives that
\bse
&&n^{-1/2} \sumi \frac{\partial \S\eff(y_i, w_i, \delta_i,
  \z_i; \bb_0, E_{10}, \ba_0)}{\partial \ba}
(\wh \ba - \ba_0)\\
&=&n^{-1/2} \sumi \frac{\partial \S\eff(y_i, w_i, \delta_i,
  \z_i; \bb_0, E_{10}, \ba_0)}{\partial \ba}\{\ba(\bb_0, \wh E_1, \wh E_2) - \ba(\bb_0, E_{10}, E_{20})\}\\
&=&n^{-1/2} \sumi \left[\frac{\partial \S\eff(y_i, w_i, \delta_i,
  \z_i; \bb_0, E_{10}, \ba_0)}{\partial \ba}\{\ba(\bb_0, \wh E_1, E_{20}) - \ba(\bb_0, E_{10}, E_{20})\}\right.\\
  &&\left.\frac{\partial \S\eff(y_i, w_i, \delta_i,
  \z_i; \bb_0, E_{10}, \ba_0)}{\partial \ba}\{\ba(\bb_0,  E_{10}, \wh E_2) - \ba(\bb_0, E_{10}, E_{20})\} + o_p(n^{-1/2})\right]\\
  &=&n^{-1/2} \sum_{j=1}^n \{\bh_{1{\rm s}2}(y_j,w_j, \delta_j, \z_j) + \bh_{1{\rm k}2}(y_j,w_j, \delta_j, \z_j) + \bh_{2{\rm s}}(y_j,w_j, \delta_j, \z_j) + \bh_{2{\rm k}}(y_j,w_j, \delta_j, \z_j)\} + o_p(1)\\
  &=& - n^{-1/2} \sum_{j=1}^n \{\bh_{1{\rm s}1}(y_j,w_j, \delta_j, \z_j) + \bh_{1{\rm k}1}(y_j,w_j, \delta_j, \z_j)\} + o_p(1),
\ese
where the last equality holds by  Proposition \ref{pro:5}. Then the same
argument as in \eqref{eq:n33} and \eqref{eq:n34} leads to 
\bse
\0 &=& n^{-1/2} \sumi \S\eff(y_i, w_i, \delta_i,
\z_i; \wh\bb, \wh E_1, \wh \ba)  \\ 
&=& n^{-1/2} \sumi \frac{\partial \S\eff\{y_i, w_i, \delta_i,
  \z_i; \bb_0, E_1(\cdot; S_{C|Y,\Z}, f_{\Delta, W,Y|\Z}), \ba_0\}}{\partial S_{C|Y,\Z}}
(\wh S_{C|Y,\Z} - S_{C|Y,\Z})\\
&&+n^{-1/2} \sumi \frac{\partial \S\eff\{y_i, w_i, \delta_i,
  \z_i; \bb_0, E_1(\cdot; S_{C|Y,\Z}, f_{\Delta, W,Y|\Z}), \ba_0\}}{\partial f_{\Delta, W,Y|\Z}}(\wh f_{\Delta, W,Y|\Z} - f_{\Delta, W,Y|\Z})\\
&&+n^{-1/2} \sumi \frac{\partial \S\eff(y_i, w_i, \delta_i,
  \z_i; \bb_0, E_{10}, \ba_0)}{\partial \ba}
(\wh \ba - \ba_0)\\
&&+ n^{-1/2} \sumi \frac{d \S\eff(y_i, w_i, \delta_i,
  \z_i; \bb_0, E_{10}, \ba_0)}{d \bb \trans}
(\wh\bb - \bb_0)  \\ 
&& + n^{-1/2}  \sumi \S\eff(y_i, w_i, \delta_i,
\z_i; \bb_0, E_{10}, \ba_0)+ o_p(n^{1/2}\|\wh\bb - \bb_0\|_2) + o_p(1)\\
&=& n^{-1/2} \sumi  \S\eff(y_i, w_i, \delta_i,
\z_i; \bb_0, E_{10}, \ba_0)+\{ \bB_0  + o_p(1)\} n^{1/2}(\wh\bb - \bb_0) +o_p(1),
\ese
where 
  \bse
\bB_0 &\equiv& E\{d \S\eff (Y,W,\Delta, \Z ; \bb_0)/d \bb \trans\} \\
&=& -E\{\S\eff (Y,W,\Delta, \Z ; \bb_0)\S_\bb\trans (Y,W,\Delta, \Z ; \bb_0)\}\\
&=& -E\{\S\eff^{\otimes 2}  (Y,W,\Delta, \Z ; \bb_0)\}.
\ese
Under Condition \ref{con:nBinv2}, $\{\bB_0 + o_p(1)\}^{-1} =
\bB_0^{-1} + o_p(1)$ by the continuous mapping theorem. Then by the
central limit theorem, the asymptotic distribution of $\wh\bb$ is
\bse
&&n^{1/2}(\wh\bb - \bb_0)\\
&=& - \bB_0^{-1}n^{-1/2} \sumi \S\eff(y_i, w_i, \delta_i,
\z_i; \bb_0, E_{10}, \ba_0) +o_p(1)\\
&\stackrel{d}{\to}& \Normal(0, [E\{\S\eff^{\otimes 2} (Y,W,\Delta, \Z;\bb_0)\}]^{-1}).
\ese
\qed
}


\end{document}